\documentclass[12pt]{article}
\usepackage[dvips]{color,graphicx}
\usepackage{hyperref}
\usepackage{amsmath,amssymb}
\usepackage{epsfig}
\usepackage[english]{babel}
\usepackage{pgf,pgfarrows,pgfnodes,pgfautomata,pgfheaps}
\usepackage{amsfonts,amsmath,amsthm,amssymb,enumerate,array,amssymb}
\usepackage[latin1]{inputenc}
\usepackage{bm}
\usepackage{fixmath}
\usepackage{amsbsy}
\usepackage{graphicx}
\usepackage{capt-of}
\usepackage{dcolumn}
\usepackage{yfonts}

\newcommand{\be}{\begin{equation}}
\newcommand{\ee}{\end{equation}}
\newcommand{\bea}{\begin{eqnarray}}
\newcommand{\eea}{\end{eqnarray}}

\begin{document}

\begin{titlepage}

\begin{flushright}
TCC-008-13\\
UTTG-11-13
\end{flushright}

\vspace{0.3cm}
\begin{center} \Large \bf Anisotropic Models for Globular Clusters,\\
Galactic Bulges and Dark Halos
\end{center}
\vspace{0.1cm}
% \vskip 0.1cm
\begin{center}
P. H. Nguyen$^{\star}$\footnote{phn229@physics.utexas.edu}
and J. F. Pedraza$^{\dagger}$\footnote{jpedraza@physics.utexas.edu}

\vspace{0.2cm}
$^{\star}$Center for Relativity, Department of Physics and Texas Cosmology Center,\\
The University of Texas, Austin, Texas 78712, USA\\
 \vspace{0.2cm}
$^{\dagger}\,$Theory Group, Department of Physics and Texas Cosmology Center,\\
The University of Texas, Austin, Texas 78712, USA\\
 \vspace{0.2cm}
\end{center}

\begin{center}
{\bf Abstract}
\end{center}
\noindent
Spherical systems with polytropic equations of state are of great interest in astrophysics. They are widely used to describe neutron stars, red giants, white dwarfs, brown dwarfs, main sequence stars, galactic halos, and globular clusters of diverse sizes. In this paper we construct analytically a family of self-gravitating spherical models in the post-Newtonian approximation of general relativity. These models present interesting cusps in their density profiles which are appropriate for the modeling of galaxies and dark matter halos. The systems described here are anisotropic in the sense that their equiprobability surfaces in velocity space are nonspherical, leading to an overabundance of radial or circular orbits, depending on the parameters of the model under consideration. Among the family of models, we find the post-Newtonian generalization of the Plummer and Hernquist models. A close inspection of their equation of state reveals that these solutions interpolate smoothly between a polytropic sphere in the asymptotic region and an inner core that resembles an isothermal sphere. Finally, we study the thermodynamics of these models and argue for their stability.

\vspace{0.2in}
\smallskip
\end{titlepage}
\setcounter{footnote}{0}

\tableofcontents

\section{Introduction}

The study of many-body astrophysical systems has been an important issue in relativistic
astrophysics for the past decades. In general, the number of constituents of such systems is enormous and it is neither practical nor worthwhile to follow the interactions and evolution of each particle in the ensemble. From the statistical point of view, most of the qualitative properties of the system can be inferred from the distribution $F(\vec{x}, \vec{v}, t)$, a quantity that determines the probability of finding a single particle in a phase-space volume $d^3xd^3v$ around the position $\vec{x}$ and the velocity $\vec{v}$.

The distribution function (or DF for short) is dynamical. It is governed by
an appropriate kinetic equation, and it in turn determines the statistical evolution of the system.
In the framework of the general theory of relativity (GR) the DF is assumed to obey the general relativistic version of the Vlasov equation or the Fokker-Planck equation \cite{kand1,kand2,kand3,kremer,chacon}. The first one is devoted to systems sufficiently smooth, so that they may be considered to be collisionless, whereas the latter one also accounts for gravitational encounters. One can actually consider systems in which a number of particle species can collide and produce different species. This is how the formation of the light elements in the big bang
nucleosynthesis is calculated (see \cite{nucleus} for a review). Even though these interactions might have important effects in the evolution of some astrophysical systems over large time scales, for most of the applications in this paper we will focus on the collisionless case, which is otherwise believed to capture the relevant physics in a wide variety of scenarios.

There are many applications of kinetic theory in relativistic astrophysics. In stellar dynamics, for instance, the systems described are halos, galaxies, or stellar clusters of diverse sizes. In all these cases the ``particles'' of the system are stars. Applications to cosmology consider galaxies or even clusters of galaxies as the basic constituents. The point here is that their
internal structure is irrelevant at cosmological scales so they can be modeled as particles. Finally, applications to the description of compact objects can also be considered. Although in these situations collisions actually play an important role, analytical solutions to the Vlasov equation are of great interest and serve as a useful starting point to develop a perturbative expansion that accounts for the internal interactions.

A number of methods to construct self-consistent stellar models have appeared in the literature over the years \cite{eddi,frick,lynden,hunt93,jiao,PRGfrac,An:2012fw}. A first approach consists in starting with known profiles for the matter distribution and gravitational fields (which can be inferred directly from photometric and kinematic observations). Since the mass density of the system is defined by the
integration of the DF over the velocity space, the problem of finding a DF is that
of solving an integral equation. This is the so-called ``$\rho$ to $F$ approach.'' Conversely, one can start by assuming a general form for the DF following symmetry considerations and a few physically reasonable assumptions. This is known as the ``$F$ to $\rho$ approach'' and is the main tool we shall adopt for the purposes of the present paper.

Now, even though for most systems under consideration Newtonian gravity is assumed to be dominant, general
relativistic corrections might play an important role in their evolution. Studying this issue in the fully relativistic context is challenging. Not only do we face technical difficulties while trying to obtain analytical solutions to the Einstein-Vlasov system, but also the comparison to the Newtonian predictions is very limited (it should also be noted that there exist a few exact solutions to the Einstein equations that become, in the Newtonian limit, well-known astrophysical models, but those solutions are not based on kinetic theory \cite{Fackerell,Buchdahl:1959zz,Glass Mashhoon}). Thus, in order to estimate the effects on the various observables we are interested in, it would be nice to have a framework to include systematically general relativistic corrections to a given Newtonian model.
The post-Newtonian approximation is perfectly suited for this
purpose. The appropriate scheme that describes the effects of the first
corrections beyond the Newtonian theory was first developed in \cite{eins1, eins2, eins3}, and it is
known as the first post-Newtonian (1PN) approximation. This approach holds if the particles in the system are moving nonrelativistically ($\bar{v}\ll c$), and it gives the corrections up to order $\bar{v}^{2}/c^{2}$, where $\bar{v}$ is a typical velocity in the system and $c$ is
the speed of light. Currently, higher order PN approximations have appeared in the literature
because of the increasing interest regarding the kinematics and associated emission of
gravitational waves by binary pulsars, neutron stars, and black holes, with
promising candidates for detectors such as LIGO, VIRGO, and GEO600 (see
\cite{futamase, bichak} and the references therein).

In a recent series of papers, the first steps towards this objective were given in \cite{Agon:2011mz}, obtaining a version of the Vlasov equation that accounts for the first general relativistic corrections. With this tool at hand, the authors obtained the 1PN version of the Eddington polytropes, starting from an ergodic DF
proportional to $E^{n}$. Further applications to galactic dynamics were presented in \cite{RamosCaro:2012rz}.
 The purpose of this paper is to implement the techniques developed in \cite{Agon:2011mz,RamosCaro:2012rz} to study the influence of relativistic corrections in different astrophysical scenarios. In particular, we will center our attention on the study of spherically symmetric systems with local anisotropy,
which seems to be very reasonable for describing the matter distribution under many circumstances, and has been proven to
be very useful in the study of relativistic systems. To date, we have encountered a large body of work with such applications in the literature \cite{Bowers,Bayin:1982vw,Singh:1992zz,Coley:1994pk,Singh:1995zz,Martinez:1996jq,Das:1997xz,Herrera Santos 1997,Herrera:1997si, Corchero:1997tc,Das:2000ym,Dev:2000gt,Herrera:2001vg,Hernandez:2001nr,Mak:2001eb,Krisch:2001ay,Herrera:2002bm,Dev:2003qd,Herrera:2004xc,Chaisi:2005rb,Naidu:2005pj,Maharaj:2005vb,Chaisi:2006sc,Barreto:2006cr,Herrera:2007kz,Herrera:2008bt,Lake:2009cd,Sharma:2012vc,Mimoso:2013iga,Herrera:2013dfa,Sharma:2013fsa,Sgro:2013tia}.

By local anisotropy, we mean systems whose DFs depend on the phase space coordinates not only through the energy (as in the polytropic case above), but also through the angular momentum. On the level of the orbit of individual stars, this translates into a bias of the system towards either circular orbits or more elongated ones. Such a bias is captured by the so-called velocity dispersion tensor, on which we will elaborate later, and is an observationally measurable quantity, besides the mass density profile. In this way, building models with local anisotropy can help us narrow down the range of possible DFs that give rise to a given density profile. Another good reason to study anisotropic systems is that the anisotropy in velocity space leads to a pressure anisotropy, which is believed to play an important role in the physics of gravitational collapse. Moreover, this pressure anisotropy may have a destabilizing effect on the system, resulting in the system evolving away from a spherically symmetric configuration. This may yield insights into the fate of self-gravitating systems for very large time scales, a subject about which little is currently known. Finally, let us mention that, for the sake of simplicity, we will only consider in this paper models with constant anisotropy. While such models are not particularly realistic, we will see that the relativistic corrections for them are analytically tractable. Also, such models can be considered as building blocks for more realistic anisotropic systems where the anisotropy varies from one part of the system to another.

The rest of the paper is organized as follows. In Sec. \ref{1pnSec} we start by reviewing the main entries of the 1PN approximation of general relativity and introducing the Einstein-Vlasov system. Then, in Sec. \ref{PolySec}, we write down a set of self-gravitation equations for the so-called generalized polytropes, which are one of the simplest anisotropic generalizations of the polytropic DFs. Sections \ref{NewtSec} and \ref{PNewtSec} are devoted to obtaining particular solutions for a family of models in the Newtonian and post-Newtonian limits, respectively. Along the way, we study in detail the main properties of the models. In particular, we study the equation of state for the Newtonian model (assuming a barotropic form) and its possible implications for stability. In the 1PN regime, we learn that the relativistic correction results in a less centrally dominated mass density profile, which in turns implies a slightly flatter galactic rotation curve. Finally, in Sec. \ref{FinalSec}, we give a brief summary of our main results and comment briefly on future work.

\section{The 1PN Approximation of General Relativity\label{1pnSec}}

The Newtonian theory of gravity is commonly used to describe a wide range of astrophysical phenomena at different scales, ranging from the celestial mechanics of a few bodies up to the description of star clusters and galaxies which are composed of billions of stars. Remarkably, most of the observations agree with a very good precision with the theory, and it is just when it comes to very precise measurements that small deviations from the Newtonian description start to appear.

Newtonian dynamics is therefore strongly expected to define an excellent starting point for an approximation scheme of general relativity,
in situations for which the velocities of the bodies are small compared to the velocity of light $(v^2 \ll c^2)$. The post-Newtonian approximation is a systematic expansion that accounts for the first general relativistic corrections over the Newtonian dynamics. It has been carefully reviewed in a number of references (for a textbook analysis see for instance \cite{WB}), but for completeness we will devote the present section to provide the reader with the basic definitions.

\subsection{The field equations}

Consider a system that is bounded by the gravitational interactions of its constituents, and let $\bar M$, $\bar v$ and $\bar r$ be the typical values of the mass, velocity and separation. In Newtonian mechanics the typical kinetic energy $\bar M \bar v^2/2$ is roughly of the same order of magnitude as the typical potential energy $G \bar M^2/\bar r$, so
\be
\bar{v}^2\sim\,\frac{G\bar M}{\bar r}.
\ee
The idea of the first post-Newtonian approximation is to express all physical quantities in terms of a series expansion of $\bar{v}/c\ll1$
and keep only the first order beyond the Newtonian theory. Although it is sometimes referred to as an expansion in inverse powers of the speed of light, it is a good idea to keep track of dimensions and perform the expansion in terms of a dimensionless parameter.

Consider for example the metric tensor. The expansion for $g_{\mu\nu}$ reads
\bea
g_{00}&=&\,\!^0\!g_{00}+\,\!^2\!g_{00}
+\,\!^4\!g_{00}+\cdots,\nonumber\\
g_{ij}&=&\,\!^0\!g_{ij}+ \,\!^2\!g_{ij} +\,\!^4\!g_{ij}+\cdots,\\
g_{0i}&=& \,\!^1\!g_{0i} +\,\!^3\!g_{0i} +\,\!^5\!g_{0i}+\cdots,\nonumber
\eea
where the symbol $\,\!^n\!g_{\mu\nu}$ refers to the term of order $({\bar v}/c)^n$ in the expansion of $g_{\mu\nu}$. In our conventions $x^\mu=(ct,x^i)$ so, for the computation of the line element $ds^{2}=g_{\mu\nu}dx^{\mu}dx^{\nu}$, we have to keep in mind that temporal indices carry an extra power of the speed of light. Odd powers of ${\bar v}/c$ appear in $g_{0i}$ because these components must change sign under a time-reversal transformation $t\rightarrow -t$.

Without loss of generality, we can say that at zeroth order $\,\!^0\!g_{00}=-1$ and $\,\!^0\!g_{ij}=\delta_{ij}$, reflecting the fact that any manifold is locally flat. By working in harmonic coordinates (i.e., coordinates such that $g^{\mu\nu}\mathrm{\Gamma}^\lambda_{\mu\nu}=0$) we can further simplify the above expansions by writing them as a function of the Newtonian potential $\phi$ and two new post-Newtonian potentials $\psi$ and $\vec{\xi}$ defined as in \cite{WB}. To our order of approximation we get
\be
\begin{split}\label{potentials}
\,\!^2\!g_{00}&=\,-2\phi/c^2,\\
\,\!^4\!g_{00}&=\,-2(\phi^2+\psi)/c^4,\\
\,\!^2\!g_{ij}&=\,-2\phi\delta_{ij}/c^2,\\
\,\!^1\!g_{0i}&=\,0,\\
\,\!^3\!g_{0i}&=\,\xi_i/c^3.
\end{split}
\ee
At 1PN order then, the line element can be written as
\begin{equation}
ds^2=-\left(1+\frac{2\phi}{c^2}+\frac{2(\phi^2+\psi)}{c^4}\right)c^2dt^2+2\left(\frac{\xi_i}{c^3}\right)cdtdx^i+
\left(1-\frac{2\phi}{c^2}\right)\delta_{ij}dx^idx^j.
\end{equation}

It is convenient to assume a similar expansion for the components of the energy-momentum tensor. In particular,
from their interpretation of energy density, momentum flux, and energy flux, we expect that\footnote{A comment regarding the units is in order here. In our conventions, the energy-momentum tensor is normalized such that $\,\!^0T^{00}=\rho$, $\,\!^2T^{ij}=\delta_{ij}p_i/c^2$, and so on.}
\bea\label{tmnexpa}
T^{00}&=&\,\!^0T^{00} +\,\!^2T^{00}+\cdots,\nonumber\\
T^{ij}&=&\,\!^2T^{ij} +\,\!^4T^{ij}+\cdots,\\
T^{0i}&=&\,\!^1T^{0i}+\,\!^3T^{0i} +\cdots.\nonumber
\eea

The above expressions lead to a consistent expansion of the Einstein field equations. At 1PN order these can be written as
\bea
\nabla^2 \phi &=& 4\pi G \,\,\!^0T^{00}, \label{ec_campo_pn1} \\
\nabla^2 \psi &=& 4\pi G c^{2}\,(\,\!^2T^{00}
+ \;\,\!^2T^{ii}) + \frac{\partial^2 \phi}{\partial t^2}, \label{ec_campo_pn2} \\
\nabla^2 \xi_i &=& 16 \pi G c \,\,\!^1T^{0i}, \label{ec_campo_pn3}
\eea
along with the coordinate condition
\begin{equation}\label{condition-phi-xi}
    4\frac{\partial\phi}{\partial t}+\nabla\cdot\vec{\xi}=0.
\end{equation}

\subsection{Particle motion and the geodesic equation}
The post-Newtonian approximation was first developed to study the problem of motion in celestial mechanics. Among other things, it gives a correct estimation for the perihelion precession of Mercury \cite{Ein}, a crucial fact that motivated the adoption of general relativity. Here, we will review a few key points that we need for the remaining part of the paper.

Consider the action for a free-falling particle,
\be
\mathcal{S}=\int d\tau\sqrt{-g_{\mu\nu}U^\mu U^\nu}\,.
\ee
The symbol $U^\mu=\partial x^\mu/\partial\tau$ represents the particle's four-velocity which is related to the three-velocity by $U^i=U^0 v^{i}/c$ (Greek indices run from 0 to 3 and Latin indices from 1 to 3). Although we are free to choose an arbitrary affine parametrization, in the expression above $\tau$ denotes the particle's proper time. For a timelike geodesic there is an additional constraint, namely the normalization of the four-velocity $g_{\mu\nu}U^{\mu} U^{\nu}=-c^2$.

The Euler-Lagrange equation for this system leads to the geodesic equation. At 1PN order and for general potentials $\phi$, $\psi$, and $\vec{\xi}$ one finds that
the free-falling particle obeys the equation
\be
\frac{d \vec{v}}{d t}=-\nabla\phi-\frac{1}{c^{2}}\left[\nabla\left(2\phi^{2}+\psi\right)+\frac{\partial \vec{\xi}}{\partial t}-\vec{v}
\times(\nabla\times\vec{\xi})-3\vec{v}\frac{\partial\phi}{\partial t}-4\vec{v}(\vec{v}\cdot\nabla\phi)
+v^{2}\nabla\phi\right],\label{ecmovimiento1PN}
\ee
which partially resembles the equation of motion for a charged particle with velocity $\vec{v}$ in the presence of electromagnetic fields. This law of
motion will determine, for instance, the rotation curve for test particles moving in equatorial circular geodesics (see \cite{RamosCaro:2012rz} for details) \footnote{Here $\{R,\varphi,z\}$ are cylindrical coordinates.},
\be\label{rotlaw}
v_c^2=
R\frac{\partial\phi}{\partial R}\left(1+\frac{4\phi}{c^{2}}+\frac{R}{c^{2}}
\frac{\partial\phi}{\partial R}\right)+\frac{R}{c^{2}}\left(\frac{\partial\psi}{\partial R}
-\sqrt{R\frac{\partial\phi}{\partial R}}\frac{\partial\xi_{\varphi}}{\partial R}\right)\Bigg|_{z=0}\,.
\ee
There is an important difference between the above relation and the usual formula for the rotation curves: in the Newtonian
case $v^2_c$ is linear in $\partial \phi/\partial R$, whereas in the 1PN case, it depends on nonlinear terms involving the potentials and their derivatives. The corrections introduced by these nonlinearities are found to be significant in some cases, especially for large radial distances,
which is surprising given that one would expect major corrections near the center, where the
mass concentration is maximum. For example, the authors of \cite{Balasin:2006cg} presented
a model in which the percentage of dark matter needed to explain flat rotation curves is around $\sim30\%$ less than that required by the Newtonian theory.

Likewise, we can get the Lagrangian by expanding $d\tau/dt$ at 1PN order. Following Weinberg \cite{WB}, we get (after some algebra)
\begin{equation}\label{lag-wein}
\mathfrak{L}=\frac{v^2}{2}-
\phi-\frac{1}{c^2}\left(\frac{\phi^2}{2}+\frac{3\phi v^2}{2}-\frac{v^4}{8}+\psi-\vec{v}\cdot\vec{\xi}\right).
\end{equation}
Given the symmetries of the problem, we can easily derive the various integrals of motion. For the purposes of the present paper we need two in particular: the energy and the angular momentum. For static spacetimes, $\xi=0$ and the potentials $\phi$ and $\psi$ are independent of time. The Hamiltonian $\mathcal{H}=\sum_i\dot{x}_i\partial\mathfrak{L}/\partial v_i-\mathfrak{L}$ is then a conserved quantity,
\be\label{ec_ene}
\mathcal{H} = \frac{v^2}{2} + \phi + \frac{1}{c^2}\left(\frac{3v^{4}}{8}-\frac{3v^{2}\phi}{2}+\frac{\phi^2}{2}+\psi\right)=E\,,
\ee
and this can be regarded as the 1PN generalization of the energy. If we further constrain further the problem by assuming spherical symmetry,
the fields $\phi$ and $\psi$ will depend on the spatial coordinates only through
$r=\sqrt{x^{2}+y^{2}+z^{2}}$, and in this case one finds that the quantities
\begin{equation}\label{momento-angular}
    L_{i}=\varepsilon_{ijk}x^{j}v^{k}\left(1-\frac{3\phi}{c^{2}}+\frac{\mathbf{v}^{2}}{2c^{2}}\right)
\end{equation}
are also integrals of motion. These are the components of the angular momentum at 1PN order.

\subsection{The Einstein-Vlasov system}

In the kinetic theory of gases, the Vlasov equation arises as an effective description of a system composed of many particles in the regime when their interactions are negligible. In particular, no collisions are included
in the model, and each particle is acted on only by smooth fields generated collectively by all the particles in the ensemble. When, in addition, the system is coupled to general relativity, then the resulting set of equations is known as the Einstein-Vlasov system (for a review on the subject, see for example \cite{Rendall:1996gx,Rendall:2002xa,Andreasson:2011ng}).

In the framework of kinetic theory, the state of the system is described from a statistical point of view. The starting point is the DF\footnote{When the number of particles is $N\gg1$, then the $N$-body distribution function is separable in the absence of collisions. Here we are dealing with the one-particle DF, which is the relevant quantity in such situations.}, $F(\vec x,\vec v,t)$, which depends on the spatial coordinates, velocity and time. For the applications we want to consider here, we will require that the DF of the system satisfies the general relativistic version of the Vlasov equation,
\begin{equation}\label{liouville}
{\cal L}_U F = \left(U^\mu \frac{\partial}{\partial x^\mu} - \mathrm{\Gamma}_{\mu\nu}^i U^\mu
U^\nu \frac{\partial}{\partial U^i}\right) F(x^\mu,U^i) = 0,
\end{equation}
where $(x^\mu,U^i)$ is the set of configuration and four-velocity coordinates\footnote{The four-velocity is normalized such that $g_{\mu\nu}U^\mu U^\nu=-c^2$, so it is enough to consider dependence on the spatial components $U^i$.}, $\mathrm{\Gamma}^i_{\mu\nu}$ are the Christoffel symbols and ${\cal L}_U$ is the Liouville operator. In the 1PN approximation, the above equation can be written as (see \cite{Agon:2011mz} for details)
\begin{equation}\label{ec_general}
\begin{split}
\qquad\qquad\frac{\partial F}{\partial t}+v^{i}\frac{\partial F}{\partial x^{i}}-
\frac{\partial \phi}{\partial x^{i}}\frac{\partial F}{\partial v^{i}}+
\frac{1}{c^2}\left(\frac{v^{2}}{2}-\phi \right)
\left(\frac{\partial F}{\partial t}+v^{i}\frac{\partial F}{\partial x^{i}}\right)\qquad\qquad\qquad\qquad\\\\
+\frac{1}{c^{2}}\left[4v^{i} v^{j}\frac{\partial \phi}{\partial x^{j}}-
\left(\frac{3v^{2}}{2}+3\phi\right)\frac{\partial \phi}{\partial x^{i}}
-v^{j}\left(\frac{\partial \xi_{i}}{\partial x^{j}}-\frac{\partial \xi_{j}}{\partial
  x^{i}}\right)+
3v^{i}\frac{\partial \phi}{\partial t}-\frac{\partial \psi}{\partial x^{i}}-\frac{\partial \xi_{i}}{\partial t}
\right]\frac{\partial F}{\partial v^i}=0.
\end{split}
\end{equation}
We have to emphasise that this expression is only valid in the collisionless regime. For cases in which encounters play a dominant role, the right-hand side of (\ref{ec_general}) must be replaced by a term of the Fokker-Planck type \cite{RamosCaro:2008zz}.

Similar to the classical case, the 1PN equation
can be expressed in various ways, each of
which is useful in different contexts \cite{Agon:2011mz}. The one that is relevant for us is in terms of Poisson brackets,
\be
\frac{\partial F}{\partial t}+\{F,\mathcal{H}\}=0,\label{form2}
\ee
where $\mathcal{H}$ is the 1PN Hamiltonian (\ref{ec_ene}). Since all integrals of motion
commute with $\mathcal{H}$, this implies that Jeans theorem \cite{jeans} is valid at 1PN order. That is, any
static solution of the Vlasov equation depends only on the integrals of
motion of the system, and any function of the integrals
yields a static solution of the Vlasov equation \footnote{It was an open question for some time whether or not Jeans theorem holds in the fully relativistic case. This issue was settled in 1999 by Schaeffer in the negative \cite{Schaeffer}, who found particular solutions that cannot be written as a function of the integrals of motion.}.

The second moment of the DF is the energy-momentum tensor. Via the Einstein equations, this establishes a connection between $F$ and the potentials $\phi$, $\psi$, and $\vec\xi$ that can be summarized as a set of coupled differential equations. These are known as the 1PN self-gravitation equations \cite{Agon:2011mz,RamosCaro:2012rz}.

The starting point is the general relativistic expression for the energy-momentum tensor,
\begin{equation}\label{tmunu}
T^{\mu\nu}(x^i,t)=\frac{1}{c}\int\frac{ U^\mu U^\nu}{U^0}F(x^i,U^i,t)\sqrt{-g}d^3U.
\end{equation}
Expanding to the various orders required by the 1PN approximation we get
\begin{eqnarray}
\,\!^0T^{00} &=&\int \,\!^0\!F d^{3}v, \label{T000} \\
\,\!^2T^{00}&=& \frac{3}{c^2}\int (v^{2}
-2\phi) \,\,\!^0\!F d^{3}v+\int \,\!^2\!F d^{3}v, \label{T002}  \\
\,\!^1T^{0i} &=& \frac{1}{c}\int v^{i} \,\,\!^0\!F d^{3}v, \label{T0j1} \\
\,\!^2T^{ij} &=& \frac{1}{c^2}\int v^{i} v^{j}\,\,\!^0\!Fd^{3}v, \label{Tij2}
\end{eqnarray}
along with $\,\!^0T_{ij}=0$, as expected. Finally, substituting in (\ref{ec_campo_pn1})-(\ref{ec_campo_pn3}), one gets the 1PN self-gravitation equations,
\begin{eqnarray}
\nabla^2 \phi &=&4\pi G\int \,\!^0\!F d^{3}v, \label{selfgrav-1}  \\
\nabla^2 \psi &=& 8\pi G\int(2v^{2}-3\phi)\,\,\!^0\!Fd^{3}v+4\pi G c^{2}\int\,\!^2\!Fd^{3}v, \label{selfgrav-2}\\
\nabla^2 \xi_{i} &=& 16\pi G\int v^{i}\,\,\!^0\!Fd^{3}v.\label{selfgrav-3}
\end{eqnarray}

Thus, any system characterized by an equilibrium DF can be written as a function of the integrals of motion. This automatically implies that the DF is a solution to the kinetic equation (\ref{ec_general}). The energy-momentum tensor can be obtained through (\ref{T000})-(\ref{Tij2}), which acts as a source of the gravitational field according to the field equations (\ref{ec_campo_pn1})-(\ref{ec_campo_pn3}). In order to have a self-consistent description, the relations (\ref{selfgrav-1})-(\ref{selfgrav-3}) must be satisfied. All of these equations are written as power expansions in $\bar{v}/c$ and, as a consequence, we can clearly distinguish between the Newtonian contribution and the post-Newtonian corrections.

\section{Generalized Polytropes\label{PolySec}}

\subsection{Distribution functions}

In astrophysics, there is a well-known family of spherical models
 characterized by ergodic DFs of the form
\begin{equation}\label{df-poli}
F(\mathcal{E}) = \begin{cases} k_{n}\mathcal{E}^{n-3/2} & \mbox{for }\mathcal{E}>0, \\ 0 & \mbox{for }\mathcal{E}\leq0. \end{cases}
\end{equation}
These models are known as ``stellar dynamical polytropes.'' In (\ref{df-poli}), $\mathcal{E}=E_0-E$ is the relative energy \footnote{In practice, $E_0$ is chosen such that $F=0$ for $\mathcal{E}<0$. For an isolated system, $E_0$ is related to the value of the potential at infinity and the
relative energy is equal to the binding energy.}, $k_{n}$ is a constant related to the total mass of the systems, and $n$ is the polytropic index. The reason why these models are of particular interest is because they lead to a simple equation of state, $p\propto \rho^{1+1/n}$, that is widely used to describe a variety of astrophysical systems. Among them are neutron stars, red giants, white dwarfs, brown dwarfs,
main sequence stars, galactic halos, globular clusters of diverse size, galaxies and galaxy clusters. A full account of gaseous polytropes can be found in \cite{Chandrasekhar,polybook}.

The post-Newtonian version of the stellar polytropes was consider recently in \cite{Agon:2011mz}. Although most of the work in that paper was numerical, it was clear that the corrections introduced by the relativistic effects can be relevant in the computation of certain observables. Here, we will consider generalizations of these models that are described by a distribution function of the form \cite{henon}
\begin{equation}\label{df-anis}
F(\mathcal{E},L) = \begin{cases} k_{\gamma\delta} L^{2\gamma}\mathcal{E}^{\delta} & \mbox{for }\mathcal{E}>0, \\ 0 & \mbox{for }\mathcal{E}\leq0, \end{cases}
\end{equation}
where $\gamma$ and $\delta$ are constants and $L$ is the magnitude of the angular momentum. These  are known as ``generalized polytropes'' and they are found to be anisotropic in the sense that their equiprobability surfaces in velocity space are nonspherical. This leads to an overabundance of radial or circular orbits depending on the value of the constant $\gamma$. For now we are going to assume arbitrary values for $\gamma$ and $\delta$, but later we will specialize to a subset of models that are analytically solvable.

We begin by splitting the energy and the angular momentum into classical and post-Newtonian contributions, $E=E_{cl}+E_{pn}$ and $L^2=L^2_{cl}+L^2_{pn}$, where
\begin{equation}
E_{cl} = \frac{v^2}{2} + \phi,\qquad E_{pn} = \frac{1}{c^2} (\frac{3}{8}v^4 - \frac{3}{2}v^2\phi + \frac{\phi^2}{2} + \psi),
\end{equation}
and
\begin{equation}
L_{cl}^{2} = (\vec{r} \times \vec{v})^{2},\qquad L_{pn}^{2} = \frac{1}{c^2} [(\vec{r} \times \vec{v})^{2} (v^{2} - 6\phi) + 2 (\vec{r} \times \vec{\xi}) \cdot (\vec{r} \times \vec{v})].
\end{equation}

We also assume that $E_0$ can be split into two contributions, a leading term and a correction of order $(\bar{v}/c)^2$, so that $\mathcal{E}=\mathcal{E}_{cl}+\mathcal{E}_{pn}$. For later convenience, we will write
\be
E_0=\phi_0+\frac{\psi_0}{c^2}\,.
\ee
Making the approximation that the post-Newtonian contributions are much smaller than the classical ones, we get
\be
F =\,\!^0\!F+\,\!^2\!F,
\ee
where $\,\!^0\!F$ and $\,\!^2\!F$ are the zeroth order and second order terms, respectively,
\begin{equation}
\,\!^0\!F = k_{\gamma\delta}L_{cl}^{2\gamma}\mathcal{E}_{cl}^{\delta}\qquad\text{and}\qquad \,\!^2\!F= k_{\gamma\delta}\left( \delta L_{cl}^{2\gamma}\mathcal{E}_{cl}^{\delta-1}\mathcal{E}_{pn} +  \gamma L_{cl}^{2(\gamma-1)}\mathcal{E}_{cl}^{\delta}L_{pn}^{2}\right).
\end{equation}

It is convenient to use spherical coordinates in velocity space (with the z axis pointing in the direction of $\vec{r}$): $v_{r} = v\cos{\eta}$, $v_{\theta} = v \sin{\eta} \cos{\chi}$ and $v_{\phi} = v \sin{\eta} \sin{\chi}$. As usual, $v\in[0,\infty]$ $\eta\in[0,\pi]$ and $\chi\in[0,2\pi]$. In these variables, the angular momentum becomes
\begin{equation}
L_{cl}^{2} = r^{2}v^{2}\sin^{2}{\eta},
\end{equation}
and
\begin{equation}
L_{pn}^{2} = \frac{1}{c^2} \left(r^{2}v^{2}\sin^{2}\eta (v^2-6\phi) + 2r^2 \xi_{\phi} v \sin{\eta} \sin{\chi} + 2r^2 \xi_{\theta} v \sin{\eta} \cos{\chi}\right).
\end{equation}
Finally, substituting into the expressions for $\,\!^0\!F$ and $\,\!^2\!F$ we obtain
\begin{equation}
\,\!^0\!F = k_{\gamma\delta}(rv\sin{\eta})^{2\gamma} (\phi_0-\frac{v^2}{2}- \phi)^{\delta}
\end{equation}
and
\begin{eqnarray}
\,\!^2\!F &\!\!=\!\!& \frac{k_{\gamma\delta}}{c^2}\left( \delta (rv\sin{\eta})^{2\gamma}(\phi_0-\frac{v^2}{2}- \phi)^{\delta-1} (\psi_0-\frac{3}{8}v^4 + \frac{3}{2}v^2\phi - \frac{\phi^2}{2} - \psi)\right.  \nonumber\\
&&\left.\quad\quad\,\,+\gamma(rv\sin{\eta})^{2(\gamma-1)}(\phi_0-\frac{v^2}{2}- \phi)^{\delta}\left(r^{2}v^{2}\sin^{2}\eta (v^2-6\phi) + 2r^2 \xi_{\phi} v \sin{\eta} \sin{\chi} + 2r^2 \xi_{\theta} v \sin{\eta} \cos{\chi}\right)\right)\,.
\end{eqnarray}

\subsection{The energy-momentum tensor and the field equations}

The distribution function (\ref{df-anis}) acts as a source for the energy-momentum distribution according to (\ref{T000})-(\ref{Tij2}). However, note that $\,\!^0\!F$ is even with respect to $v$, so $\,\!^1T^{0i}=0$ and $\,\!^2T^{ij}=0$ for $i\neq j$. In particular, the first relation implies that the vector potential $\xi_{i}$ is sourceless,
\be
\nabla^2 \xi_{i}=0,
\ee
which agrees with the fact that for any static and spherically symmetric system the only physical solution to the coordinate condition (\ref{condition-phi-xi}) is that $\xi_{i}=0$. The remaining components of the energy-momentum tensor are
\begin{equation}
\,\!^0T^{00} = \int_{0}^{v_e} dv \int_{0}^{\pi} d\eta \int_{0}^{2\pi}d\chi\, v^{2}\sin{\eta}\,\,\!^0\!F\,,
\end{equation}
\begin{equation}
\,\!^2T^{00} = \int_{0}^{v_e} dv \int_{0}^{\pi} d\eta \int_{0}^{2\pi}d\chi\, v^{2}\sin{\eta}\left( \left(\frac{3v^2}{c^2} - \frac{6\phi}{c^2}\right)\,\!^0\!F + \,\!^2\!F\right),
\end{equation}
and
\begin{equation}\label{Tii}
\,\!^2T^{ii} = \frac{1}{c^2}\int_{0}^{v_e} dv \int_{0}^{\pi} d\eta \int_{0}^{2\pi}d\chi\, v^{4}\sin{\eta}\,\,\!^0\!F \,.
\end{equation}
Here $v_{e}$ denotes the escape velocity, i.e.
the velocity at which a particle reaches its maximum value of energy, $\mathcal{E}=0$,
so that it is confined to the distribution of matter. Also, repeated indices in Eq. (\ref{Tii}) stand for summation. The escape velocity can be computed from (\ref{ec_ene}) and the result is
\begin{equation}\label{vel-escape}
v_{e}=\sqrt{2(\phi_0-\phi)}+\mathcal{O}\left((\bar{v}/c)^2\right).
\end{equation}
Note that we are keeping just the zeroth order term. This is fine because even when the integrand is $\,\!^0\!F$, the correction due to the escape velocity is proportional to an integral of the form
\begin{equation}\label{intscape}
\int_{\sqrt{-2\mathrm{\Phi}}}^{\sqrt{-2\mathrm{\Phi}}+X/c^{2}} v^{2\gamma+2}\bigg(-\frac{v^2}{2}-\mathrm{\Phi}\bigg)^{\delta} dv\,,
\end{equation}
where $\mathrm{\Phi}=\phi-\phi_0$ and $X$ is a function of $\phi$ and $\psi$ of order $1$. The change of variable $v \rightarrow c^{2}(v-\sqrt{-2\mathrm{\Phi}})$ then reveals that this integral is of order $\mathcal{O}((\bar{v}/c)^{2+2\delta})$ after integration, which is negligible in comparison with the post-Newtonian corrections.

To evaluate these integrals, the following abbreviation is useful:
\begin{equation}\label{intconv}
I(\alpha,\beta) = \int_{0}^{\sqrt{-2\mathrm{\Phi}}} v^{\alpha} \bigg( -\frac{v^2}{2}-\mathrm{\Phi} \bigg)^{\beta} dv\,.
\end{equation}
For $\alpha > -1$ and $\beta > -1$, this evaluates to
\begin{equation}\label{integral}
I(\alpha,\beta) = 2^{\frac{1}{2}(\alpha-1)}(-\mathrm{\Phi})^{\frac{1}{2}(1+\alpha+2\beta)} \frac{\mathrm{\Gamma}(\frac{1+\alpha}{2})\mathrm{\Gamma}(1+\beta)}{\mathrm{\Gamma}(\frac{3+\alpha}{2}+\beta)}\,.
\end{equation}

In terms of the function $I(\alpha,\beta)$, the components of the stress-energy tensor are
\begin{equation}
\,\!^0T^{00} = 2\pi^{3/2} k_{\gamma\delta} r^{2\gamma} \frac{\mathrm{\Gamma}(\gamma+1)}{\mathrm{\Gamma}(\gamma+\frac{3}{2})} I(2+2\gamma,\delta)\,,
\end{equation}
\begin{eqnarray}
\,\!^2T^{00} &\!\!=\!\!& \frac{2 \pi^{3/2} k_{\gamma\delta}}{c^2} r^{2\gamma} \frac{\mathrm{\Gamma}(\gamma+1)}{\mathrm{\Gamma}(\frac{3}{2}+\gamma)} \bigg[(3+\gamma)I(4+2\gamma,\delta) - 6\phi(1+\gamma)I(2+2\gamma,\delta) \\
&&- \frac{3}{8}\delta I(6+2\gamma,\delta-1) + \frac{3}{2}\delta\phi I(4+2\gamma,\delta-1) - \delta(\frac{\phi^2}{2}+\mathrm{\Psi})I(2+2\gamma,\delta-1)\bigg]\,, \nonumber
\end{eqnarray}
and
\begin{equation}
\,\!^2T^{ii} = \frac{2\pi^{3/2} k_{\gamma\delta} r^{2\gamma}}{c^2} \frac{\mathrm{\Gamma}(1+\gamma)}{\mathrm{\Gamma}(\gamma+\frac{3}{2})} I(4+2\gamma,\delta).
\end{equation}
We have also defined $\mathrm{\Psi} = \psi - \psi_{0}$. Using (\ref{integral}), these expressions reduce to
\begin{equation}\label{0T00poly}
\,\!^0T^{00} = 2^{\gamma+\frac{3}{2}}\pi^{3/2}k_{\gamma\delta}\frac{\mathrm{\Gamma}(1+\delta)\mathrm{\Gamma}(1+\gamma)}{\mathrm{\Gamma}(\gamma+\delta+\frac{5}{2})}r^{2\gamma}(-\mathrm{\Phi})^{\gamma+\delta+\frac{3}{2}},
\end{equation}
\begin{eqnarray}\label{2T00}
\,\!^2T^{00} &\!\!=\!\!& \frac{1}{c^{2}}2^{\gamma-\frac{3}{2}}k_{\gamma\delta}\pi^{3/2}r^{2\gamma}(-\mathrm{\Phi})^{\gamma+\delta+\frac{1}{2}}\frac{\mathrm{\Gamma}{(1+\delta)}\mathrm{\Gamma}{(1+\gamma)}}
{\mathrm{\Gamma}{(\gamma+\delta+\frac{7}{2})}}\bigg[\mathrm{\Phi}^{2}(3+2\gamma)(9+2\gamma)+ 6\phi\mathrm{\Phi}(1+2\gamma)(5+2\gamma+2\delta) \nonumber\\
&&\quad\quad\quad\quad\quad\quad- 30 \mathrm{\Psi}- 8(\gamma+\delta)(4+\gamma+\delta)\mathrm{\Psi}- \phi^{2}(3+2\gamma+2\delta)(5+2\gamma+2\delta) \bigg]\,,
\end{eqnarray}
and
\begin{equation}\label{2Tii}
\,\!^2T^{ii} = \frac{1}{c^2}2^{\gamma+\frac{3}{2}}k_{\gamma\delta}\pi^{3/2}(3+2\gamma)r^{2\gamma}(-\mathrm{\Phi})^{\gamma+\delta+\frac{5}{2}}
\frac{\mathrm{\Gamma}{(1+\delta)}\mathrm{\Gamma}{(1+\gamma)}}{\mathrm{\Gamma}(\gamma+\delta+\frac{7}{2})}\,.
\end{equation}
Einstein's field equations (\ref{ec_campo_pn1})-(\ref{ec_campo_pn2}) then take the form
\begin{equation}\label{eins1}
\nabla^{2} \mathrm{\Phi} = 2^{\gamma+\frac{7}{2}}\pi^{5/2}Gk_{\gamma\delta}\frac{\mathrm{\Gamma}(1+\delta)\mathrm{\Gamma}(1+\gamma)}{\mathrm{\Gamma}(\gamma+\delta+\frac{5}{2})}r^{2\gamma}(-\mathrm{\Phi})^{\gamma+\delta+\frac{3}{2}}\,,
\end{equation}
and
\begin{eqnarray}\label{eins2}
\nabla^{2}\mathrm{\Psi} &\!\!=\!\!&2^{\gamma+\frac{1}{2}} \pi^{5/2} G k_{\gamma\delta} \frac{\mathrm{\Gamma}{(1+\delta)}\mathrm{\Gamma}{(1+\gamma)}}
{\mathrm{\Gamma}{(\gamma+\delta+\frac{7}{2})}} r^{2\gamma}(-\mathrm{\Phi})^{\gamma+\delta+\frac{1}{2}}\bigg[\mathrm{\Phi}^{2}(3+2\gamma)(9+2\gamma)+ 6\phi\mathrm{\Phi}(1+2\gamma)(5+2\gamma+2\delta)\\
&&\quad\quad- 30 \mathrm{\Psi}- 8(\gamma+\delta)(4+\gamma+\delta)\mathrm{\Psi}- \phi^{2}(3+2\gamma+2\delta)(5+2\gamma+2\delta) \bigg]\nonumber\\
&&\quad\quad+2^{\gamma+\frac{7}{2}} \pi^{5/2} G k_{\gamma\delta}(3+2\gamma)
\frac{\mathrm{\Gamma}{(1+\delta)}\mathrm{\Gamma}{(1+\gamma)}}{\mathrm{\Gamma}(\gamma+\delta+\frac{7}{2})} r^{2\gamma}(-\mathrm{\Phi})^{\gamma+\delta+\frac{5}{2}}\,.\nonumber
\end{eqnarray}

\section{Newtonian Limit\label{NewtSec}}

\subsection{Solving the field equations at leading order}

Let us start with the equation for the Newtonian potential (\ref{eins1}). Assuming spherical symmetry, the equation for $\mathrm{\Phi}(r)$ becomes
\begin{equation}\label{poissoneq}
\frac{1}{r^2}\frac{d}{dr}\left(r^{2}\frac{d\mathrm{\Phi}}{dr}\right) = \alpha_{\gamma\delta} r^{2\gamma}(-\mathrm{\Phi})^{\gamma+\delta+\frac{3}{2}}\,,
\end{equation}
with
\be
\alpha_{\gamma\delta}=2^{\gamma+\frac{7}{2}}\pi^{5/2}G k_{\gamma\delta}\frac{\mathrm{\Gamma}(1+\delta)\mathrm{\Gamma}(1+\gamma)}{\mathrm{\Gamma}(\gamma+\delta+\frac{5}{2})}\,.
\ee
Now, let $\tilde{\mathrm{\Phi}} = -r\mathrm{\Phi}$. With this change of variables, the above equation reduces to
\begin{equation}\label{eqtilde}
\frac{d^{2}\tilde{\mathrm{\Phi}}}{dr^2} = -\alpha_{\gamma\delta} r^{\gamma-\delta-\frac{1}{2}}\tilde{\mathrm{\Phi}}^{\gamma+\delta+\frac{3}{2}},
\end{equation}
which after some redefinitions takes the general form $y''(x) = A\, x^p\, y^q$. This is known as the Emden-Fowler differential equation. All known solutions are listed, for example, in \cite{Polyanin}, and among them, there are a few one-parameter families and some isolated points (in the space of $p$ and $q$). To have a physically sensible model we have to impose a further constraint, namely the convergence of (\ref{integral}). We thus focus our attention on the family
\be
\gamma=\frac{1}{4}(m-5)\quad\text{and}\quad\delta=\frac{1}{4}(3m-1),
\ee
with $m>-1$. Other physically sound models are discussed in Appendix \ref{appother}. A few comments are in order here. First note that the DF (\ref{df-anis}) becomes
\begin{equation}\label{dfm}
F(\mathcal{E},L) = \begin{cases} k_{m} L^{\frac{1}{2}(m-5)}\mathcal{E}^{\frac{1}{4}(3m-1)} & \mbox{for }\mathcal{E}>0, \\ 0 & \mbox{for }\mathcal{E}\leq0. \end{cases}
\end{equation}
This family of DFs is known to be related to the hypervirial potential-density pairs presented in \cite{Evans:2005zv}. The models all possess the remarkable property that the virial theorem holds locally, from which they earn their name as the hypervirial family. Moreover, it is found that some members present cosmologically interesting cusps at the center and are appropriate for the modeling of galaxies and dark matter halos. Our goal here is to study further properties of these models in the Newtonian theory and to construct their post-Newtonian generalizations in order to investigate the effect of the relativistic corrections.

The only isotropic model in this family corresponds to $m=5$ and it is known as the Plummer model \cite{plummer} \footnote{Another analytic family of anisotropic DFs containing the Plummer model was considered in \cite{Dejonghe87}.}. This is one of the few polytropic models that is analytically solvable \cite{BT}. For $m<5$ the power of the angular momentum is negative. This means that there is a huge probability of finding a particle with small angular momentum, leading to an overabundance of radial orbits. Also, for these models we expect most of the matter distribution to be located near the center of the system. For $m>5$ the situation is exactly the opposite. In this case, the probability of finding a particle in phase space grows with the angular momentum, which would lead to an overabundance of circular orbits. For these models, we do not expect a large mass concentration in the inner region. We will come back to the discussion of these properties in Sec. \ref{dispersions}.

The authors of \cite{Evans:2005zv} used the global properties of the potential-density pairs to infer the corresponding DFs. The purpose of this section is to use the direct method to check their results and to gain some insight on the behavior of the models.  To begin with, note that in terms of $m$ Eq. (\ref{eqtilde}) becomes
\begin{equation}
\frac{d^{2}\tilde{\mathrm{\Phi}}}{dr^2} = -\alpha_{m} r^{-\frac{1}{2}(m+3)}\tilde{\mathrm{\Phi}}^{m}\,.
\end{equation}
From \cite{Polyanin}, the solution is given in parametric form by
\begin{equation}\label{parametric1}
r(\tau) = aC_{2}^{2} \exp{\bigg[2\int\bigg(\frac{8}{m+1}\tau^{m+1}+\tau^{2}+C_{1}\bigg)^{-1/2}d\tau\bigg]}\,,
\end{equation}
\begin{equation}\label{parametric2}
\tilde{\mathrm{\Phi}}(\tau) = bC_{2} \tau \exp{\bigg[\int\bigg(\frac{8}{m+1}\tau^{m+1}+\tau^{2}+C_{1}\bigg)^{-1/2}d\tau\bigg]}\,,
\end{equation}
where $C_{1}$ and $C_{2}$ are integration constants, and $a$ and $b$ are related by
\begin{equation}
-\alpha_{m} = \bigg(\frac{a}{b^2}\bigg)^{\frac{m-1}{2}}\,.
\end{equation}
We wish to have a solution that is well behaved at $r\to\infty$ and at $r=0$. This can be achieved by tuning the constants $C_{1}$ and $C_{2}$. In particular, for $C_1=0$ and $C_2=1$, one finds that
\begin{equation}
\int \bigg(\frac{8}{m+1}\tau^{m+1}+\tau^{2}\bigg)^{-1/2}d\tau = \bigg(\frac{2}{1-m}\bigg)\sinh^{-1}{\bigg(\sqrt{\frac{m+1}{8}}\tau^{\frac{1-m}{2}}\bigg)}\,.
\end{equation}
Solving for $\tau$, we obtain
\begin{equation}
\tau^{1-m} = \bigg(\frac{8}{m+1}\bigg) \sinh^{2}{\bigg[\bigg(\frac{1-m}{4}\bigg)\log{\left(\,\frac{r}{a}\,\right)}\bigg]}\,.
\end{equation}
On the other hand, from (\ref{parametric1})-(\ref{parametric2}) it follows that
\begin{equation}
\frac{r}{\tilde{\mathrm{\Phi}}^{2}} = \frac{a}{b^{2}\tau^{2}}\,,
\end{equation}
or equivalently
\begin{equation}
\bigg(\frac{r}{\tilde{\mathrm{\Phi}}^{2}}\bigg)^{\frac{m-1}{2}} = \frac{-\alpha_{m}}{\tau^{m-1}}\,.
\end{equation}
Substituting the solution for $\tau$ and solving for $\tilde{\mathrm{\Phi}}$, we find \footnote{In the process, we make use of the identity $\sinh^{-1}{(\log{z})}=(z^{2}+1)/(2z)$.}
\begin{equation}\label{phitilde}
\tilde{\mathrm{\Phi}}(r)=\sqrt{a}\left(\frac{1+m}{2\alpha_{m}}\right)^{\frac{1}{m-1}}\left(\,\frac{r}{a}\,\right)\left(1+\left(\,\frac{r}{a}\,\right)^{\frac{m-1}{2}}\right)^{-\frac{2}{m-1}}\,,
\end{equation}
which leads to
\begin{equation}\label{phisln}
\mathrm{\Phi}(r)=-\sqrt{a}\left(\frac{1+m}{2\alpha_{m}}\right)^{\frac{1}{m-1}}\left(a^{\frac{m-1}{2}}+r^{\frac{m-1}{2}}\right)^{-\frac{2}{m-1}}\,.
\end{equation}
The constant $a$ is a dimensionful parameter of the solution that fixes a length scale. In particular, we shall call the regions $r<a$ and $r>a$ the inner and outer (or asymptotic) regions, respectively. Also, note that in order to have a finite potential at $r=0$ we must restrict ourselves to the range $m>1$. For all these cases we obtain that $\mathrm{\Phi}\to0$ as $r\to\infty$, so from now on we set $\phi_0=0$ \footnote{Thus, only particles with $E<0$ are allowed in the ensemble.}, which implies that $\mathrm{\Phi}(r)=\phi(r)$.

\begin{figure}
$$
\begin{array}{cc}
  \includegraphics[width=7cm]{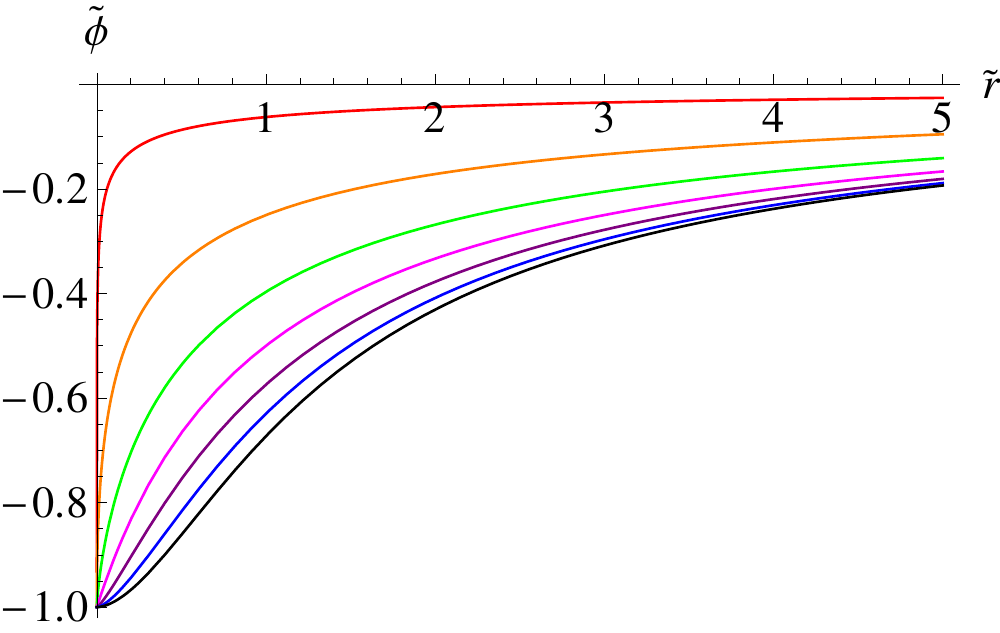} & \includegraphics[width=7cm]{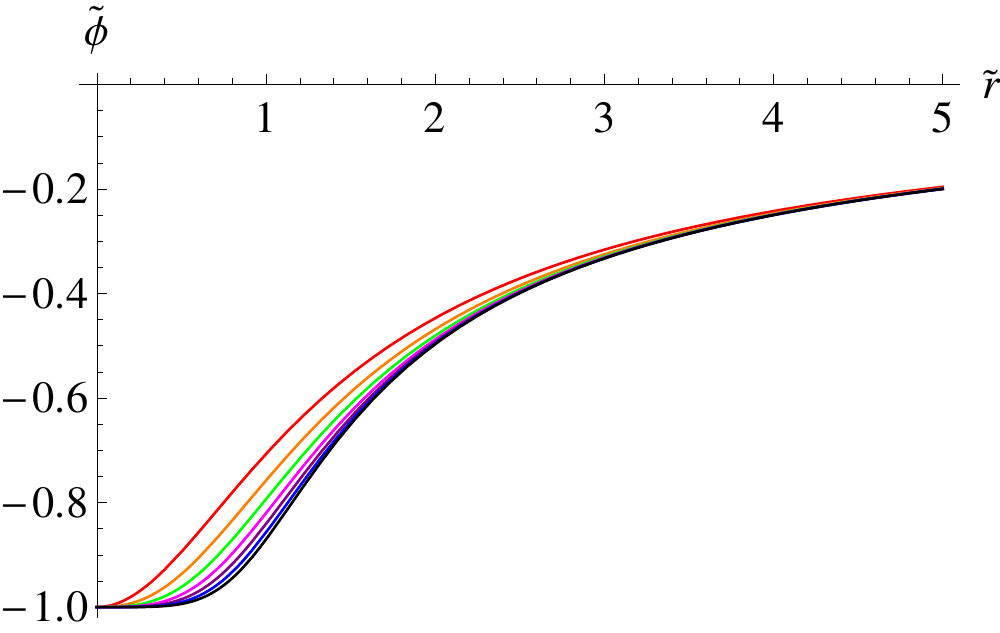}
\end{array}
$$
\caption{Dimensionless potential $\tilde{\phi}$ as a function of the dimensionless radius $\tilde{r}$ for $m<5$ and $m\geq5$ respectively. Left panel: $m\in\{\frac{3}{2},2,\frac{5}{2},3,\frac{7}{2},4,\frac{9}{2}\}$ from top to bottom. Right panel: $m\in\{5,6,7,8,9,10,11\}$ from top to bottom.\label{fig1}}
\end{figure}

This family of potentials was also considered in \cite{Mahdavi:2002ds}. In that paper it was shown that these models can successfully describe the temperature and density profiles often seen in cooling flow clusters (from x-ray data), and some potential relevance for the modeling of collisionless dark matter halos was suggested \footnote{The author of \cite{Mahdavi:2002ds} mentioned the possibility of obtaining analytic DFs by means of inverting the density as a function of the potential. Although this process is, in principle, possible, it would lead to \emph{isotropic} DFs depending only on the energy. It would be interesting to compute them and compare the results with the findings in this paper.}. Among these models, the $m=5$ case corresponds to the well-known Plummer potential \cite{plummer},
\begin{equation}
\phi(r) \propto \frac{1}{\sqrt{r^{2}+a^{2}}}\,,
\end{equation}
and the $m=3$ case corresponds to the Hernquist potential \cite{Hernquist:1990be,Baes:2002tw},
\begin{equation}
\phi(r) \propto \frac{1}{r+a}\,.
\end{equation}
Now, in order to show graphically the behavior of these models, we first define a dimensionless (and normalized) potential through
\be
\tilde{\phi}(\tilde{r})=\sqrt{a}\left(\frac{1+m}{2\alpha_{m}}\right)^{-\frac{1}{m-1}}\phi(\tilde{r})=-\left(1+\tilde{r}^{\frac{m-1}{2}}\right)^{-\frac{2}{m-1}}\,,
\ee
where $\tilde{r}=r/a$. In Fig. \ref{fig1} we show $\tilde{\phi}$ as a function of $\tilde{r}$ for some particular models. For the cases with $m>3$ we have that $\phi'(0)=0$, whereas for $1<m<3$ we have $\phi'(0)\to\infty$. The model with $m=3$ is the only case with a finite inner slope. On the other hand, for all cases the asymptotic region of the potential has Coulombic behavior for $r\gg a$, i.e. $\phi\sim-1/r$, and this relation becomes more exact as we increase $m$.

\subsection{Physical properties of the models}

\subsubsection{Normalization and mass density}

As a first step towards the analysis of the physical properties, we have to fix the overall constant $k_m$ that appears in front of the DF. This can be done by imposing that the integral over phase space of the DF is the total mass of the system $M$. At lowest order, this implies that
\begin{equation}\label{Nmass}
4\pi\int_{0}^{\infty} \rho\, r^{2} dr = M\,,
\end{equation}
with $\rho=\!\,^0T^{00}$. From (\ref{0T00poly}) and (\ref{phisln}) we find
\begin{equation}
k_{m} = \frac{a^{(m-1)/2}}{2^{(m+13)/4}\pi^{5/2}G^{m}M^{m-1}} \frac{\mathrm{\Gamma}(m+2)}{\mathrm{\Gamma}\left(\frac{1}{4}(m-1)\right)\mathrm{\Gamma}\left(\frac{3}{4}(m+1)\right)}\,.
\end{equation}
Substituting into the solution, we find
\begin{equation}\label{phifinal}
\phi(r) = -\frac{GM}{\left(a^{\frac{m-1}{2}}+r^{\frac{m-1}{2}}\right)^{\frac{2}{m-1}}}\,.
\end{equation}
Note that this result is intuitive. The overall factor $GM$ is independent of $m$ because in the limit of large radius, $r\gg a$, (\ref{phifinal}) should reduce to the Coulomb potential regardless of the value of $m$ (this applies for any potential sourced by a matter distribution that is confined or that decays sufficiently fast for large distances).

The mass density can be obtained by means of the Poisson equation (\ref{poissoneq}). Using our result for the potential, we get
\be\label{denfinal}
\rho(r)=\frac{(m+1)M a^{\frac{m-1}{2}}r^{\frac{m-5}{2}}}{8\pi\left(a^{\frac{m-1}{2}}+r^{\frac{m-1}{2}}\right)^{\frac{2m}{m-1}}}\,.
\ee
As we anticipated in the paragraph below (\ref{dfm}), there are markedly different behaviors for the density depending on the value of $m$ (which follows directly from the dependence of the DF on the angular momentum). The case $m=5$ is the only one with finite density at the origin. This corresponds to the Plummer model. For $m<5$ the density profile diverges near the center as a power law, despite the fact that the total mass is finite. Until recently, such a density dependence was considered unphysical, but it is now known that dark matter halos and early-type galaxies always have power-law density cusps \cite{Merritt:1995,Merritt:1995hp,Gebhardt:1996ik}. The case $m=3$ is of particular interest, as it resembles the well-known Navarro-Frenk-White profile for small radius \cite{Navarro:1995iw,Navarro:1996gj}. For $m>5$ the density profile vanishes at the origin. Distributions of matter in the form of shells have been a useful tool in astrophysics, often providing simplified but analytically tractable models in cosmology, gravitational collapse and supernovae \cite{Vogt:2009gs}. Finally, one can always consider a superposition of various potential-density pairs or even the gluing of two different models at some radius by means of the appropriate junction conditions. This last possibility is briefly discussed in Appendix \ref{appglue}.

\begin{figure}
$$
\begin{array}{cc}
  \includegraphics[width=7cm]{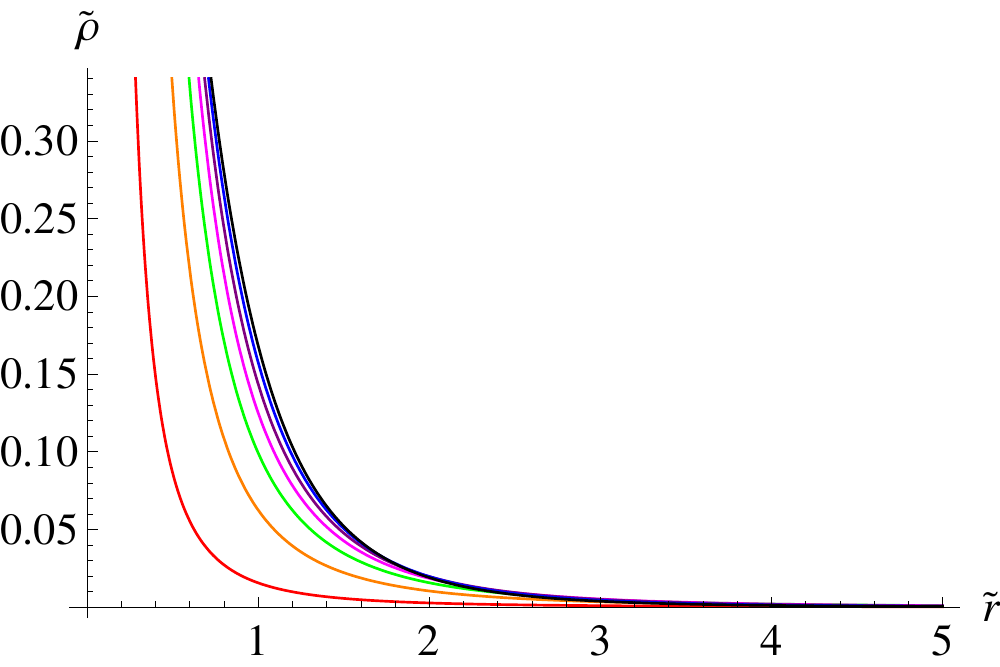} & \includegraphics[width=7cm]{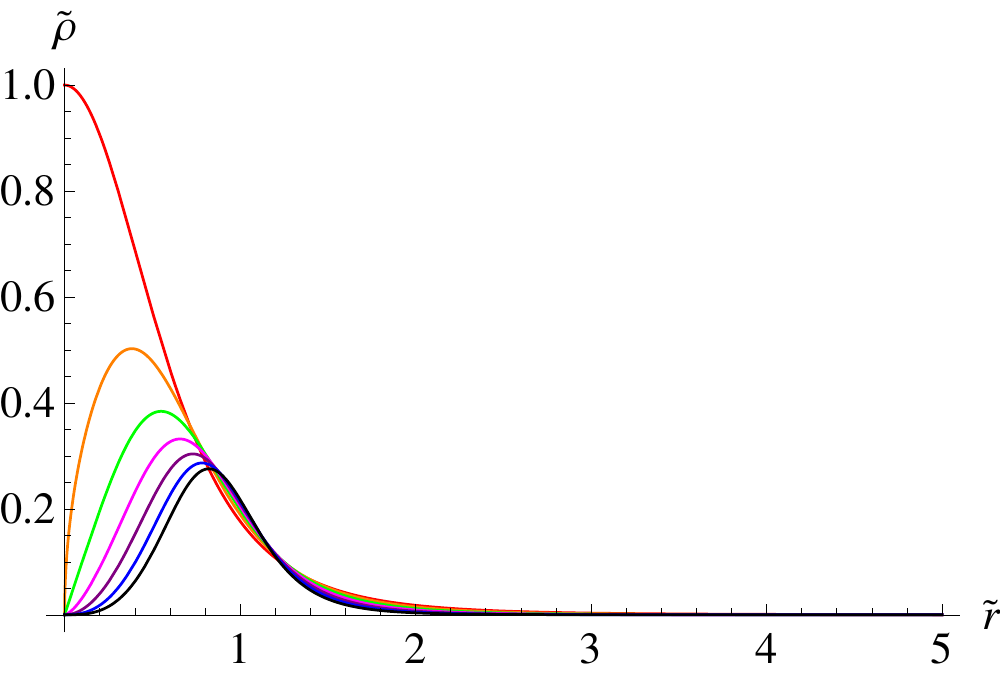}
\end{array}
$$
\caption{Dimensionless density $\tilde{\rho}$ as a function of the dimensionless radius $\tilde{r}$ for $m<5$ and $m\geq5$ respectively. Left panel: $m\in\{\frac{3}{2},2,\frac{5}{2},3,\frac{7}{2},4,\frac{9}{2}\}$ from left to right. Right panel: $m\in\{5,6,7,8,9,10,11\}$ from top to bottom.\label{fig2}}
\end{figure}

To show graphically the behavior of (\ref{denfinal}) we define the dimensionless density as
\be
\tilde{\rho}(\tilde{r})=\frac{8\pi a^3}{(m+1)M}\,\rho(\tilde{r})=\frac{\tilde{r}^{\frac{m-5}{2}}}{\left(1+\tilde{r}^{\frac{m-1}{2}}\right)^{\frac{2m}{m-1}}}\,,
\ee
where, again, $\tilde{r}$ is the dimensionless radius $\tilde{r}=r/a$. In Fig. \ref{fig2} we plot $\tilde{\phi}$ for some particular models. For $m>5$ the density has a maximum at some point $\bar{a}$ that depends on the value of $m$,
\begin{equation}\label{abar}
\bar{a} = \bigg(\frac{m-5}{m+5}\bigg)^{\frac{2}{m-1}} a\,.
\end{equation}
It is worth noticing that as we increase $m$ the matter distribution becomes more and more localized around $r=\bar{a}$ (the explicit limit $m\to\infty$ represents a shell-like configuration at $\bar{a}\to a$, with a potential that vanishes in the interior and becomes Coulombic for $r>a$). Otherwise, the behavior of the density for the different values of $m$ agrees with our previous discussion.

 \subsubsection{Velocity dispersion and the anisotropic parameter\label{dispersions}}
Instead of working with the distribution function, a somewhat less powerful approach is to work with the density of stars $\nu{(r)}$ (obtained by integrating the DF over all velocities and dividing by the total mass $M$) and velocity dispersion tensor $\sigma_{ij}$ defined by
\begin{equation}
\sigma_{ij}{(r)} = \frac{1}{\nu{(r)}} \int d^{3}v\, v_{i}v_{j} F{(r,\vec{v})}\,.
\end{equation}
This is, of course, the second moment of the Newtonian distribution function. To compute $\sigma_{ij}$ we will not, however, use the definition above. Instead, we follow a less direct route by first deriving a relation between the mass density $\rho$ in terms of the potential $\phi$ and the radial distance $r$. To do this, we perform a change of coordinate in velocity space, from the spherical coordinates $(v,\eta,\chi)$ to $(E,v_{t}^{2},\chi)$, where $E = \frac{v^2}{2} + \phi$ is the energy and $v_{t} = v \sin{\eta}$ is the tangential velocity. In terms of the new coordinates, the angular momentum and the volume element in velocity space become
\begin{equation}
L = rv_{t}\,,
\end{equation}
\begin{equation}
d^{3}v = 2\pi \frac{dE dv_{t}^{2}}{\sqrt{2(E-\phi)-v_{t}^{2}}}\,.
\end{equation}
Integrating over all of velocity space then yields the function $\rho{(r,\phi)}$. Of course, upon substituting $\phi{(r)}$ into this function we must recover the mass density profile $\rho{(r)}$. Explicitly, we find for our family
\begin{equation}
\rho(r,\phi) =\frac{(m+1)a^{(m-1)/2}}{8\pi G^m M^{m-1}} r^{\frac{m-5}{2}}(-\phi)^{m}\,,
\end{equation}
which can also be obtained by combining (\ref{phifinal}) and (\ref{denfinal}). The velocity dispersions can be computed from $\rho{(\phi,r)}$ as follows \cite{Dejonghe87}
\begin{equation}\label{disp1}
\sigma_{r}^{2}{(r)} \equiv \left\langle v_r^2\right\rangle=-\frac{1}{\rho{(\phi,r)}} \int_{0}^{\phi} \rho{(\phi',r)}d\phi' = -\frac{\phi}{m+1}\,,
\end{equation}
\begin{equation}\label{disp2}
\sigma_{\varphi}^{2}{(r)} \equiv \left\langle v_\varphi^2\right\rangle  = -\frac{1}{\rho{(\phi,r)}} \int_{0}^{\phi}\partial_{r^{2}}[r^{2}\rho{(\phi',r)}]d\phi' = -\frac{m-1}{4(m+1)}\phi\,,
\end{equation}
and $\sigma_{\theta}^{2}(r)=\sigma_{\varphi}^{2}(r)$. Here, it is understood that we are working in spherical coordinates. The anisotropic parameter $\beta$ as defined in \cite{BT} is
\begin{equation}
\beta = 1 - \frac{\sigma_{\varphi}^{2}}{\sigma_{r}^{2}} = \frac{5-m}{4}\,.
\end{equation}
In particular, $\beta$ is constant. This is true for all DFs of the form $f(E,L) = L^{\delta}g(E)$. When $\beta > 0$, near-radial orbits are preferred. This happens when $m < 5$, and these models are possibly subject to the radial-orbit instability. By contrast, $\beta < 0$ implies that near-circular orbits are preferred, and this happens for $m > 5$. Moreover, as $m \rightarrow \infty$, circular orbits become more and more dominant. When $\beta$ vanishes, both kinds are equally probable (therefore the model is isotropic). In the context of velocity dispersion and the density of stars, the Vlasov equation becomes the Jeans equation:
\begin{equation}
\frac{d}{dr}{(\nu \sigma_{r}^{2})} + \frac{2\beta}{r}\nu \sigma_{r}^{2} + \nu \frac{d\phi}{dr} = 0\,.
\end{equation}
There are some quantities computed from the second moments that are useful when comparing with observations. Among them, the surface densities, surface brightness (assuming a constant mass-to-light ratio) and the line-of-sight velocities were computed for the hypervirial family in \cite{Evans:2005zv}. However, these quantities alone do not completely determine a particular model; hence, an important thing to do is to study the higher order moments. The next nonzero moments are the fourth order ones, which in this case reduce to
\begin{equation}
\tau_{rr}^{4}{(r)} = -\frac{1}{\rho{(\phi,r)}} \int_{0}^{\phi}(\phi'-\phi)\rho{(\phi',r)}d\phi' = \frac{\phi^{2}}{2+3m+m^{2}}\,,
\end{equation}
\begin{equation}
\tau_{r\varphi}^{4}{(r)} = \tau_{r\theta}^{4}{(\phi,r)} = -\frac{1}{3\rho{(\phi,r)}} \int_{0}^{\phi}(\phi'-\phi)\partial_{r^{2}}[r^{2}\rho{(\phi',r)}]d\phi' = \frac{m-1}{12(2+3m+m^2)}\phi^{2}\,,
\end{equation}
\begin{eqnarray}
\tau_{\varphi\varphi}^{4}{(r)} = \tau_{\theta\theta}^{4}{(\phi,r)} = 3 \tau_{\varphi \theta}^{4}{(\phi,r)} &\!\!=\!\!& -\frac{1}{2\rho{(\phi,r)}} \int_{0}^{\phi}(\phi'-\phi) \partial^2_{r^{2}}[r^{4}\rho{(\phi',r)}]d\phi'\,, \nonumber \\
&\!\!=\!\!& \frac{(m-1)(m+3)}{32(2+3m+m^{2})}\phi^{2}\,.
\end{eqnarray}
The analog of the Jeans equation for the fourth order moments has been derived in the literature \cite{Merrifield1990} and used extensively to investigate the degeneracy in projected quantities for anisotropic systems (see for instance \cite{Gerhard1993,Lokas:2003ks}).

\subsubsection{Pressure and the equation of state}
In this subsection and the next one, we will study properties of our stellar system by approximating the system as a fluid. Consequently, we will use concepts from hydrodynamics and thermodynamics such as pressure, temperature, equation of state or hydrostatic equilibrium to describe the system. While many of these concepts could be defined in a more or less natural way in the context of stellar systems, it is important to keep in mind that the ``microscopic'' pictures are quite different: on one hand, the pressure inside a self-gravitating fluid counteracts gravity, resulting in hydrostatic equilibrium; on the other hand, what sustains a stellar system against gravitational collapse is the angular momentum of the individual stars.

That said, the analogy between stellar systems and self-gravitating fluids is a fruitful one, especially with regards to systems described by isotropic distribution functions. For example, there exist theorems which state that the stability of an isotropic system may be inferred from the stability of a barotropic fluid with the same density and pressure \cite{BT}. The stability of fluids is often a simpler problem as there exist simple criteria for the stability of fluids (for example using the adiabatic index). For anisotropic systems, however, the analogy with a barotropic fluid is problematic: nothing guarantees that the mass density is an invertible function of $r$. Nevertheless we will pursue this analogy as far as we can in this section.

The first thermodynamical quantity we are interested in is the pressure of the system. However, in anisotropic systems the different dispersions in velocity space lead to different pressures along the different directions. In our case, a straightforward computation leads to
\be\label{pradial}
p_r(r)=\rho(r)\sigma_{r}^{2}(r)=\frac{a^{\frac{m-1}{2}}G M^2r^{\frac{m-5}{2}}}{8\pi\left(a^{\frac{m-1}{2}}+r^{\frac{m-1}{2}}\right)^{\frac{2(m+1)}{m-1}}}\,,
\ee
\be\label{pperp}
p_\theta(r)=\rho(r)\sigma_{\theta}^{2}(r)=\frac{m-1}{4}p_r(r)\,,
\ee
and $p_\varphi(r)=p_\theta(r)\equiv p_\perp(r)$. We can also define an average pressure as
\be\label{pfinal}
p(r)\equiv\frac{\rho}{3}\left\langle v^2\right\rangle=\frac{\rho}{3}\left\langle v_r^2+v_\varphi^2+v_\theta^2\right\rangle=\frac{(m+1)a^{\frac{m-1}{2}}G M^2r^{\frac{m-5}{2}}}{48\pi\left(a^{\frac{m-1}{2}}+r^{\frac{m-1}{2}}\right)^{\frac{2(m+1)}{m-1}}}\,.
\ee
This quantity is useful in the sense that it gives us a notion of average speed which will ultimately be related to the temperature of the system (see Sec. \ref{thermo} for details).

Of course, these pressures are related to the spatial components of the energy-momentum tensor. In the nonrelativistic limit however, the pressure is not expected to appear at leading order in the energy-momentum tensor because $\rho\gg p/c^2$. At next-to-leading order we have that $\,\!^2T^{ij}=p_{ij}/c^2$, where $p_{ij}=\langle v_i v_j\rangle$ (although the off-diagonal terms vanish in our case by symmetry arguments) \footnote{Note, in particular, that, at leading order, none of the spatial components of the energy-momentum tensor depends of the post-Newtonian potential $\psi$, as expected.}. For instance, the average pressure (\ref{pfinal}) could also be obtained by substituting (\ref{phifinal}) in (\ref{2Tii}). Finally, as a consistency check we also notice that (\ref{0T00poly}) and (\ref{2Tii}) imply that $p=-\rho\phi/6$, which agrees with (\ref{disp1})-(\ref{disp2}). This simple relation will be helpful in the next section.

\begin{figure}
$$
\begin{array}{cc}
  \includegraphics[width=7cm]{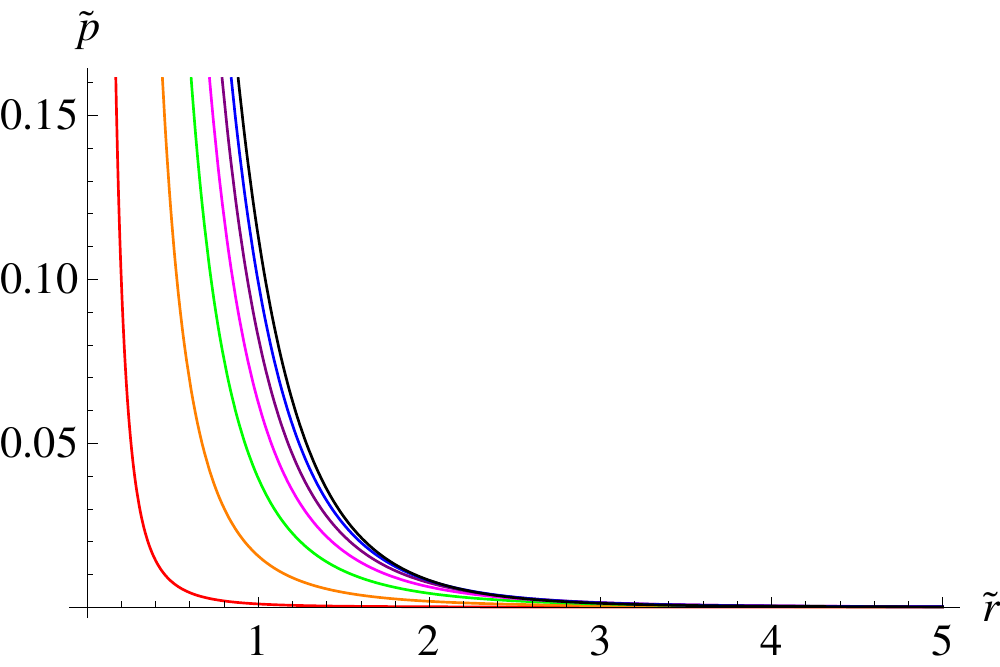} & \includegraphics[width=7cm]{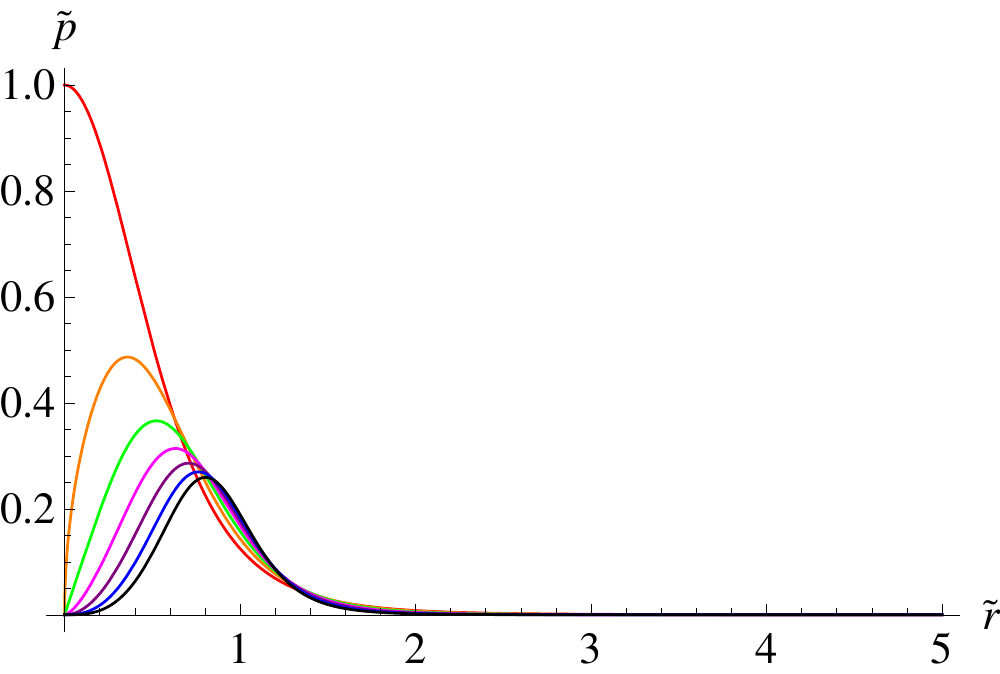}
\end{array}
$$
\caption{Dimensionless pressure $\tilde{p}$ as a function of the dimensionless radius $\tilde{r}$ for $m<5$ and $m\geq5$ respectively. Left panel: $m\in\{\frac{3}{2},2,\frac{5}{2},3,\frac{7}{2},4,\frac{9}{2}\}$ from left to right. Right panel: $m\in\{5,6,7,8,9,10,11\}$ from top to bottom.\label{fig3}}
\end{figure}

Now, we define a dimensionless pressure through
\be
\tilde{p}(\tilde{r})=\frac{48\pi a^4}{(m+1)G M^2}p(\tilde{r})=\frac{\tilde{r}^{\frac{m-5}{2}}}{\left(1+\tilde{r}^{\frac{m-1}{2}}\right)^{\frac{2(m+1)}{m-1}}}\,,
\ee
where $\tilde{r}=r/a$. In Fig. \ref{fig3} we plot $\tilde{p}$ for some values of $m$. In general, the behavior for this quantity is very similar to the density. The case $m=5$ is the only one whose pressure is finite and decreasing. For $m<5$ the pressure diverges at the origin but is still a decreasing function of the radius. For $m>5$ the pressure vanishes at the center, then increases to a maximum at $r= \bar{a}$ and finally goes back to zero as the radius increases. In the limit $m\to \infty$ the pressure becomes sharply localized at $r=a$.

We can relate the pressure and density through an appropriate equation of state $p=p(\rho,s)$ or
$p=p(\rho,T)$. Nevertheless, for the purposes of the present section it is sufficient to consider the simple case of
a barotropic equation of state, where the pressure is determined by the density, $p = p(\rho)$. To do this, we first have to invert (\ref{denfinal}) to get $r(\rho)$ and then plug it into (\ref{pfinal}). A few comments are in order here. First note that for $m>5$, $r(\rho)$ would be multivalued given that for such cases the density $\rho(r)$ is nonmonotonic. For these models we divide the equation of state into two parts, one that is valid for $r\leq \bar{a}$ and the other that is valid for $r\geq\bar{a}$. For $m\leq5$ the equation of state can be defined globally.

In practice, we can only analytically invert $r(\rho)$ for $m=5$ \footnote{The case $m=3$ can be inverted as well, but the solution is quite complicated. We will refrain from writing out the result here, since it is not particularly illuminating.}. In this case we get
\be
\tilde{r}(\tilde{\rho})=\sqrt{\tilde{\rho}^{-2/5}-1},
\ee
which leads to the simple result
\be
\tilde{p}(\tilde{\rho})=\tilde{\rho}^{\,6/5}\,.
\ee
This is the equation of state of a polytrope. For other values of $m$ we can expand $\tilde{\rho}(\tilde{r})$ in the regimes $\tilde{r}\gg1$ and $\tilde{r}\ll1$, and then solve the equation perturbatively. For $\tilde{r}\gg1$ we obtain
\be\label{expansion2}
\tilde{\rho}(\tilde{r})=\tilde{r}^{-\frac{m+5}{2}}\left[1-\frac{2m}{m-1}\tilde{r}^{-\frac{(m-1)}{2}}+\mathcal{O}\left(\tilde{r}^{-(m-1)}\right)\right]\,,
\ee
whereas for $\tilde{r}\ll1$ we get
\be\label{expansion1}
\tilde{\rho}(\tilde{r})=\tilde{r}^{\frac{m-5}{2}}\left[1-\frac{2m}{m-1}\tilde{r}^{\frac{m-1}{2}}+\mathcal{O}\left(\tilde{r}^{\,m-1}\right)\right]\,.
\ee

We must proceed with certain care in order to invert these relations. Let us first consider the regime $\tilde{r}\gg1$. In this case $\tilde{\rho}\ll1$ regardless of the value of $m$. At leading order $\tilde{r}\simeq\tilde{\rho}^{-2/(m+5)}$, so from (\ref{expansion2}) we can write
\be
\tilde{r}\simeq\left(\frac{\tilde{\rho}}{1-\frac{2m}{m-1}\tilde{\rho}^{\frac{m-1}{m+5}}}\right)^{-\frac{2}{m+5}}\simeq\tilde{\rho}^{-\frac{2}{m+5}}\left[1-\frac{4m}{(m-1)(m+5)}\tilde{\rho}^{\frac{m-1}{m+5}}+\mathcal{O}\left(\tilde{\rho}^{\frac{2(m-1)}{m+5}}\right)\right]\,.
\ee
The expansion parameter is $\tilde{\rho}^{(m-1)/(m+5)}$, which is small for all values of $m$. Now, consider the regime $\tilde{r}\ll1$.
Notice that in this case $\tilde{\rho}\gg1$ for $m<5$ whereas $\tilde{\rho}\ll1$ for $m>5$. For $m\neq 5$ at leading order we have $\tilde{r}\simeq\tilde{\rho}^{2/(m-5)}$, so from (\ref{expansion1}) it follows that
\be
\tilde{r}\simeq\left(\frac{\tilde{\rho}}{1-\frac{2m}{m-1}\tilde{\rho}^{\frac{m-1}{m-5}}}\right)^{\frac{2}{m-5}}\simeq\tilde{\rho}^{\frac{2}{m-5}}\left[1+\frac{4m}{(m-1)(m-5)}\tilde{\rho}^{\frac{m-1}{m-5}}+\mathcal{O}\left(\tilde{\rho}^{\frac{2(m-1)}{m-5}}\right)\right]\,.
\ee
In this case the expansion parameter is $\tilde{\rho}^{(m-1)/(m-5)}$. In particular, the power of $\tilde{\rho}$ is negative for $m<5$ and positive for $m>5$ so we have a consistent perturbative expansion in both cases.

The next step is to substitute these expressions in (\ref{pfinal}) to obtain the equation of state in the two regimes. For $\tilde{r}\gg1$ we obtain
\be
\tilde{p}(\tilde{\rho})=\tilde{\rho}^{\,1+1/n}\left[1+\frac{2(m-5)}{(m-1)(m+5)}\tilde{\rho}^{\frac{m-1}{m+5}}+\mathcal{O}\left(\tilde{\rho}^{\frac{2(m-1)}{m+5}}\right)\right]\,,\qquad n\equiv\frac{m+5}{2}\,.
\ee
At leading order, this is the equation of state for a polytrope with index $n$, but it has corrections that appear when one goes to higher densities. On the other hand, for $\tilde{r}\ll1$ and $m\neq5$ we get
\be
\tilde{p}(\tilde{\rho})=\tilde{\rho}\left[1-\frac{2}{(m-1)}\tilde{\rho}^{\frac{m-1}{m-5}}+\mathcal{O}\left(\tilde{\rho}^{\frac{2(m-1)}{m-5}}\right)\right]\,,
\ee
which at leading order behaves like the equation of state of an isothermal gas \footnote{This can be derived by starting with a general polytropic model and then taking a suitable limit as $n\to\infty$ \cite{Hunter2001}.}. Incidentally, this is also the famous equation of state usually considered in cosmology, $p=\omega\rho c^2$, with $\omega$ being a dimensionless constant. This is closely related to the thermodynamic equation of state of an ideal gas law, which may be written as
\be
p=\rho RT\,.
\ee
Here $R$ is a constant that depends on the gas, $T$ is the temperature and $\bar{v}=\sqrt{RT}$ is the characteristic thermal speed of the molecules. Thus, in order to have a consistent post-Newtonian expansion we require that
\be\label{omegall}
\omega = \left(\frac{\bar{v}}{c}\right)^2\ll1\,,
\ee
which means that we are dealing with ``cold gases.'' In our case, we get
\be
\omega=\frac{GM}{6ac^2}\,,
\ee
and the characteristic speed turns out to be
\be
\bar{v}=\sqrt{\frac{GM}{6a}}\,.
\ee

\begin{figure}
$$
\begin{array}{cc}
  \includegraphics[width=7cm]{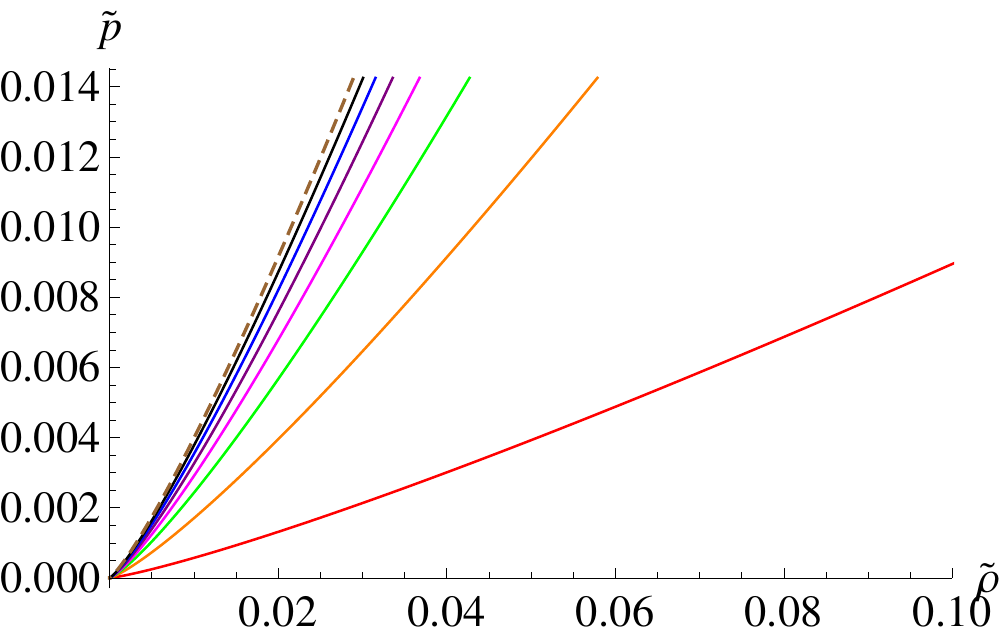} & \includegraphics[width=7cm]{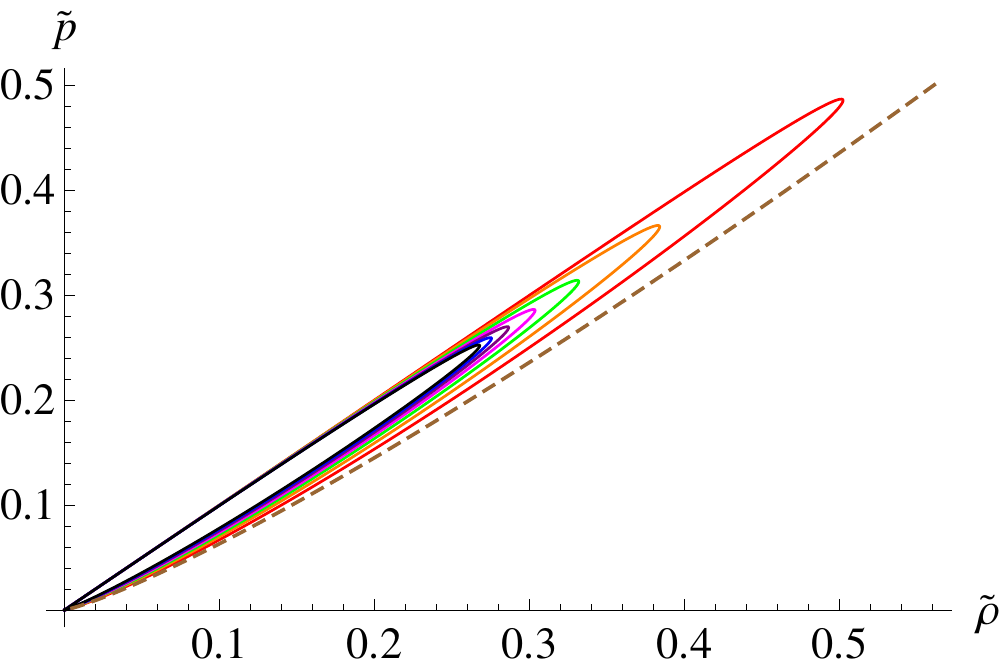}
\end{array}
$$
\caption{Dimensionless pressure $\tilde{p}$ as a function of the dimensionless density $\tilde{\rho}$ for $m<5$ and $m>5$ respectively. Left panel: $m\in\{\frac{3}{2},2,\frac{5}{2},3,\frac{7}{2},4,\frac{9}{2}\}$ from right to left. Right panel: $m\in\{6,7,8,9,10,11,12\}$ from right to left. In both plots, the case $m=5$ is shown in dashed lines for comparison.\label{fig4}}
\end{figure}

Now, for any value of $m$ one can always invert $r(\rho)$ numerically and then plug it into the expression for the pressure (\ref{pfinal}) in order to obtain the equation of state. The result is shown in Fig. \ref{fig4}. For $m>5$ the equation of state $p(\rho)$ is multivalued and forms a loop. One part increases linearly with $\rho$ and is valid in the inner region, $r< \bar{a}$. The other part increases like a power law and is valid for the outer region, $r>\bar{a}$. For $m\leq5$ the equation of state is well defined globally and it behaves as a power law for small $\rho$ ($r\gg a$). For large $\rho$, on the other hand, it behaves linearly but it is difficult to see it graphically because the slope depends strongly on the value of $m$, which leads to a great dispersion. We thus truncated the plotting range in these cases to focus on the first regime.

\begin{figure}
$$
\begin{array}{cc}
  \includegraphics[width=7cm]{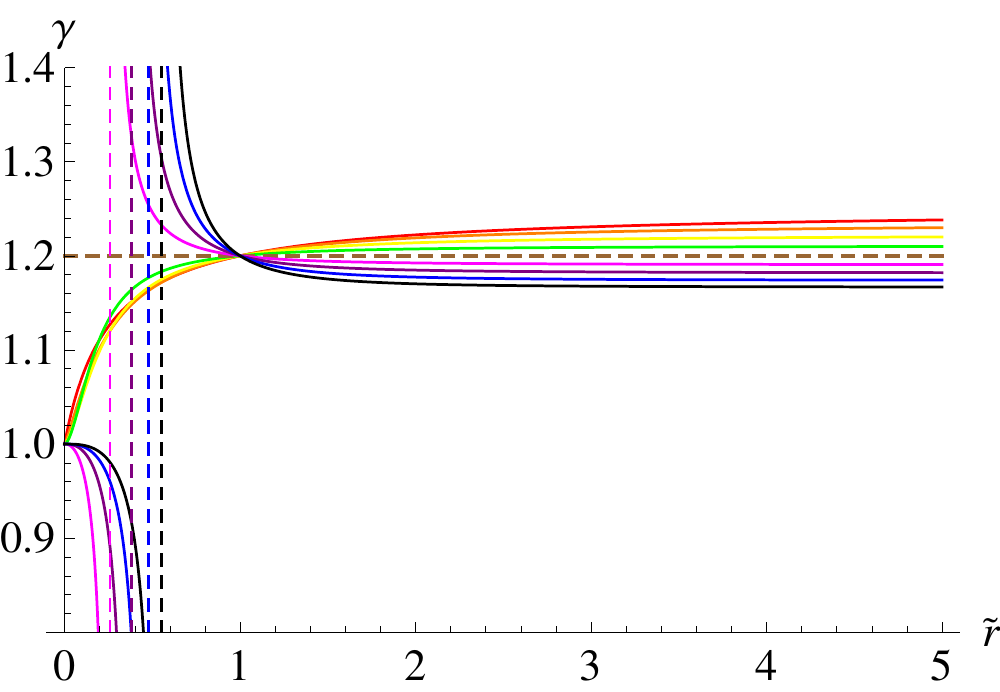} & \includegraphics[width=7cm]{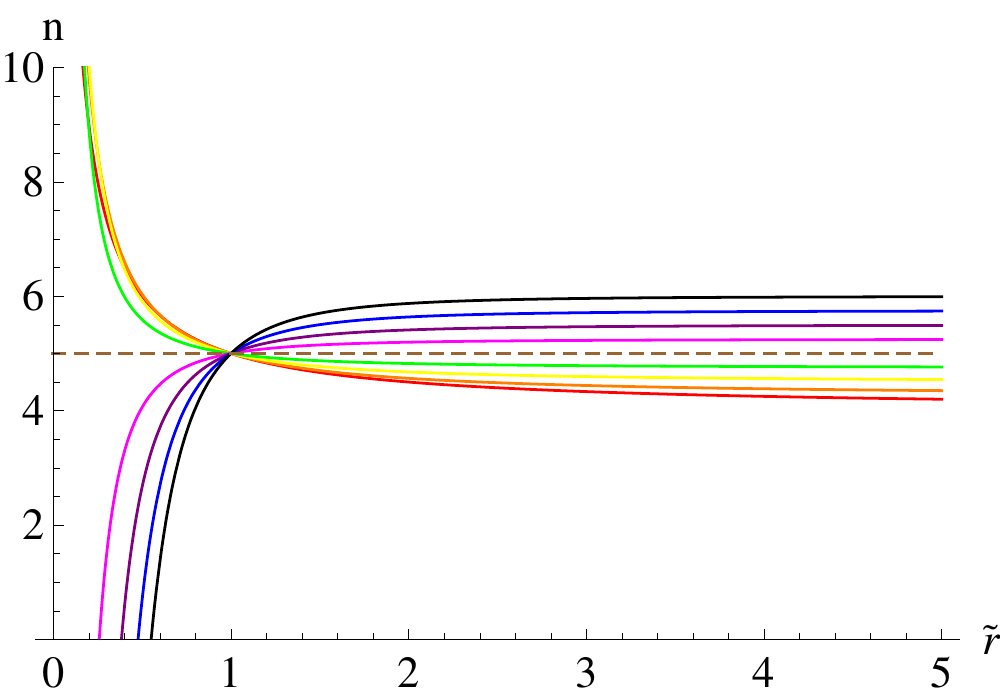}
\end{array}
$$
\caption{Left panel: Adiabatic index $\mathrm{\Gamma}_1$ as a function of the dimensionless radius $\tilde{r}$ for $m\in\{3,\frac{7}{2},4,\frac{9}{2},5,\frac{11}{2},6,\frac{13}{2},7\}$ from top to bottom (in the region $\tilde{r}\gg1$). Right panel: Polytropic index $n$ as a function of the dimensionless radius $\tilde{r}$ for $m\in\{3,\frac{7}{2},4,\frac{9}{2},5,\frac{11}{2},6,\frac{13}{2},7\}$ from bottom to top (in the region $\tilde{r}\gg1$). In both plots, the case $m=5$ is shown as dashed lines.\label{fig5}}
\end{figure}

Before closing this section let us perform a final computation that might help to clarify the above discussion. We are interested in the behavior of the adiabatic index $\mathrm{\Gamma}_1$ as a function of the radius, which can be computed as
\be
\mathrm{\Gamma}_1(r)=\frac{d\ln p}{d\ln \rho}=\frac{\rho}{p}\frac{dp/dr}{d\rho/dr}\,.
\ee
In general, this is identified as the ratio of heat capacities, $\mathrm{\Gamma}_1=c_p/c_v$, and for the particular case of polytropes the adiabatic index turns out to be a constant, $\mathrm{\Gamma}_1=1+1/n$. For our models, a straightforward computation shows that
\be
\mathrm{\Gamma}_1(\tilde{r})=\frac{(m+7) \tilde{r}^{m/2}-(m-5) \tilde{r}^{1/2}}{(m+5) \tilde{r}^{m/2}-(m-5) \tilde{r}^{1/2}}=1+\frac{1}{n(\tilde{r})}\,,
\ee
where
\be
n(\tilde{r})=\frac{m+5}{2}-\frac{m-5}{2} \tilde{r}^{-(m-1)/2}\,.
\ee
For $m=5$ the last term vanishes and we recover the expected result for $n$. For $m\neq5$ we still have a polytropic index $n=(m+5)/2$ for large radius but it gets corrections for any finite $r$. In particular, in the limit $\tilde{r}\to 0$ we get $n\to\infty$ and we recover the isothermal result, $\mathrm{\Gamma}_1=1$. Notice, however, that the second term can be positive or negative depending on the value of $m$. To see it explicitly we plot in Fig. \ref{fig5} both the adiabatic index $\mathrm{\Gamma}_1$ and the polytropic index $n$ as a function of the radius. For $m\leq5$ the plots are generally well behaved. For $m>5$, $\mathrm{\Gamma}_1$ presents a discontinuity at $r=\bar{a}$, exactly where $n$ vanishes (and changes sign). These models also present a small region near $r\lesssim \bar{a}$ for which $\mathrm{\Gamma}_1<0$ and therefore seem to be thermodynamically unstable. We will come back to this point in Sec. \ref{Nstability}.

\subsubsection{Hydrostatic equilibrium and thermodynamics \label{thermo}}
Conservation equations in general relativity come from the conservation of the energy-momentum tensor,
\be
\mathbf{\nabla}_{\!\nu}T^{\mu\nu}=0\,.
\ee
There are four equations, one for each value of the free index $\mu$. However, for spherically symmetric spacetimes with anisotropic pressures, only one of these does not vanish identically: the one for which $\mu=r$. It implies \footnote{This equation is fully relativistic \cite{Bowers}. In particular, it assumes a metric of the form $$ds^2=-e^{2\phi/c^2}c^2dt^2+e^{2\Lambda/c^2}dr^2+r^2d\mathrm{\Omega}^2,$$
and a stress-energy tensor of the form $T^{\mu\nu}\!=\mathrm{diag}(\rho,p_r/c^2,p_\perp/c^2,p_\perp/c^2)$. At lowest order, it is clear that we can identify $\phi$, $\rho$ and $p_i$ as the Newtonian potential, density and pressures, respectively.}
\be
\frac{d p_r}{dr}=-\left(\rho+\frac{p_r}{c^2}\right)\frac{d \phi}{d r}+\frac{2}{r}(p_\perp-p_r)\,.
\ee
This equation tells us what pressure gradient is needed to keep the fluid static in the gravitational field, an effect that depends on $d\phi/dr$. If we use the field equations to eliminate $\phi$ we recover the so-called Tolman-Oppenheimer-Volkoff (TOV) equation \cite{Tolman:1939jz,Oppenheimer:1939ne} for anisotropic spheres. The important point here is that, when supplemented with an equation of state $F(\rho, p)=0$, the TOV equation completely determines the structure of a spherically symmetric body in equilibrium.

If we consider matter that is nonrelativistic, the terms of order $(\bar{v}/c)^2$ can be neglected and the TOV equation becomes the Newtonian equation for hydrostatic equilibrium,
\be\label{hydro}
\frac{d p_r}{dr}=-\rho\frac{d \phi}{d r}+\frac{2}{r}(p_\perp-p_r)\,.
\ee
This equation is commonly used to find the equilibrium structure of a spherically symmetric body with anisotropic pressures when general-relativistic corrections are not important. In our case, using (\ref{phifinal})-(\ref{denfinal}) and (\ref{pradial})-(\ref{pperp}), we can see that Eq. (\ref{hydro}) is always satisfied, thus implying equilibrium. This result is not surprising. From (\ref{tmunu}) one can immediately see that the energy-momentum tensor is divergence-free, and then, even the relativistic version of (\ref{hydro}) must be satisfied.

Now, by analogy with an ideal gas, we can define the temperature $T$ (or thermal energy) of a self-gravitating system through the relation \cite{BT}
\be
\frac{1}{2}\left\langle v^2\right\rangle=\frac{3}{2}k_\mathrm{B}T\,,
\ee
where $k_{\mathrm{B}}$ is Boltzmann's constant. In general, the mean-square velocity and hence the temperature depend on position. For our models, at leading order we get
\be\label{temp}
k_{\mathrm{B}}T=\frac{\int \!^0\!Fv^2 d^3v}{\int \!^0\!F d^3v}=\frac{p}{\rho}=-\frac{\phi}{6}\,.
\ee
The temperature profiles can be easily inferred from Fig. \ref{fig1}. In general, the temperature reaches a maximum at the center (even for the $m>5$ cases) and then decreases with the radius. For $m<5$, although the densities are divergent, the temperatures are finite everywhere.

Although this notion of temperature is a mere analogy, it has been proven to be useful in the context of understanding relaxation processes within the system and the so-called gravothermal catastrophe \cite{Lynden Bell Eggleton, Spitzer Thuan}. Indeed, the mean-square velocity and hence the temperature are found to be position dependent, so if we follow the analogy to its logical end, Fourier's law for thermal conduction predicts a heat flow inside the system due to a nonzero temperature gradient. In this way, even though we are working with an equilibrium solution of the Vlasov system, we may gain insight into what happens to the system beyond the assumption of staticity.

For a general spherical system, the expression for the thermal conductivity as a function of $r$ is given by \cite{Lynden Bell Eggleton}
\begin{equation}\label{conduc}
\kappa{(r)} = \frac{\nu(r)r^{2}k_{B}}{t_{relax}}\,,
\end{equation}
where $\nu(r)$ is the number density and $t_{relax}$ is the relaxation time. Substituting (\ref{temp}) and (\ref{conduc}) into Fourier's law $q = -\kappa \nabla T$, we obtain the heat flux as
\begin{equation}
q{(r)} \propto r^{2}\rho{(r)} \frac{\partial\phi}{\partial r}\propto \frac{r^{m-2}}{(a^{\frac{m-1}{2}}+r^{\frac{m-1}{2}})^{\frac{3m+1}{m-1}}}\,.
\end{equation}
As explained in \cite{Lynden Bell Eggleton}, it is tempting to identify this heat flow as the relaxation process itself.

Finally, we can also interpret the TOV equation in terms of thermodynamic quantities by recalling that (\ref{hydro}) is just a restatement of the conservation laws. In particular, plugging (\ref{temp}) into Eq. (\ref{hydro}) we arrive at
\be\label{TOVnew}
\frac{1}{\rho}\frac{d (T\rho)}{dr}=-\frac{m+1}{6}\frac{d \phi}{d r}+\frac{m-5}{2r}T\,.
\ee
This equation resembles a local version of the first law of thermodynamics. Identifying $u\sim T\rho$ as the internal energy density for an infinitesimal fluid element with volume $V\sim 1/\rho$, then we can write
\be
du=dq-dw,
\ee
where
\be
dq=\rho\,\frac{m-5}{2r}T\, dr=T\,ds
\ee
is the heat transfer to the infinitesimal volume, $s$ is the entropy density, and
\be
dw=\rho\,\frac{m+1}{6}\frac{d \phi}{d r}dr=-\rho\,\frac{m+1}{6} F_r\, dr=-\rho\,\frac{m+1}{6} p_r\, dV=-\rho\, p\,dV
\ee
is the work (per unit volume) done by the system on its surroundings. Thus, Eq. (\ref{TOVnew}) reveals an interesting feature that is present in anisotropic models: for $m\neq5$ there is entropy production and thus the system is dissipative.

\section{Post-Newtonian Corrections\label{PNewtSec}}

\subsection{Solving the field equations at the next-to-leading order}
Now that we have studied in certain detail the properties of the Newtonian models, it is time to move on and compute the relativistic corrections.
Substituting the solution obtained in (\ref{phifinal}) for the Newtonian potential into the next leading order equation (\ref{eins2}), we obtain
\begin{equation}
\frac{1}{r^{2}}\frac{d}{d r}\bigg(r^{2}\frac{d \mathrm{\Psi}}{d r}\bigg) + \frac{\frac{1}{2}m(m+1)a^{\frac{m-1}{2}} r^{\frac{m-5}{2}}}{(a^{\frac{m-1}{2}}+r^{\frac{m-1}{2}})^{2}}\mathrm{\Psi} = C_{m} r^{\frac{m-5}{2}}a^{\frac{m+1}{2}} (a^{\frac{m-1}{2}} + r^{\frac{m-1}{2}})^{\frac{2(m+1)}{1-m}}\,,
\end{equation}
where we defined constants
\be
C_{m} = \beta_{m} \left(\frac{1+m}{2\alpha_{m}}\right)^{\frac{m+1}{m-1}}\,.
\ee
This is an inhomogeneous, second order differential equation, and thus the general solution can be written as
\be
\mathrm{\Psi}(r)=C_1\mathrm{\Psi}_1(r)+C_2\mathrm{\Psi}_2(r)+\mathrm{\Psi}_p(r)\,,
\ee
where $\mathrm{\Psi}_{1,2}(r)$ are solutions to the homogeneous equation and $\mathrm{\Psi}_p(r)$ is a particular solution of the inhomogeneous one. To solve this equation, we perform a change of both the independent and the dependent variables, as follows:
\begin{equation}
z = -(r/a)^{\frac{m-1}{2}}\,,
\end{equation}
\begin{equation}
\tilde{\mathrm{\Psi}}(r) = \mathrm{\Psi}(r)(a^{\frac{m-1}{2}}+r^{\frac{m-1}{2}})^{\frac{m+1}{m-1}} = a^{\frac{m-1}{2}}\mathrm{\Psi}(z)(1-z)^{\frac{m+1}{m-1}}\,.
\end{equation}
Then, substituting into the equation for $\mathrm{\Psi}$ we get
\begin{equation}\label{eqinz}
z(1-z)\frac{d^{2}\tilde{\mathrm{\Psi}}}{dz^{2}} + \bigg(\frac{m+1}{m-1}\bigg)(1+z)\frac{d\tilde{\mathrm{\Psi}}}{dz} - \bigg(\frac{m+1}{m-1}\bigg)\tilde{\mathrm{\Psi}}= -\frac{4C_{m}}{(m-1)^{2}}a^{\frac{m-3}{2}}(1-z)^{\frac{2}{1-m}}\,.
\end{equation}
The corresponding homogeneous equation can be cast as a hypergeometric differential equation,
\be
z(1-z)\frac{d^{2}w}{dz^{2}}+[c-(a+b+1)z]\frac{dw}{dz}-abw=0\,,
\ee
with $c = -a = (m+1)/(m-1)$ and $b=-1$. If $c$ is not an integer (i.e. if $m \neq 3,2$), the two independent solutions are ${}_{2}F_{1}{(a,b,c,z)}$ and $z^{1-c}{}_{2}F_{1}{(1+a-c,1+b-c,2-c,z)}$. In terms of $m$ these are \footnote{We use the fact that if one of the first two arguments of ${}_{2}F_{1}{(a,b,c,z)}$ is a negative integer, the series truncates.}
\begin{equation}
\tilde{\mathrm{\Psi}}_{1}(z) = 1 + z\,
\end{equation}
and
\begin{equation}
\tilde{\mathrm{\Psi}}_{2}(r) = z^{\frac{2}{1-m}} {}_{2}F_{1}\left(-\frac{m+3}{m-1},-\frac{m+1}{m-1},\frac{m-3}{m-1},z\right)\,.
\end{equation}
We will consider the special cases $m=3$ and $m=2$ separately later in this paper. To construct the particular solution, we follow the standard Wronskian method. Using the formula for the derivative of a hypergeometric function\footnote{$\frac{d}{dz} {}_{2}F_{1}{(a,b,c,z)}=\frac{ab}{c} {}_{2}F_{1}{(a+1,b+1,c+1,z)}$}, the Wronskian is found to be
\begin{eqnarray}
W[\tilde{\mathrm{\Psi}}_{1},\tilde{\mathrm{\Psi}}_{2}]{(z)} =&&\!\!\!\!\!\!\!\! z^{\frac{2}{1-m}}\frac{(m+3)(m+1)}{(m-3)(m-1)}(1+z) {}_{2}F_{1}{\bigg(\frac{4}{1-m},\frac{2}{1-m},\frac{2m-4}{m-1},z\bigg)} \nonumber \\
&&\!\!\!\!\!\!\!\! -z^{\frac{m+1}{1-m}}\bigg[\frac{2 + z(m+1)}{m-1}\bigg] {}_{2}F_{1}{\bigg(\frac{m+3}{1-m},\frac{m+1}{1-m},\frac{m-3}{m-1},z\bigg)}\,.
\end{eqnarray}
The particular solution is obtained from the homogeneous solutions and the Wronskian by evaluating the following integrals:
\begin{equation}\label{psipf}
\mathrm{\tilde{\Psi}}_{p}{(z)} = \mathrm{\tilde{\Psi}}_{1} \int \mathrm{\tilde{\Psi}}_{2} \frac{g}{fW} dz - \mathrm{\tilde{\Psi}}_{2} \int \mathrm{\tilde{\Psi}}_{1}\frac{g}{fW} dz\,,
\end{equation}
where $g=g(z)$ is the inhomogeneous term on the right-hand side of (\ref{eqinz}) and $f=f(z)$ is the coefficient of the second derivative term. The integrals in (\ref{psipf}) are challenging. To proceed, we first simplify the Wronskian using a sequence of identities of hypergeometric functions, starting with the Euler transformation \footnote{${}_{2}F_{1}{(a,b,c,z)}=(1-z)^{c-a-b}{}_{2}F_{1}{(c-a,c-b,c,z)}$}:
\begin{eqnarray}
(m-1) z^{\frac{m+1}{m-1}}(1-z)^{\frac{2m+2}{1-m}}W =&&\!\!\!\!\!\!\!\! z(1+z)\frac{(m+3)(m+1)}{m-3} {}_{2}F_{1}{\bigg(\frac{2m}{m-1},2,\frac{2m-4}{m-1},z\bigg)} \nonumber \\
&&\!\!\!\!\!\!\!\! -[2+z(m+1)](1-z){}_{2}F_{1}{\bigg(\frac{2m}{m-1},2,\frac{m-3}{m-1},z\bigg)}\,.
\end{eqnarray}
Next, using the Gauss contiguous relations \footnote{First, we use the identity $F(a,b-1,c,z) = \frac{z}{c}(c-a)F(a,b,c+1,z) + (1-z)F(a,b,c,z)$ with $a = \frac{2m}{m-1}$, $b = 2$, and $c = \frac{m-3}{m-1}$, and then we use the identity $b (1-z) {}_{2}F_{1}(a,b+1,c,z)-(c-b){}_{2}F_{1}(a,b-1,c,z) = (2b-c+(a-b)z){}_{2}F_{1}(a,b,c,z)$ with $a=\frac{2m}{m-1}$, $b = 1$, and $c = \frac{m-3}{m-1}$. We also use the fact that ${}_{2}F_{1}(a,0,c,x)=1$.}, the right-hand side above can be shown to be equal to $-2$, reducing the Wronskian to a simple form:
\begin{equation}
W{(z)} = -\bigg(\frac{2}{m-1}\bigg)z^{\frac{m+1}{1-m}}(1-z)^{\frac{2m+2}{m-1}}\,.
\end{equation}
The integrals can be evaluated in terms of the Meijer G function \cite{BatemanBook}, and the particular solution is found to be
\begin{eqnarray}
\mathrm{\Psi}_{p}{(z)} =&&\!\!\!\!\!\!\!\! \frac{C_{m}z}{a (1-z)^{\frac{m+1}{m-1}}} {}_{2}F_{1}\bigg(\frac{m+3}{1-m},\frac{m+1}{1-m},\frac{m-3}{m-1}, z\bigg) \times \nonumber \\
&&\!\!\!\!\!\!\!\! \bigg[\bigg(\frac{2}{m+1}\bigg){}_{2}F_{1}\bigg(\frac{m+1}{m-1},\frac{3(m+1)}{m-1},\frac{2m}{m-1},z\bigg)+ \frac{z}{m} {}_{2}F_{1}\bigg(\frac{2m}{m-1},\frac{3(m+1)}{m-1},\frac{3m-1}{m-1},z\bigg)\bigg] \nonumber \\
&&\!\!\!\!\!\!\!\! +\frac{2C_{m}}{a(m-1)}\frac{1+z}{(1-z)^{\frac{m+1}{m-1}}} \frac{\mathrm{\Gamma}(\frac{m-3}{m-1})}{\mathrm{\Gamma}(\frac{m+3}{1-m})\mathrm{\Gamma}(\frac{m+1}{1-m})\mathrm{\Gamma}(\frac{2m}{m-1})} G_{3,3}^{2,3} \!\left( \left. \begin{matrix} 1, \frac{2}{1-m}, \frac{4}{1-m} \\ \frac{2m+4}{1-m}, \frac{m-3}{m-1}, 0 \end{matrix}\; \right| \, 1 - z \right)\,.
\end{eqnarray}
Putting this all together we find that, for $m\neq3,2$,
\begin{equation}
\mathrm{\Psi}{(z)} = C_{1} \frac{1+z}{(1-z)^{\frac{m+1}{m-1}}} + C_{2}\frac{z^{\frac{2}{1-m}}}{(1-z)^{\frac{m+1}{m-1}}} {}_{2}F_{1}{\bigg(\frac{m+3}{1-m},\frac{m+1}{1-m},\frac{m-3}{m-1},z\bigg)} + \mathrm{\Psi}_{p}{(z)}\,.
\end{equation}

\subsubsection{Special cases}
In the case $m=5$, the homogeneous solutions can be expressed in terms of the trigonometric functions \footnote{We used the Pfaff transformation ${}_{2}F_{1}(a,b,c,z) = (1-z)^{-a}{}_{2}F_{1}(a,c-b,c,\frac{z}{z-1})$ and the formula ${}_{2}F_{1}(-a,a,\frac{1}{2},-z^{2}) = \frac{1}{2}[(\sqrt{1+z^{2}}+z)^{2a}+({\sqrt{1+z^{2}}-z})^{2a}]$.}:
\begin{equation}
\mathrm{\Psi}_{1}(r) = \frac{\sqrt{a^{2}+r^{2}}}{4r}\sin{\left(4\arctan{\left(\frac{r}{a}\right)}\right)}\,,
\end{equation}
\begin{equation}
\mathrm{\Psi}_{2}(r) = \frac{\sqrt{a^{2}+r^{2}}}{4r}\cos{\left(4\arctan{\left(\frac{r}{a}\right)}\right)}\,.
\end{equation}
To obtain the particular solution, notice that both the Meijer G function and the factor $\mathrm{\Gamma}{(-(m+3)(m-1))}$ in the denominator diverge as $m \rightarrow 5$.
To take this limit, we replace the Meijer G function by generalized hypergeometric functions as follows:
\be
\begin{split}
G_{3,3}^{2,3} \!\left( \left. \begin{matrix} a_{1}, a_{2}, a_{3} \\ b_{1}, b_{2}, b_{3} \end{matrix} \; \right| \, z \right) = &\,\, \mathcal{K}\,z^{a_{1}-1}{}_{3}F_{2} {\!\left( \left.\begin{matrix} 1-a_{1}+b_{1}, 1-a_{1}+b_{2}, 1-a_{1}+b_{3} \\ 1-a_{1}+a_{2}, 1-a_{1}+a_{3} \end{matrix} \; \right| \, \frac{1}{z} \right)}\\
&\,\, +(a_1\leftrightarrow a_2)+(a_1\leftrightarrow a_3)\,,
\end{split}
\ee
where
\be
\mathcal{K}=\frac{\mathrm{\Gamma}{(a_{1}-a_{2})}\mathrm{\Gamma}{(a_{1}-a_{3})}\mathrm{\Gamma}{(1-a_{1}+b_{1})}\mathrm{\Gamma}{(1-a_{1}+b_{2})}}{\mathrm{\Gamma}{(a_{1}-b_{3})}}\,.
\ee
This formula is valid when $|z| \geq 1$ and none of the $a_{i}$ differ by an integer. The final result simplifies to
\begin{equation}
\mathrm{\Psi}_{p}(r) = \frac{3G^{2}M^{2}}{112(r^{2}+a^{2})} - \frac{G^{2}M^{2}}{280a^{2}}\,.
\end{equation}

For $m=3$ the homogeneous solutions are
\begin{equation}
\mathrm{\Psi}_{1}(r) =\frac{\frac{r}{a}-1}{\left(1+\frac{r}{a}\right)^{2}}\,,
\end{equation}
\begin{equation}
\mathrm{\Psi}_{2}(r) = \frac{\left(\frac{r}{a}\right)^{2}-\frac{r}{a}+6\left(\frac{r}{a}-1\right)\log{\left(\frac{r}{a}\right)}+\frac{a}{r}-17}{\left(1+\frac{r}{a}\right)^{2}}\,,
\end{equation}
which lead to the particular solution
\begin{equation}
\mathrm{\Psi}_{p}(r) = \frac{G^{2}M^{2}}{60\left(a+r\right)^{2}}\left[6\left(\frac{r}{a}-1\right)\log{\left(1+\frac{a}{r}\right)}-\frac{a}{r}-6\right]\,.
\end{equation}

Finally, for $m=2$ the homogeneous solutions are
\begin{equation}
\mathrm{\Psi}_{1}(r) = \frac{\sqrt{\frac{r}{a}}-1}{\left(1+\sqrt{\frac{r}{a}}\right)^{3}}\,,
\end{equation}
\begin{equation}
\mathrm{\Psi}_{2}(r) = \frac{\frac{1}{2}\left(\frac{r}{a}\right)^{3/2} + \frac{15}{2}\left(\frac{r}{a}\right)- 8\sqrt{\frac{r}{a}} + 15\left(\sqrt{\frac{r}{a}}-1\right)\log{\left(\frac{r}{a}\right)}+ \frac{15}{2}\sqrt{\frac{a}{r}} +\frac{1}{2}\left(\frac{a}{r}\right)  - 72 }{\left(1+\sqrt{\frac{r}{a}}\right)^{3}}\,,
\end{equation}
and the particular solution is
\be
\mathrm{\Psi}_{p}(r) = \frac{5G^2M^2}{896(\sqrt{a}+\sqrt{r})^{4}}\left[
60\left(\frac{r}{a}-1\right)\log{\left(\sqrt{\frac{a}{r}}+1\right)}-60\sqrt{\frac{r}{a}}-16\sqrt{\frac{a}{r}}-\frac{a}{r}+30\right]\,.
\ee

\subsection{Fixing the integration constants}
To fix one of the integration constants, we will require that $\mathrm{\Psi}$ remains finite at the origin. For $m \neq 3,2$ this means setting $C_{2} = 0$. For $m=3$ we set $C_{2} = \frac{G^{2}M^{2}}{60a^{2}}$, whereas for $m=2$ we set $C_{2} = \frac{5G^{2}M^{2}}{448a^{2}}$. We also have to subtract the value at infinity of $\mathrm{\Psi}$ to obtain $\psi$. This value turns out to be independent of the remaining integration constant, and it is given by
\begin{eqnarray}
\psi_{0} = -\lim_{r\to\infty} \mathrm{\Psi}{(r)} =&&\!\!\!\!\!\!\!  \frac{(9m-43)G^{2}M^{2}}{64ma^{2}}\frac{\mathrm{\Gamma}{\left(\frac{2}{m-1}\right)}\mathrm{\Gamma}{\left(\frac{m-3}{m-1}\right)}}{\mathrm{\Gamma}{\left(\frac{3(m+1)}{m-1}\right)}\mathrm{\Gamma}{\left(\frac{2m}{m-1}\right)}\mathrm{\Gamma}{\left(\frac{m+1}{1-m}\right)}} \times \\
&&\!\!\!\!\!\!\!  \left[2m\mathrm{\Gamma}{\left(\frac{2(m+1)}{m-1}\right)}\mathrm{\Gamma}{\left(\frac{2m}{m-1}\right)}-(m+1)\mathrm{\Gamma}{\left(\frac{3m-1}{m-1}\right)}\mathrm{\Gamma}{\left(\frac{m+3}{m-1}\right)}\nonumber \right]\,.
\end{eqnarray}
For $m=5$ the above formula leads to $\psi_{0} = \frac{G^{2}M^{2}}{280a^{2}}$. For $m=3$ and $m=2$ we get $\psi_{0} = -\frac{G^{2}M^{2}}{60a^{2}}$ and $\psi_{0} = -\frac{5G^{2}M^{2}}{896a^{2}}$, respectively. The fact that we have to shift the Newtonian cutoff energy $E_{0}$ by a small amount in order for the solution to be asymptotically flat was numerically confirmed in the fully relativistic case by \cite{Rein}. Our results show that this shift in $E_{0}$ is already needed at the 1PN order.

As for the other constant of integration, we will leave it as a free parameter and express it in terms of the value of $\psi$ at the origin, which we will denote by $\psi{(0)}$. For $m\neq3,2$ we get
\begin{equation}
C_{1} = -\frac{2C_m\mathrm{\Gamma}(\frac{m-3}{m-1})}{a(m-1)\mathrm{\Gamma}(\frac{m+3}{1-m})\mathrm{\Gamma}(\frac{m+1}{1-m})\mathrm{\Gamma}(\frac{2m}{m-1})} G_{3,3}^{2,3}\!\left( \left. \begin{matrix} 1, \frac{2}{1-m}, \frac{4}{1-m} \\ \frac{2m+4}{1-m}, \frac{m-3}{m-1}, 0 \end{matrix}\; \right| \, 1 \right)\, + \psi{(0)} - \psi_{0}\,.
\end{equation}
For $m=5$ we get the simple result
$C_{1} = -\frac{3G^{2}M^{2}}{112a^{2}} + \psi{(0)}$, and as for $m=3,2$ we obtain
$C_{1} = -\frac{2G^{2}M^{2}}{5a^{2}} - \psi{(0)}$ and
$C_{1} = -\frac{125G^{2}M^{2}}{224a^{2}} - \psi{(0)}$, respectively.

In Fig. \ref{figpsi} (left panel) we plot the dimensionless potential $\tilde{\psi}=\psi a^{2} / G^{2} M^{2}$ as a function of $\tilde{r}$. For concreteness, we have fixed the value of $\psi$ at the center such that $\tilde{\psi}(0)=\tilde{\phi}(0)^2=1$ so that we can compare our results directly with the numerical polytropes presented in \cite{Agon:2011mz}. For this choice of parameters the behavior of the post-Newtonian potential is somehow universal, depending very little on the value of $m$; it is maximum at the center, it reaches a minimum around $r\sim a$, and it goes to zero at infinity.

\subsection{Effect on some physical observables}

\subsubsection{Corrections to the energy density}
In general relativity we can write the energy-momentum tensor of a general anisotropic fluid in the form
\be
T^{\mu\nu}\!=\varepsilon\, u^\mu u^\nu+p\, h^{\mu\nu}+\mathrm{\Pi}^{\mu \nu}\,,
\ee
where $\varepsilon$ and $p$ are the energy density and isotropic pressure
along a four-velocity field $u^\mu$, $h^{\mu\nu} = g^{\mu\nu}+u^\mu u^\nu$, and $\mathrm{\Pi}^{\mu\nu}$ is the anisotropic and
traceless stress tensor,
\be
\mathrm{\Pi}^\mu_{\,\,\nu} \ = \mathrm{diag}(0,-2\mathrm{\Pi},\mathrm{\Pi},\mathrm{\Pi})\,.
\ee
The radial and tangential pressures are then related to $p$ and $\mathrm{\Pi}=\mathrm{\Pi}(r)$ through
\be
p_\perp-p_r = 3\mathrm{\Pi},\qquad 2p_\perp+p_r=3p\,.
\ee
For a comoving observer and to our order of approximation, we are left with $T^{\mu\nu}\!=\,\!^0T^{\mu\nu} +\,\!^2T^{\mu\nu}$, which can be written in terms of $\phi$ and $\psi$ as in (\ref{0T00poly})-(\ref{2Tii}). For further purposes, it will be convenient to represent the energy density as $\varepsilon=\,\!^0\varepsilon +\,\!^2\varepsilon$, where $\,\!^0\varepsilon=\rho$ is the rest-mass energy density and $\,\!^2\varepsilon=\rho_2$ is the first relativistic correction. This is the only observable that gets corrected in the 1PN approximation at the level of the energy-momentum tensor.

\begin{figure}
$$
\begin{array}{cc}
  \includegraphics[width=7cm]{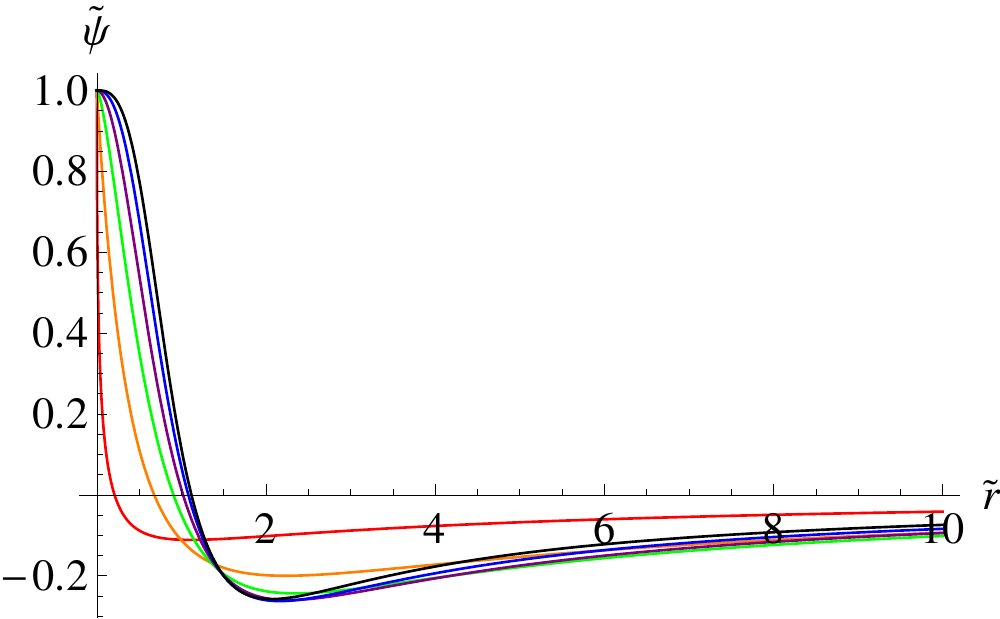} & \includegraphics[width=7cm]{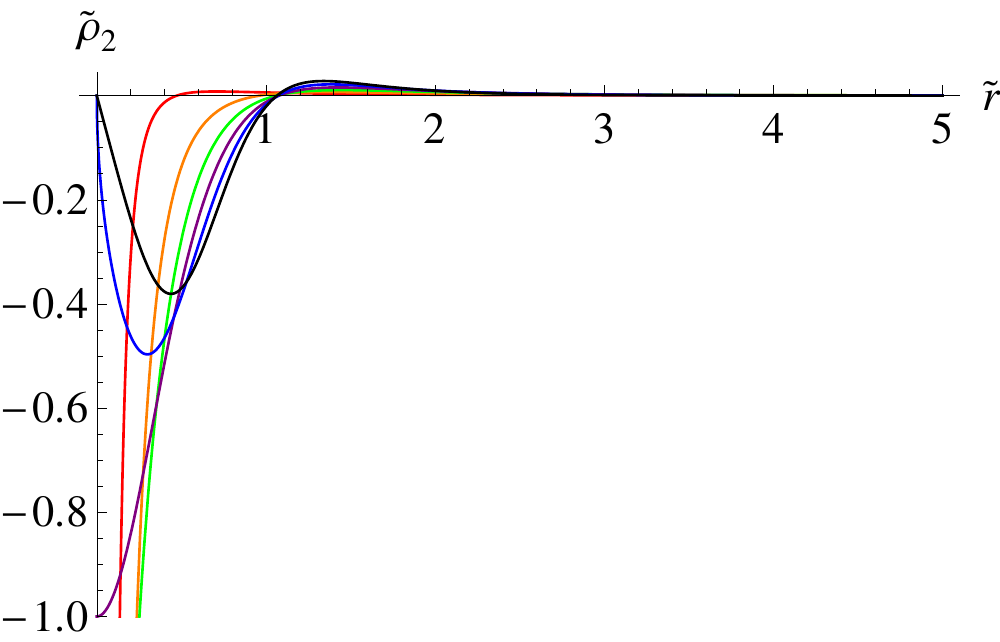}
\end{array}
$$
\caption{Left panel: Post-Newtonian potential $\tilde{\psi}$ as a function of the dimensionless radius $\tilde{r}$ for $m\in\{2,3,4,5,6,7\}$. Right panel: Correction to the energy density $\tilde{\rho}_2$ as a function of the dimensionless radius $\tilde{r}$ for $m\in\{2,3,4,5,6,7\}$. For all the plots we have set $\tilde{\psi}(0)=1$.\label{figpsi}}
\end{figure}

From (\ref{2T00}) we get
\begin{equation}
{}^{(2)}T^{00} \equiv \rho_2 = -\frac{a^{\frac{m-1}{2}}}{8\pi c^{2} G^{m}M^{m-1}} r^{\frac{m-5}{2}}\bigg[\frac{59-9m}{32}(-\phi)^{m+1}+m\mathrm{\Psi}(-\phi)^{m-1}\bigg]\,.
\end{equation}
In Fig. \ref{figpsi} (right panel) we plot the dimensionless $\tilde{\rho}_2$ as a function of $\tilde{r}$. We have also fixed $\tilde{\psi}(0)=1$ as discussed in the previous section. The correction to the energy density is somehow surprising; it is negative in the inner core but becomes positive for $r\geq a$. We believe this feature is essential to improve the behavior of the rotation curves as predicted by the Newtonian theory. This effect was already observed in \cite{Agon:2011mz}, but for the sake of comparison we will devote the next section to the study of this observable. As a final comment, note that the behavior of $\rho_2$ near the center is very similar to $\rho$ itself, i.e. $\rho_2\to-\infty$ for $m<5$, it is finite for $m=5$, and it vanishes for $m>5$. In fact, it is easy to verify that the inner slopes are the same (although with opposite signs), which makes the total energy density, $\tilde{\varepsilon}=\tilde{\rho}+\lambda\tilde{\rho}_2$, positive everywhere for $\lambda\equiv\frac{GM}{ac^{2}}\ll 1$ \footnote{This is the range of validity of the 1PN approximation.}.

It is important to note that the value of $\psi(0)$ fixes the post-Newtonian correction to the total mass of the system. Although this quantity is difficult to compute for general $m$, we were able to handle the integrals for the simplest case, the post-Newtonian Plummer model. If we set $m=5$, then
\begin{eqnarray}
\rho_{2}{(r)} =&&\!\!\!\!\!\!\! -\frac{GM^{2}}{8\pi c^{2} a^{4}}\bigg[\frac{5a^4}{4r\left(r^2+a^2\right)^{3/2}} \left(\frac{\psi{(0)}a^{2}}{G^{2}M^{2}} - \frac{13}{560}\right) \sin{\left(4 \arctan{\left(\frac{r}{a}\right)}\right)} \nonumber \\
&&\!\!\!\!\!\!\!\qquad\qquad\quad-\frac{5a^4}{280\left(r^2+a^2\right)^2}+\frac{4a^6}{7\left(r^2+a^2\right)^3}\bigg]\,.
\end{eqnarray}
The integral of the DF over all phase space must be normalized to the \emph{total} mass of the system,
\begin{equation}
\int \sqrt{h}T^{00} d^{3}x = M_{\mathrm{total}}={}^{0}\!M + {}^{2}\!M + \cdots\,,
\end{equation}
where ${}^{0}\!M$ is just the Newtonian mass given by (\ref{Nmass}) and $h$ is the determinant of the induced metric on a constant-$t$ hypersurface. At second order we obtain
\begin{equation}
\sqrt{h} = \bigg(1-\frac{2\phi}{c^2}\bigg)^{3/2} \approx 1 - \frac{3\phi}{c^2}\,
\end{equation}
and thus the 1PN correction to the total mass can be computed via
\begin{equation}
{}^{2}\!M = 4\pi\int_{0}^{\infty} \bigg[\rho_{2}-\frac{3\phi}{c^{2}}\rho \bigg]  r^{2} dr\,.
\end{equation}
For the Plummer model we find that
\begin{equation}
{}^{2}\!M = \frac{GM^{2}}{ac^{2}}\bigg(\frac{35\pi}{64} - \frac{13}{3360} + \frac{\psi{(0)}a^{2}}{6G^{2}M^{2}}\bigg)\,,
\end{equation}
or, in terms of the dimensionless quantities,
\begin{equation}
{}^{2}\!M = M \lambda \bigg(\frac{35\pi}{64} - \frac{13}{3360} + \frac{\tilde{\psi}{(0)}}{6}\bigg)\,.
\end{equation}
For other models one can perform the integral numerically for different values of $\psi(0)$, but the result is always of order $\mathcal{O}(\lambda)$, which is a small correction in the range of parameters allowed in the 1PN approximation.

\subsubsection{Rotation curves}

The galaxy rotation problem is the discrepancy between the observed galaxy rotation curves and the Newtonian prediction assuming a centrally dominated mass associated with the observed luminous material \cite{Persic:1995ru,Begeman:1991iy}. Even though dark matter is by far the most accepted explanation for the resolution to the galaxy rotation problem, there have been other proposals with varying degrees of success. Among them, the most popular ones involve certain modification of the laws of gravity, starting with the seminal works \cite{Milgrom:1983ca,Bekenstein:1984tv} and continuing with a large body of work that includes \cite{Mannheim:1988dj,Mannheim:1992vj,Sofue:2000jx,Nucamendi:2000jw,Lake:2003tr,Mak:2004hv,Capozziello:2004us,Reuter:2004nv,Moffat:2004bm,Martins:2007uf,Chang:2008yv} and the recent additions \cite{Alvarez:2012gx,Lu:2013bt,Stabile:2013jon,Alexandre:2013ura,Meierovich:2013jha,Chang:2013twa}.

On the other hand, while some authors argue that by including relativistic corrections the inclusion of a dark matter halo is
unnecessary at galactic scales \cite{Cooperstock-Tieu,Pro-Cooperstock1,Pro-Cooperstock2,Pro-Cooperstock3,Pro-Cooperstock4}, several
publications have pointed out that this is not entirely true \cite{Anticoperstok1,Anticoperstok2,Anticoperstok3,Anticoperstok4,Anticoperstok5}. The purpose of this section is then to investigate this issue in the 1PN approximation of general relativity. Here, the idea is not to argue whether GR is enough or not enough to overcome the galaxy rotation problem but to estimate the importance of the first corrections over the Newtonian gravity. In fact, one of the advantages of our framework is that it gives us the possibility to compare directly with the Newtonian theory, given that our models are direct generalizations of classical ones.

\begin{figure*}
$$
\begin{array}{cc}
 \includegraphics[width=7cm]{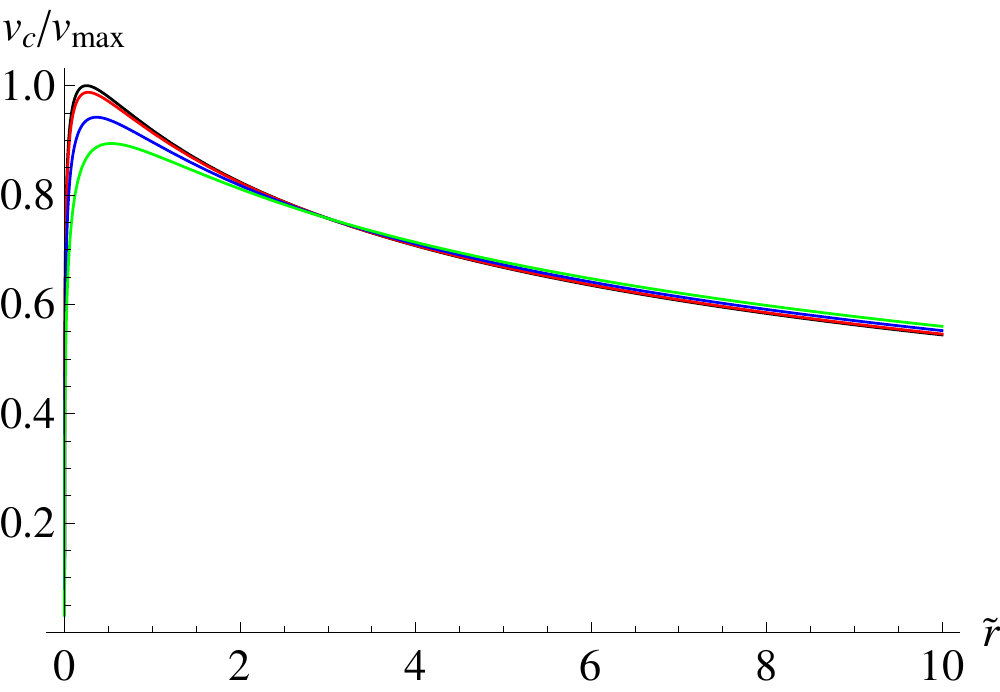} & \includegraphics[width=7cm]{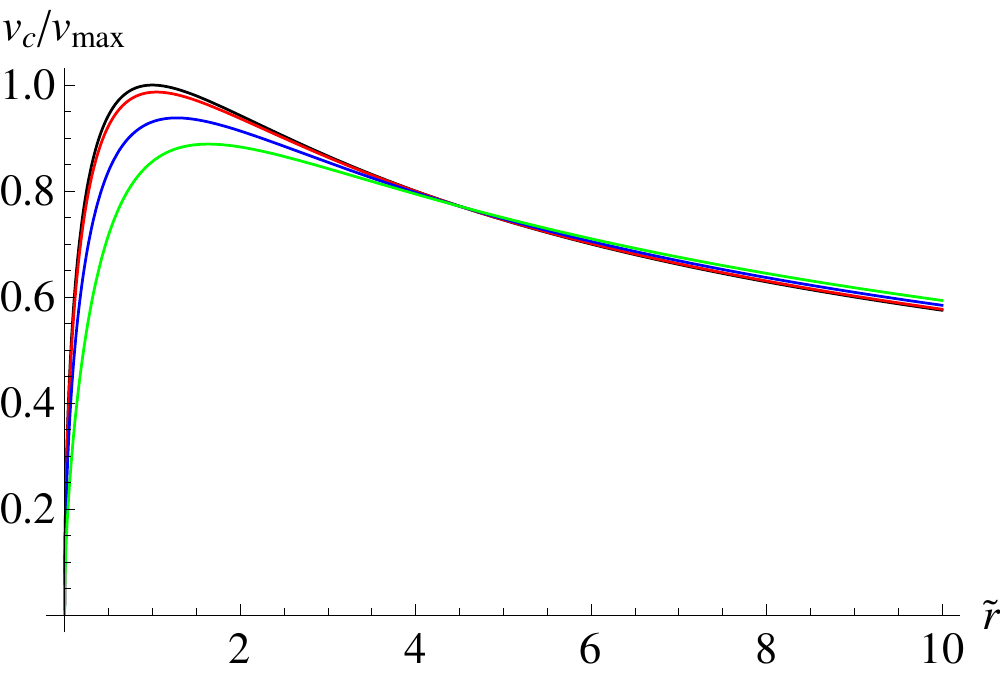}\\
 \quad\quad(a) & \quad\quad(b)\\
  \includegraphics[width=7cm]{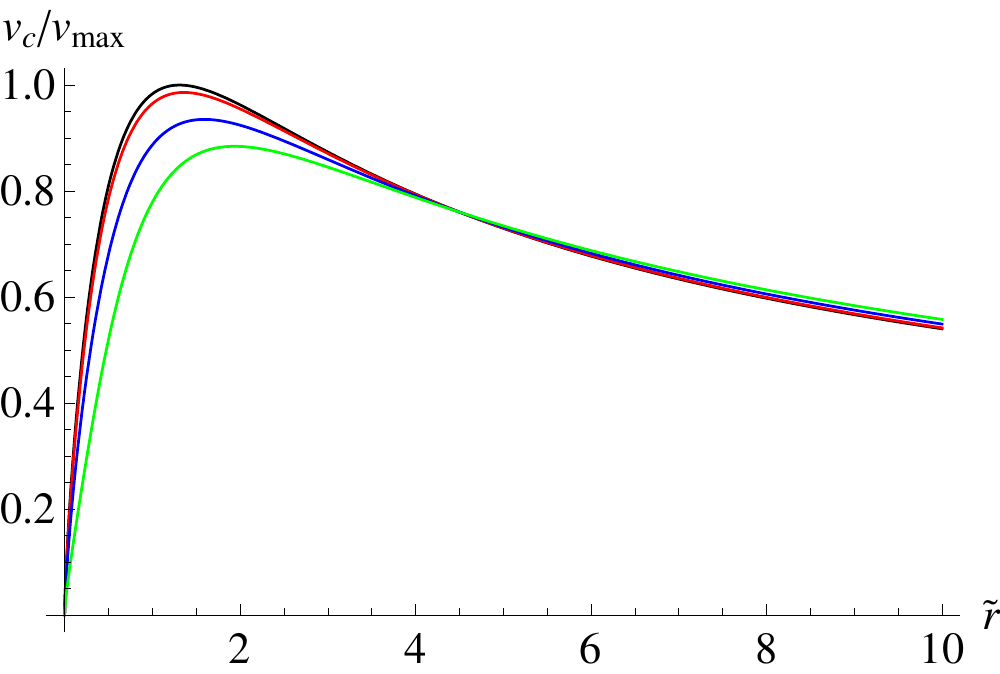} & \includegraphics[width=7cm]{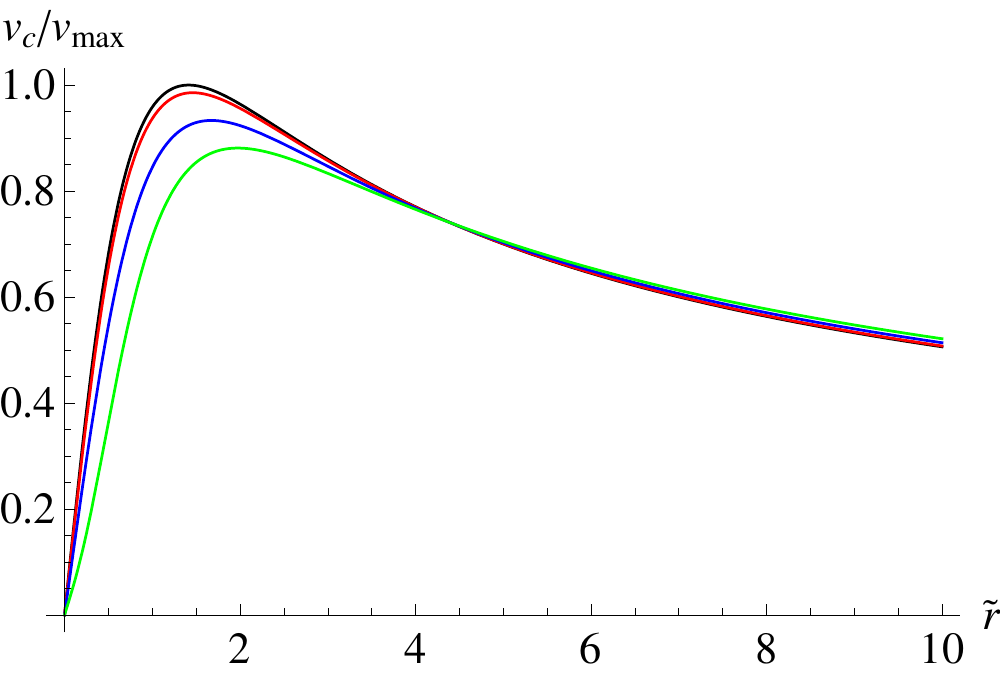}\\
 \quad\quad(c) & \quad\quad(d)\\
\includegraphics[width=7cm]{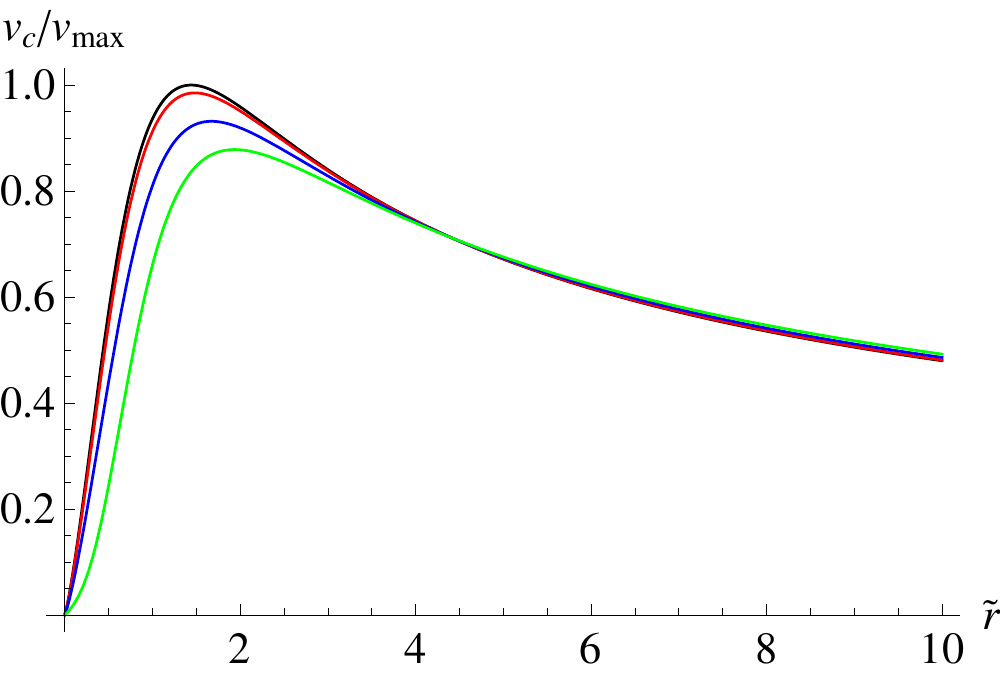} & \includegraphics[width=7cm]{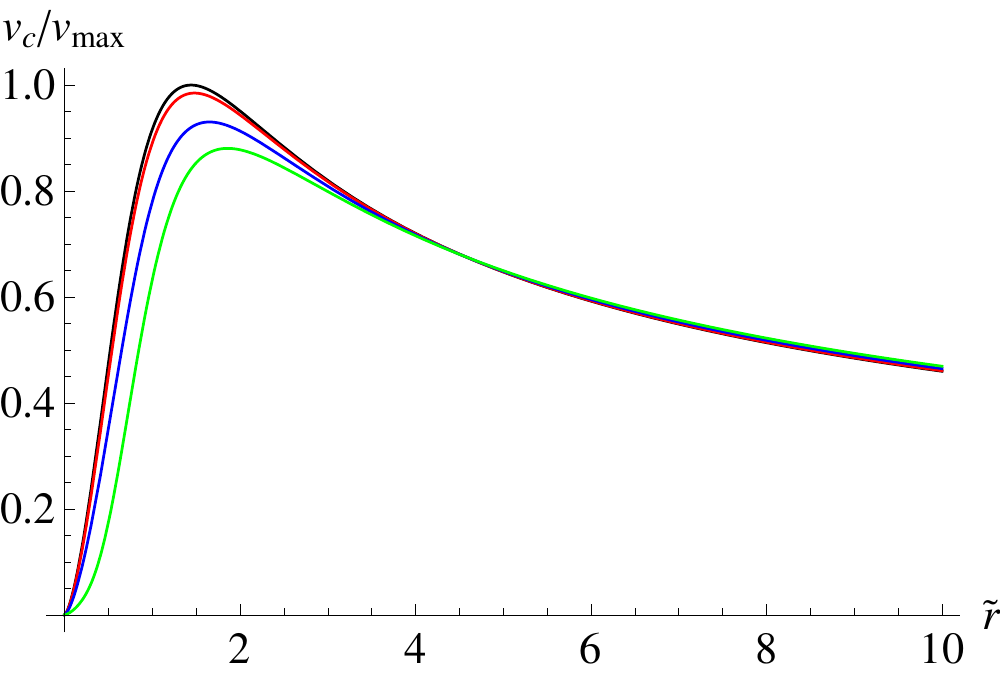}\\
 \quad\quad(e) & \quad\quad(f)
\end{array}
$$
\caption{Rotation curves $v_c$ as a function of $\tilde{r}$ for $m\in\{2,3,4,5,6,7\}$ and different values of $\lambda=0$ (black), $10^{-2}$ (red), $5\times10^{-2}$ (blue) and $10^{-1}$ (green). In order to make these plots we have chosen $\psi\sim\phi^2$ at $r\to0$. For $\lambda\sim10^{-2}$ or more, the differences with the Newtonian case become significant. In general, the relativistic corrections tend to flatten the curves, making $v_c$ smaller near the center and bigger for large distances.}
\label{figrot}
\end{figure*}

From Eq. (\ref{rotlaw}) we can obtain an expression for $v_c$ in terms of dimensionless quantities:
\be
v_c^2=
\tilde{r}\frac{\partial\tilde{\phi}}{\partial \tilde{r}}+\lambda\left[4r\tilde{\phi}\frac{\partial\tilde{\phi}}{\partial \tilde{r}}+\tilde{r}^2\left(\frac{\partial\tilde{\phi}}{\partial \tilde{r}}\right)^2+\tilde{r}\frac{\partial\tilde{\psi}}{\partial \tilde{r}}\right]\,.
\ee
The parameter $\lambda$ (defined in the previous section) is a measure
of how important the 1PN corrections are.  In Fig. \ref{figrot},
we plot the circular velocity for various models when $\lambda=0,10^{-2},5\times10^{-2}$, and $10^{-1}$. The 1PN corrections become important
as we increase $\lambda$; in general, the profile decreases in the inner region and increases far from the center, giving a flatter distribution in comparison to the standard case. As we can see, the general trend implies that general relativistic corrections actually improve the behavior of rotation curves, a phenomenon that was anticipated in \cite{Agon:2011mz,RamosCaro:2012rz}.

\subsubsection{Comments on stability\label{Nstability}}

While any positive, normalizable function of the integrals of motion represents an equilibrium solution of the Poisson-Vlasov or Einstein-Vlasov system, only a handful of density profiles are observed in real astrophysical systems. It is therefore important to narrow down the range of acceptable models to those which are stable against perturbations. Also, even if a particular model is unstable, the study of stability provides insights into the evolution of such systems over large time scales, a topic of great relevance in astrophysics.

Stability can be studied on several levels: most commonly, one linearizes the dynamics around an equilibrium solution of the Vlasov-Poisson system and studies the energy cost incurred by small deviations around the equilibrium (the energy method). Applying this methodology in Newtonian theory, we have learned a lot about the stability of stellar systems with respect to radial perturbations. In particular, a generalization of the Doremus-Feix-Baumann theorem asserts that all radial modes of a spherical stellar system with $\partial F(E,L)/\partial E < 0$ are stable \cite{Kandrup & Sygnet}. Applying this result to our family, we conclude that for all $m$ the Newtonian models are stable to radial perturbations. On the other hand, very little is known about nonradial perturbations of an anisotropic system, except by means of numerical simulations. A type of instability associated with nonradial perturbations commonly encountered in simulations is the radial-orbit instability, which occurs in systems where radial orbits are predominant \cite{Barnes,Merritt & Aguilar}. Comparing with our model, radial orbits are preferred for values of $m$ smaller than 5, and in the limit $m \rightarrow \infty$, radial orbits are suppressed. Thus, we expect that, in general, the models become more and more stable to nonradial perturbations with increasing $m$.

In addition to the energy method described above, other methods to study the stability of self-gravitating systems are available. For example, we may not linearize the perturbed equation of motion (nonlinear stability). We may decompose the perturbations into a linear combination of modes (spectral stability). Two studies of the stability of (Newtonian) generalized polytropes using such methods are \cite{Guo} and \cite{Polyachenko}. For the remainder of this section, however, we will focus on the stability of circular orbits since the tools available to us can be easily generalized to the relativistic setting.

In the theory of central potentials of Newtonian dynamics, the stability of a circular orbit in a potential $\phi(r)$ can be inferred from the so-called effective potential $\phi_{\mathrm{eff}}$:
\begin{equation}
\phi_{\mathrm{eff}} = \phi + \frac{L^{2}}{2mr^{2}}\,,
\end{equation}
where $L$ is the angular momentum of the particle and $m$ is its mass. From the kinematics of uniform circular motion, we have
\begin{equation}
\phi' = m\frac{v^2}{r} = \frac{L^2}{r^{3}m^{2}}\,,
\end{equation}
from which we find the energy and angular momentum of a circular orbit at radius $r$ to be
\begin{equation}
E = \frac{1}{2}r\phi' + \phi\,,
\end{equation}
\begin{equation}
L^{2} = m^{2}r^{3}\phi'\,.
\end{equation}
Substituting $L^{2}$ into $\phi_{\mathrm{eff}}$ and requiring the second derivative of the effective potential to be positive in order for the orbit to be stable, we find the criterion
\begin{equation}
\frac{3}{r}\phi' + \phi'' > 0\,.
\end{equation}
Alternatively, by differentiating $L^{2}$ with respect to $r$, we can recast the criterion in the following equivalent form:
\begin{equation}
L\frac{dL}{dr} > 0\,.
\end{equation}
The latter is known as Rayleigh's criterion \cite{Lord}, and it can be justified by the following reasoning: suppose we perturb a circular orbit of radius $a$ in the radially outward direction, to a new radius $r > a$, while keeping its angular momentum the same. Then if the initial orbit is to be stable, the gravitational force at $r$ must be greater than the centrifugal force (in the noninertial reference frame of the moving particle) so that the particle comes back to the initial radius $a$. This implies
\begin{equation}
L^{2}{(r)} > L^{2}{(a)}\,,
\end{equation}
or, by expanding $L^{2}$ around $a$, we recover the Rayleigh criterion. In the relativistic theory, this line of reasoning works for any static axisymmetric line element. In spherical coordinates $(t, r, \vartheta, \varphi)$, such a metric takes the form
\be
 ds^2=g_{tt} c^{2} dt^2 + g_{rr}dr^2 +  g_{\vartheta \vartheta }d\vartheta^2
 +g_{\varphi\varphi}d\varphi^2\,, \label{metsph}
\ee
where the components $g_{\mu\nu}$ depend only on the variables $r$ and $\vartheta$. The geodesic equation for a circular motion on the plane $\vartheta=\pi/2$ is then
\be
g_{tt,r} c^{2} \dot{t}^2+g_{\varphi\varphi,r}\dot{\varphi}^2=0\,. \label{geo1}
\ee
We also have the constants of motion:
\bea
&&-1= g_{tt} c^{2}\dot{t}^2 + g_{\varphi\varphi}\dot{\varphi}^2\,, \label{const1}    \\
&&E= g_{tt} c^{2} \dot{t}\,,  \label{const2}  \\
&&L= g_{\varphi\varphi}\dot{\varphi}\,,\label{const3}
\eea
corresponding to the square of the four-velocity, and the conserved quantities of the Killing vector field $\partial_{t}$ and of the Killing vector field $\partial_{\phi}$, respectively (here the overdot denotes a derivative with respect to the proper time $s$). The constant $E$ can be identified as the relativistic specific energy and $L$ as the specific angular momentum. Note that the equation of motion (\ref{geo1}) can be cast as a balance equation valid on the plane $\vartheta=\pi/2$:
\be
\frac{g_{tt,r}E^2}{g_{tt}^2 c^{2}} = - \frac{g_{\varphi\varphi,r}L^2}{g_{\varphi\varphi}^2}\,.
\label{bal}
\ee
So, as in the Newtonian case, we have a balance between the ``gravitational force'' and the ``centrifugal force.'' Following the same reasoning as for the Newtonian case, we obtain the criterion for stability,
\begin{equation}
LL_{,r}>0\,.
\end{equation}
We claim that this expression is equivalent to $EE_{,r} > 0$. Indeed, from (\ref{geo1})-(\ref{const3}) we find
\bea
&&L^2= \frac{g_{\varphi\varphi}^2   g_{tt,r} }{  g_{tt} g_{\varphi\varphi,r} -
 g_{tt,r} g_{\varphi\varphi}  }\,, \label{hh}  \\
&&E^2=-\frac{c^{2}g_{tt}^2 g_{\varphi\varphi,r}}{  g_{tt} g_{\varphi\varphi,r} -
 g_{tt,r} g_{\varphi\varphi}  }\,. \label{EE}
\eea
From (\ref{hh}) and (\ref{EE}) we get
\be
LL_{,r} =-\frac{g_{\varphi\varphi}  }{g_{tt}c^{2}}EE_{,r}\,.  \label{consc}
\ee
Since $-g_{\varphi\varphi}/ g_{tt}$ is identically positive, the claim is established. Next, we substitute the following components of the metric into (\ref{EE}) and expand to 1PN order:
\begin{equation}
g_{tt} = -c^{2} - 2\phi - \frac{2}{c^2}(\phi^{2}+\psi)\,,
\end{equation}
\begin{equation}
g_{\varphi\varphi} = \bigg(1 - \frac{2\phi}{c^{2}}\bigg)r^{2}\,.
\end{equation}
The result is
\begin{equation}
E = c^{2} + \phi + \frac{1}{2}r\phi' + \frac{1}{8c^2}\left(4\phi^{2} + 8\psi + 4r\phi\phi' + 3r^{2}\phi'^{2} + 4r\psi'\right)\,.
\end{equation}
To first order, we recover, as expected, Newtonian energy in addition to the rest mass. The Rayleigh criterion becomes
\begin{equation}
\frac{1}{2}c^{2}\left(3\phi' + r\phi''\right) + \left(3\phi\phi' + 2r\phi'^{2} + \frac{3}{2}\psi' + r\phi\phi'' + r^{2}\phi'\phi'' + \frac{1}{2}r\psi''\right) > 0\,.
\end{equation}
Notice that if we keep only the lowest order term, we recover the Newtonian Rayleigh criterion in terms of $\phi$. Using the dimensionless quantities, this is
\begin{equation}
\frac{1}{2}\left(3\tilde{\phi}' + \tilde{r}\tilde{\phi}''\right) + \lambda\left(3\tilde{\phi}\tilde{\phi}' + 2\tilde{r} \tilde{\phi}'^{2} + \frac{3}{2}\tilde{\psi}' + \tilde{r}\tilde{\phi}\tilde{\phi}'' + \tilde{r}^{2}\tilde{\phi}'\tilde{\phi}'' + \frac{1}{2}\tilde{r}\tilde{\psi}''\right) > 0\,.
\end{equation}

\begin{figure*}
$$
\begin{array}{cc}
 \includegraphics[width=7cm]{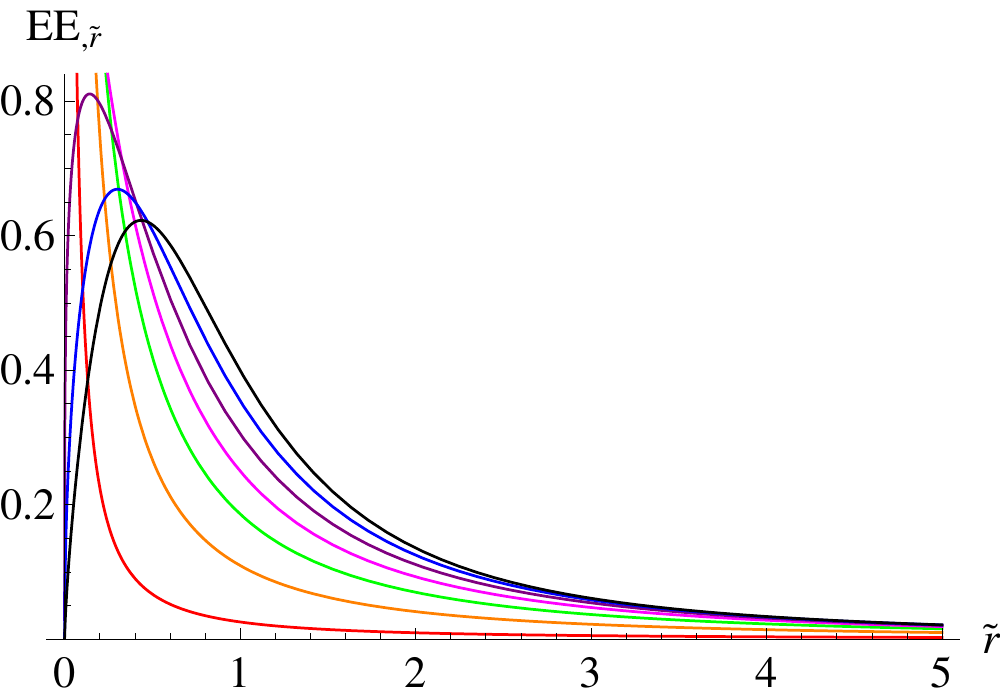} & \includegraphics[width=7cm]{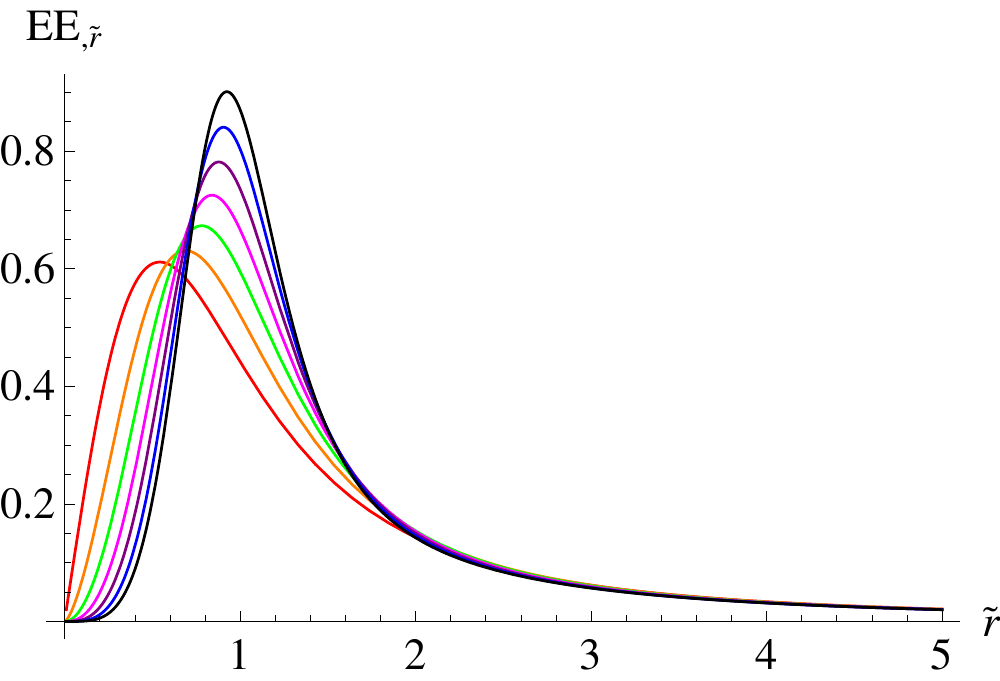}\\
 \quad\quad(a) & \quad\quad(b)\\
  \includegraphics[width=7cm]{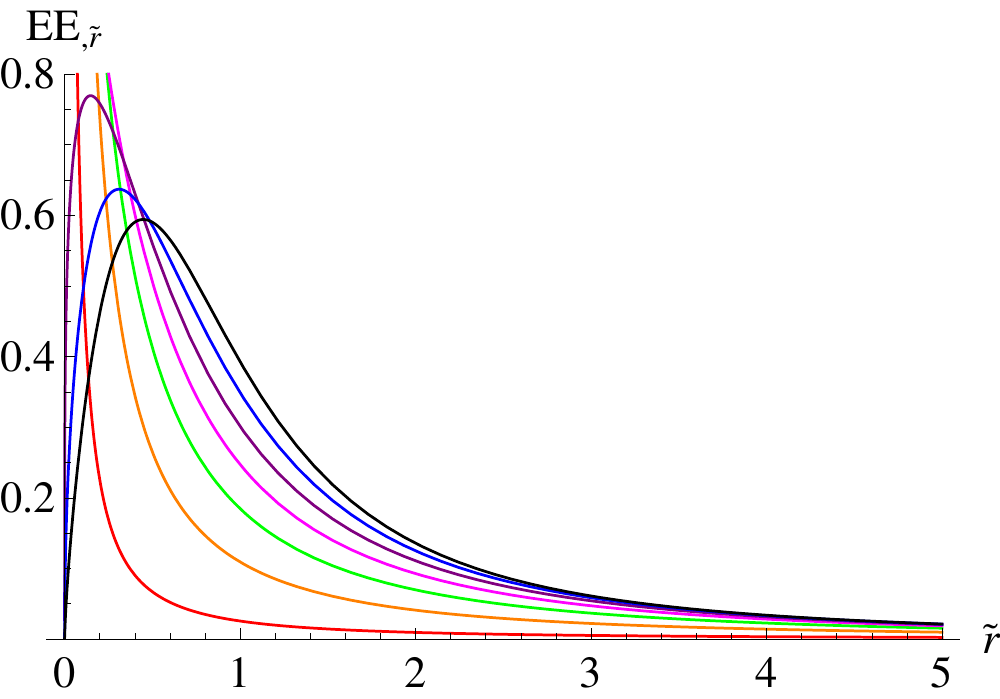} & \includegraphics[width=7cm]{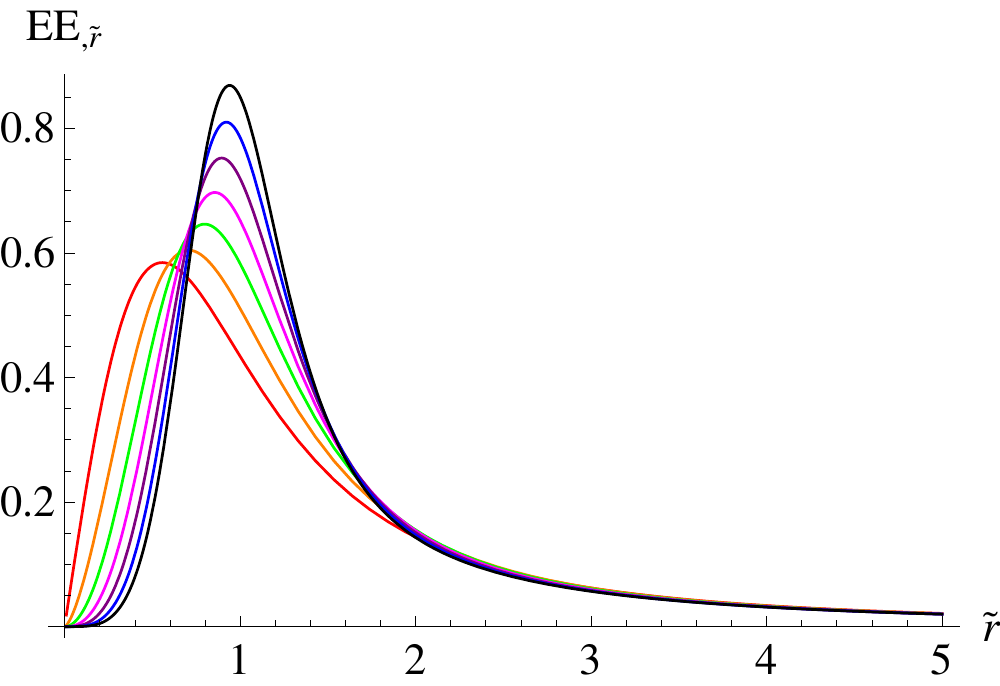}\\
 \quad\quad(c) & \quad\quad(d)\\
\includegraphics[width=7cm]{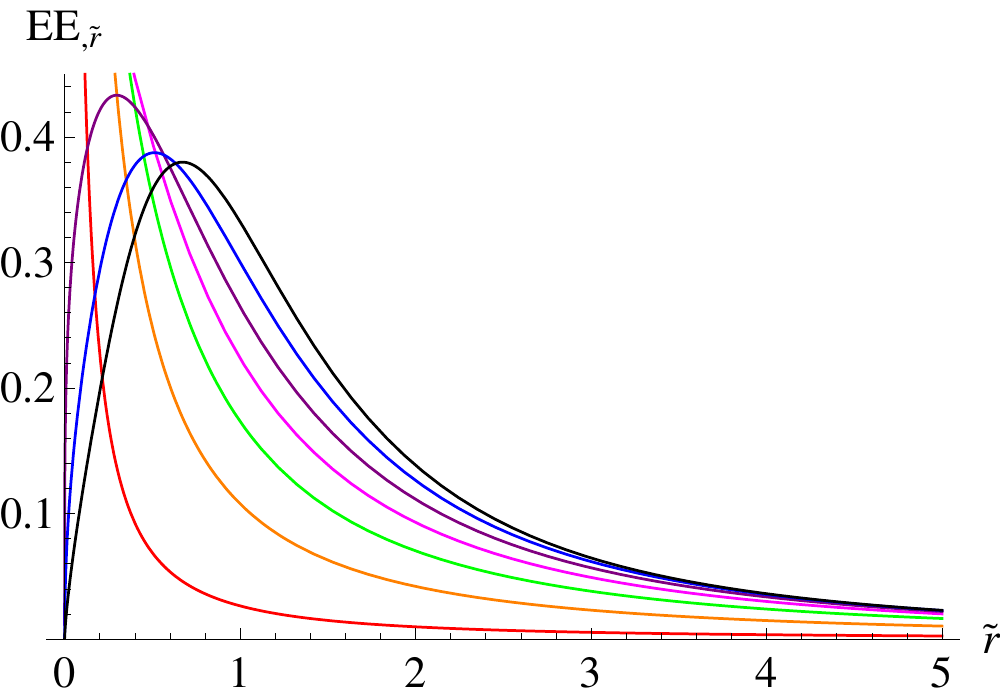} & \includegraphics[width=7cm]{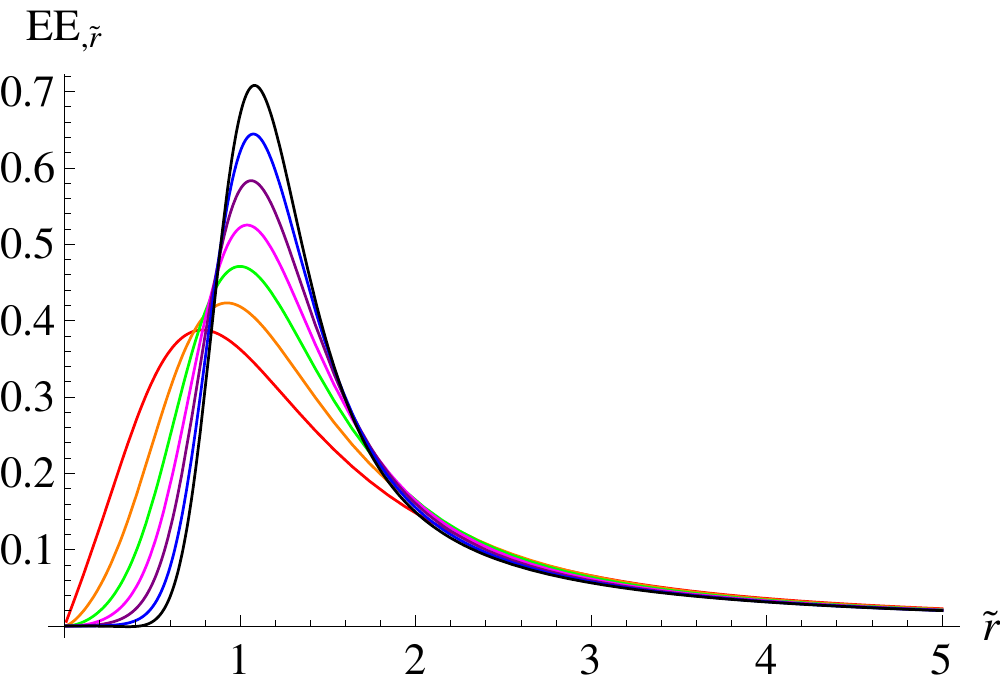}\\
 \quad\quad(e) & \quad\quad(f)
\end{array}
$$
\caption{We show $EE_{,\tilde{r}}$ as a function of $\tilde{r}$ for $m\in\{\frac{3}{2},2,\frac{5}{2},3,\frac{7}{2},4,\frac{9}{2}\}$ (left column), $m\in\{5,6,7,8,9,10,11\}$ (right column), and different values of $\lambda=0$ (first row), $10^{-2}$ (second row), $10^{-1}$ (last row). Rayleigh's criterion for stability is satisfied in all these cases, but for models with $m>5$ a small region of instability appears in the inner core for sufficiently large $\lambda$. However, for this range of parameters the 1PN approximation is no longer valid.}
\label{figstab}
\end{figure*}

In Fig. \ref{figstab} we plot this quantity as a function of $\tilde{r}$ for different values of $m$ and $\lambda$. Although for the range of parameters allowed in the 1PN approximation we find that the system is absent of instabilities, the general trend suggests that for models with $m>5$ and sufficiently large $\lambda$ there might be a small region near the center for which $EE_{,r}<0$, indicating unstable modes. Extrapolating our results to larger values of $\lambda$ we find that this phenomenon starts to happen when $\lambda\gtrsim5\times10^{-1}$. This fact seems to agree to a very good approximation with Buchdahl's theorem \cite{Buchdahl:1959zz,Chandra64}, a result that was derived in the context of spherically symmetric fluids in general relativity, and according to which a star with radius $R$ is stable when $GM/c^2R<4/9$.

\section{Final Remarks\label{FinalSec}}

We have constructed a one-parameter family of self-consistent star clusters that are spherically symmetric but anisotropic in velocity space. The model was constructed first in the Newtonian limit; then the first post-Newtonian corrections were computed. By self-consistent, we mean that the collective potential and the distribution function that gives rise to the density profile solve the Einstein-Vlasov system (in the 1PN approximation); i.e. they are simultaneous solutions of the Einstein field equation and the Vlasov equation.

The Newtonian distribution function is a particular case of a broader class of models known in the literature as generalized polytropes, which are the simplest anisotropic generalizations of polytropes. The family is labeled by $m$, which ranges from $1$ to infinity, and includes two commonly used models in astrophysics as particular cases, namely the Hernquist model for $m=3$ and the Plummer model for $m=5$. We found that the mass density profile is qualitatively different depending on whether $m<5$, $m=5$ or $m>5$. On one hand, the models with $m>5$ are unphysical due to nonmonotonic mass profiles, the ones with $m<5$, on the other hand, have density profiles that decrease monotonically with increasing radius. Moreover, the density profiles for $m<5$ diverge near the center as power laws, a feature believed to always be present in early-type galaxies.

While the equation of state for polytropes can be readily computed thanks to ergodicity of the distribution (which can be shown to imply that the mass density is monotonic), this is no longer the case if we introduce anisotropy into the model. Nevertheless, it still makes sense to talk about an equation of state for $m \leq 5$. Proceeding in analogy with the virial expansion for an interacting many-particle system at equilibrium, we compute the equation of state perturbatively in the two limits of small radius and large radius. Near the center of the system, we find that, to leading order, the pressure is proportional to the mass density, where the factor of proportionality is proportional to the expansion parameter $\lambda$ and therefore has to remain small. This is the equation of state used in cosmology for a matter-dominated universe (i.e. where matter moves nonrelativistically). At large distances from the center, the equation of state is approximately polytropic. We also calculated the adiabatic index and found that this quantity is well behaved for $m \leq 5$ but is negative in the inner region for $m > 5$. Because of anisotropy, the equation of hydrostatic equilibrium picks up an extra term. When identified as the first law of thermodynamics, the extra term can be interpreted as entropy production. Thus, anisotropy introduces dissipation into the system.

By solving for the post-Newtonian corrections, we found that generally 1PN corrections in the inner region differ qualitatively from the corrections in the outer region. For example, for all values of $m$, the post-Newtonian potential $\psi$ is a decreasing function in the inner core, assumes a minimum around $r \approx a$, and then increases to zero at infinity. This behavior of $\psi$ results in a flatter rotation curve compared to the Newotnian result. This fact can also be understood from the 1PN correction to the mass density: it is negative in the inner region and positive in the outer region. As a consequence, the total mass density is less centrally dominated than the Newtonian profile.

Finally, by studying the stability of circular orbits to radial perturbations, we discovered that, as the system becomes more and more relativistic, circular orbits in the inner core become unstable as $\lambda$ approaches $1/2$. This is in good agreement with the predictions of Buchdahl's theorem regarding the stability of relativistic fluid spheres. Of course, the stability of the model is best studied by perturbing the distribution function and does not follow directly from the stability of the orbits of the stars. Unfortunately, we expect such an investigation to be rather difficult (even in the Newtonian theory very little is known analytically about the stability of anisotropic models), but we leave these studies for future work.

\section*{Acknowledgements}
We are grateful to M. Lingam, R. Matzner, P. Morrison, J. Ramos-Caro, T. Rindler-Daller and L. Shepley for helpful discussions and comments on the manuscript. This material is based upon work supported by NASA under Grant No. NNX09AU86G and NSF Grant No. PHY-0969020. We also thank the Texas Cosmology Center, which is supported by the College of Natural Sciences, the Department of Astronomy at the
University of Texas at Austin, and the McDonald Observatory.

\appendix

\section{Another Family of Generalized Polytropes\label{appother}}
In this appendix, we present a second one-parameter family of generalized polytropes that also includes a polytrope as a particular case. We consider a distribution function of the form
\begin{equation}
F(\mathcal{E},L) = K_{n} L^{n} \mathcal{E}^{-\frac{n}{2}-\frac{1}{2}}\,,
\end{equation}
restricting ourselves to the range $-2 < n < 1$ so that the integral (\ref{intconv}) converges. For $n=0$, we obtain the polytrope with polytropic index $1$, one of the few analytically solvable polytropic models.
\subsection{The Newtonian limit}
Integrating over velocity space we get that the Newtonian mass density is given by
\begin{equation}
\rho{(r)} = 2^{\frac{n+3}{2}}\pi^{3/2}K_{n}r^{n}\mathrm{\Gamma}{\left(\frac{n}{2}+1\right)}\mathrm{\Gamma}{\left(-\frac{n}{2}+\frac{1}{2}\right)}\left(-\mathrm{\Phi}\right)\,.
\end{equation}
The Poisson equation for $\tilde{\mathrm{\Phi}}=-r\mathrm{\Phi}$ in Emden-Fowler form is
\begin{equation}
\frac{d^{2}\tilde{\mathrm{\Phi}}}{dr^{2}} = -\alpha_{n} r^{n} \tilde{\mathrm{\Phi}}\,,
\end{equation}
where in this case
\begin{equation}
\alpha_{n} = 2^{\frac{n+7}{2}}\pi^{5/2}K_{n}G \mathrm{\Gamma}{\left(\frac{n}{2}+1\right)}\mathrm{\Gamma}{\left(-\frac{n}{2}+\frac{1}{2}\right)}\,.
\end{equation}
The general solution can be expressed in terms of the Bessel functions:
\begin{equation}
\tilde{\mathrm{\Phi}}{(r)} = C_{1}\sqrt{r}J_{\frac{1}{n+2}}{\bigg(\frac{2\sqrt{\alpha_n}}{n+2}r^{1+\frac{n}{2}}\bigg)} + C_{2}\sqrt{r}Y_{\frac{1}{n+2}}{\bigg(\frac{2\sqrt{\alpha_n}}{n+2}r^{1+\frac{n}{2}}\bigg)}\,,
\end{equation}
or equivalently,
\begin{equation}
\mathrm{\Phi}{(r)} =\frac{C_{1}}{\sqrt{r}}J_{\frac{1}{n+2}}{\bigg(\frac{2\sqrt{\alpha_n}}{n+2}r^{1+\frac{n}{2}}\bigg)}+\frac{C_{2}}{\sqrt{r}}Y_{\frac{1}{n+2}}{\bigg(\frac{2\sqrt{\alpha_n}}{n+2}r^{1+\frac{n}{2}}\bigg)}\,.
\end{equation}
Here $J$ and $Y$ are Bessel functions of the first and second kinds, respectively. Because of the oscillatory nature of the Bessel functions, it is not possible to choose the constants $C_{1}$ and $C_{2}$ so that the mass density is everywhere nonnegative. Therefore, we will have to truncate the system at some radius, and the constants of integration must be chosen such that $\rho$ is nonnegative everywhere inside the cutoff radius. Also, the potential diverges as $r \rightarrow 0$ unless we choose $C_{2} = 0$.

Like the family of models labeled by $m$, this family of models has adjustable inner slopes and outer slopes that depend on the value of the parameter $n$. Using the asymptotic expansions of the Bessel functions for small arguments \footnote{For $z \rightarrow 0$, we have $J_{\alpha}{(z)} \sim \frac{1}{\Gamma{(\alpha+1)}}(z/2)^{\alpha}$ and $Y_{\alpha}{(z)} \sim -\frac{\Gamma{(\alpha)}}{\pi} (2/z)^{\alpha}$.}, we find that, close to the center,
\begin{equation}
\rho \sim K_{1} r^{n} + K_{2} r^{n-1}\,,
\end{equation}
where $K_{1}$ and $K_{2}$ are unimportant constant factors. Similarly, using the asymptotic expansions of the Bessel functions for large arguments \footnote{For $z \rightarrow \infty$, we have $J_{\alpha}{(z)} \sim \sqrt{\frac{2}{\pi z}}\cos{(z-\frac{\alpha\pi}{2}-\frac{\pi}{4})}$ and $Y_{\alpha}{(z)} \sim \sqrt{\frac{2}{\pi z}}\sin{(z-\frac{\alpha\pi}{2}-\frac{\pi}{4})}$.}, we find:
\begin{equation}
\rho \sim r^{\frac{3n}{4}-1}\,,
\end{equation}
where we averaged out the oscillatory piece in the asymptotic expansion. The case $n=-1$, corresponding to the DF $F(\mathcal{E},L) \propto L^{-1}$, is particularly interesting: the outer cusp is $\rho \sim r^{-7/4}$, which is the power law dependence of the Bahcall-Wolf cusp. As for the inner slope, it transitions from the $r^{-1}$ dependence to the steeper $r^{-2}$ dependence, a possible indication of the presence of a central black hole \cite{Le Delliou}\cite{Bahcall Wolf}.

For the rest of this section, we will set $C_{2} = 0$ so that $\mathrm{\Phi}$ remains finite as $r \rightarrow 0$. We will truncate the system at the first zero of the potential, i.e. at the radius $R$ given by
\begin{equation}
R^{1+\frac{n}{2}} = \bigg(\frac{n+2}{2\sqrt{\alpha_{n}}}\bigg)\gamma_{\frac{1}{n+2}}\,,
\end{equation}
where $\gamma_{\frac{1}{n+2}}$ is the first positive zero of the Bessel function $J_{\frac{1}{n+2}}$. This choice of truncation is justified by the fact that the mass density and the pressure are continuous across the boundary (since they have to vanish outside the system); thus, what we have is a dynamical analogue of a gaseous sphere, with a DF that is discontinuous at the radius of truncation. Also, we choose the constant $\phi_{0}$ such that the potential across the $r=R$ surface matches with the exterior Coulombic potential,
\begin{equation}
\phi_{0} = -\frac{GM}{R}\,.
\end{equation}
Then
\begin{equation}
\phi{(r)} = \frac{C_{1}}{\sqrt{r}} J_{\frac{1}{n+2}}{\bigg[\gamma_{\frac{1}{n+2}} \bigg(\frac{r}{R}\bigg)^{1+\frac{n}{2}} \bigg]} - \frac{GM}{R}\,,
\end{equation}
where we expressed $\alpha_{n}$ in terms of the cutoff radius $R$.
Substituting this into the mass density profile and normalizing to the total mass\footnote{We use the formula $\int_{0}^{\gamma_{\nu}} x^{\nu+1}J_{\nu}{(x)}dx = \gamma_{\nu}^{\nu+1}J_{\nu+1}{(\gamma_{\nu})}$.}, we find
\begin{equation}
\rho{(r)} = \frac{M}{4\pi R^{3}}\frac{\gamma_{\frac{1}{n+2}}(1+n/2)}{J_{\frac{n+3}{n+2}}{(\gamma_{\frac{1}{n+2}})}} \bigg(\frac{R}{r}\bigg)^{\frac{1}{2}-n}J_{\frac{1}{n+2}}{\bigg[\gamma_{\frac{1}{n+2}}\bigg(\frac{r}{R}\bigg)^{1+\frac{n}{2}}\bigg]}\,.
\end{equation}
The mass density is qualitatively different depending on whether $n<0$, $n=0$, or $n>0$. For $n>0$, $\rho$ vanishes at the origin and increases with distance in the inner region before decreasing to zero at some finite value of $r$. For $n=0$, $\rho$ is finite at $r=0$ and decreases with distance, while for $n<0$, $\rho$ is infinite at $r=0$ and decreases with distance. Substituting the value of $C_1$ in terms of $M$ back into the potential, we find
\begin{equation}
\phi{(r)} = -\frac{2GM}{R}\frac{1}{(n+2)\gamma_{\frac{1}{n+2}}J_{\frac{n+3}{n+2}}{(\gamma_{\frac{1}{n+2}})}}\sqrt{\frac{R}{r}}
J_{\frac{1}{n+2}}{\bigg[\gamma_{\frac{1}{n+2}}\bigg(\frac{r}{R}\bigg)^{1+\frac{n}{2}}\bigg]}- \frac{GM}{R}\,.
\end{equation}
As a consistency check, note that if we set $n=0$ (and use $\gamma_{1/2} = \pi$) we get
\begin{equation}
\phi{(r)} = -\frac{GM}{\pi r}\sin{\bigg(\frac{\pi r}{R}\bigg)} - \frac{GM}{R}\,,
\end{equation}
\begin{equation}
\rho{(r)} = \frac{M}{4r R^{2}}\sin{\bigg(\frac{\pi r}{R}\bigg)}\,,
\end{equation}
which is the familiar solution for the polytrope with polytropic index 1 \cite{polybook}. Next, we compute the velocity dispersion and the pressure in the radial and transverse directions:
\begin{equation}
\sigma_{r}^{2} = -\frac{1}{2}\phi = \frac{p_{r}}{\rho}\,,
\end{equation}
\begin{equation}
\sigma_{\theta}^{2} = -\frac{1}{2}\bigg(1+\frac{n}{2}\bigg)\phi = \frac{p_{\perp}}{\rho}\,.
\end{equation}
The anisotropy parameter is then
\begin{equation}
\beta = -\frac{n}{2}\,.
\end{equation}
Thus, nearly radial orbits are preferred for $n < 0$ and nearly circular orbits are preferred for $n > 0$.

\subsection{Comments on the post-Newtonian corrections}
At this point, we could try to solve for $\psi$ from Eq. (\ref{eins2}), with the Newtonian potential derived in the previous subsection. Unfortunately, Eq. (\ref{eins2}) was derived with the assumption that the correction due to the post-Newtonian escape velocity, given in (\ref{intscape}), is negligible. Unlike for the family labeled by $m$, this assumption no longer holds for this family of models.

To see this, recall that the quantity in (\ref{intscape}) is of the order of magnitude of $(\bar{v}/c)^{2+2\delta}$. In order for this quantity to be ignorable at 1PN order, we need $\delta > 0$, i.e. the DF is a decreasing function of energy. But the DF for the present family of models is an increasing function of energy; hence, Eq. (\ref{eins2}) is not valid. In particular, for the polytropic case ($n=0$), the quantity (\ref{intscape}) goes like $(\bar{v}/c)$, which is of order $\mathrm{0.5PN}$. For the case $n=-1$, this quantity is of order 1PN. This is problematic because it is inconsistent with the expansions (\ref{tmnexpa}) for $T^{00}$, which contains only even powers of $(\bar{v}/c)$.

In fact, the correction (\ref{intscape}) is the only place where powers of $(\bar{v}/c)$ other than even powers could be potentially introduced into our calculation. Let us come back to the family labeled by $m$ and compute the order of magnitude of (\ref{intscape}) for the two most important models: Hernquist and Plummer. For the Hernquist model, we have $\delta = 2$ and (\ref{intscape}) goes like$(\bar{v}/c)^6$. For the Plummer model, we have $\delta = \frac{7}{2}$ and (\ref{intscape}) goes like $(\bar{v}/c)^9$. Thus, the correction (\ref{intscape}) consists of even powers of $(\bar{v}/c)$ for the Hernquist model but introduces odd powers of $(\bar{v}/c)$ in the case of the Plummer model (similarly to the polytrope case of the family labeled by $n$).

We suggest that the presence of odd powers of $(\bar{v}/c)$ (corresponding to half-integer post-Newtonian orders) should be interpreted as indicating dissipative effects such as collision or radiation reaction. For example, gravitational radiation usually appear at $2.5$PN order. As a result, there seems to be dissipation of some kind in the Plummer model at $4.5$PN order. As for the polytropic case of the family labeled by $n$, the fact that dissipation makes its appearance at the lowest order possible is expected since the model is that of a compact object, where collisions cannot be ignored.

\section{Models with Multiple Components\label{appglue}}
The mass density profile for the models with $m > 5$ increases with distance in the inner region. While this is generally considered unphysical, we can eliminate this feature by forming composite models. In this appendix, we present a simple example where we replace the inner region of such models by a constant mass density. A uniform density sphere of radius $\bar{a}$ and mass $M_{1}$ is described by the potential
\begin{equation}
\phi_{1}{(r)} = \frac{GM_{1}}{2\bar{a}^{3}}r^{2}- K\,,
\end{equation}
where $K$ is an arbitrary constant shift, to be determined by continuity of the composite potential. We will glue this model to a hypervirial model with mass $M_{2}$, length scale $a$, and some $m > 5$ (we will do this for all such values of $m$):
\begin{equation}
\phi_{2}{(r)} = -\frac{GM_{2}}{(a^{\frac{m-1}{2}}+r^{\frac{m-1}{2}})^{\frac{2}{m-1}}}\,.
\end{equation}
The gluing takes place at the radius where the mass density of $\phi_{2}$ is maximal. Doing so not only makes the composite profile nonincreasing with distance, but the first derivative of $\rho$ will also be continuous. The two length scales $\bar{a}$ and $a$ are then related by Eq. (\ref{abar}). Requiring continuity of the first derivative of the potential fixes the ratio of the total masses to be
\begin{equation}
\frac{M_{1}}{M_{2}} = \bigg(\frac{m-5}{2m}\bigg)^{\frac{m+1}{m-1}}\,.
\end{equation}
Finally, requiring continuity of the potential then fixes $K$ to be
\begin{equation}
K = \frac{5GM_{2}}{4r_{0}}\bigg(\frac{m-5}{2m}\bigg)^{\frac{2}{m-1}}\bigg(1-\frac{1}{m}\bigg)\,.
\end{equation}

Another possible two-component model consists of gluing together two models in our family, with radii $a$ and $b$, total masses $M_{1}$ and $M_{2}$, and parameters $m_{1}$ and $m_{2}$, say, at the radius $r=a$. We require the potential and its first derivative to be continuous across the junction. Continuity of the potential implies the following relation between the ratio of the total masses and the ratio of the radii:
\begin{equation}
\frac{M_{1}}{M_{2}} = \frac{2^{\frac{2}{m_{1}-1}}}{[1+(b/a)^{\frac{m_{2}-1}{2}}]^{\frac{2}{m_{2}-1}}}\,.
\end{equation}
Continuity of the first derivative of the potential implies
\begin{equation}
\frac{M_{1}}{M_{2}} = \frac{2^{\frac{m_{1}+1}{m_{1}-1}}}{[1+(b/a)^{\frac{m_{2}-1}{2}}]^{\frac{m_{2}+1}{m_{2}-1}}}\,.
\end{equation}
Solving the system above, we finally find
\begin{equation}
a = b\,
\end{equation}
and
\begin{equation}
\frac{M_{1}}{M_{2}} = 2^{\frac{2}{m_{1}-1}-\frac{2}{m_{2}-1}}\,.
\end{equation}


\begin{thebibliography}{9999}


\bibitem{kand1} W.~Israel and H.~Kandrup,
 ``Nonequilibrium statistical mechanics in the general theory of relativity I. A general formalism,''
 Ann. Phys. \textbf{152}, 30 (1984).

\bibitem{kand2} H.~Kandrup,
``Non-equilibrium statistical mechanics in the general theory of relativity : II. Linear fields in a kinetic approximation,''
 Ann. Phys. \textbf{153}, 44 (1984).

\bibitem{kand3} H.~Kandrup,
``Non-equilibrium statistical mechanics in the general theory of relativity. III. Collisional stellar dynamics,''
 Ann. Phys. \textbf{169}, 352 (1986).

\bibitem{kremer} C.~Cercignani and G.~M.~Kremer, {\em The Relativistic Boltzmann Equation: Theory and Applications} (Birkh\"{a}user Verlag, Basel,2002).

\bibitem{chacon} G.~Chac\'on-Acosta and G.~M.~Kremer,
``Fokker-Planck-type equations for a simple gas and for a semirelativistic Brownian motion from a relativistic kinetic theory,''
Phys. Rev. E \textbf{76} 021201 (2007).

\bibitem{nucleus}
  G.~Steigman,
``Primordial nucleosynthesis: successes and challenges,''
  Int.\ J.\ Mod.\ Phys.\ E {\bf 15}, 1 (2006)
 [astro-ph/0511534].

\bibitem{eddi}
A.~S.~Eddington,
``The distribution of stars in globular clusters,''
Mon. Not. Roy. Astron. Soc. \textbf{76}, 572 (1916).

\bibitem{frick} W.~Fricke,
``Dynamische Begründung der Geschwindigkeitsverteilung im Sternsystem,''
Astron. Nachr. \textbf{280}, 193 (1952).

\bibitem{lynden} D.~Lynden-Bell,
``Stellar dynamics: Exact solution of the self-gravitation equation,''
Mon. Not. Roy. Astron. Soc. \textbf{123}, 447 (1962).

\bibitem{hunt93} C.~Hunter and E.~Qian,
``Two-integral distribution functions for axisymmetric galaxies,''
Mon. Not. Roy. Astron. Soc. \textbf{262}, 401 (1993).

\bibitem{jiao} Z.~Jiang and L.~Ossipkov,
``Two-integral distribution functions for axisymmetric systems,''
Mon. Not. Roy. Astron. Soc. \textbf{379}, 1133 (2007)
[arXiv:0708.3640 [math-ph]].

\bibitem{PRGfrac}
  J.~F.~Pedraza, J.~F.~Ramos-Caro, and G.~A.~Gonzalez,
``Fractional Derivative Approach to the Self-gravitation Equation,''
  Mon.\ Not.\ Roy.\ Astron.\ Soc.\  {\bf 391}, L24 (2008)
 [arXiv:0807.0119 [astro-ph]].

\bibitem{An:2012fw}
  J.~An, E.~Van Hese, and M.~Baes,
``Phase-space consistency of stellar dynamical models determined by separable augmented densities,''
  Mon.\ Not.\ Roy.\ Astron.\ Soc.\  {\bf 422}, 652 (2012)
[arXiv:1202.0004 [astro-ph.CO]].

\bibitem{Fackerell}
  E. D. Fackerell,
``Relativistic, Spherically Symmetric Star Clusters. V. a. Relativistic Version of Plummer's Model,''
  Astrophys. J. {\bf 165}, 489 (1971).

\bibitem{Buchdahl:1959zz}
  H.~A.~Buchdahl,
``General Relativistic Fluid Spheres,''
  Phys.\ Rev.\  {\bf 116}, 1027 (1959).

\bibitem{Glass Mashhoon}
  E. N. Glass and B. Mashhoon,
``On a spherical star system with a collapsed core''
  Astrophys. J. {\bf 205}, 570 (1976).

\bibitem{eins1}
  A.~Einstein, L.~Infeld and B.~Hoffmann,
``The Gravitational equations and the problem of motion,''
  Ann. Math.\  {\bf 39}, 65 (1938).

\bibitem{eins2}
  A.~Einstein and L.~Infeld,
``The Gravitational equations and the problem of motion. 2.,''
  Ann. Math.\  {\bf 41}, 455 (1940).

\bibitem{eins3} A.~Einstein and L.~Infeld L,
``On the motion of particles in general relativity theory,''
 Canad. J. Math. \textbf{1}, 209 (1949).

\bibitem{futamase}
  T.~Futamase and Y.~Itoh,
``The post-Newtonian approximation for relativistic compact binaries,''
  Living Rev.\ Rel.\  {\bf 10}, 2 (2007).

\bibitem{bichak}
  T.~Ledvinka, G.~Schaefer and J.~Bicak,
``Relativistic Closed-Form Hamiltonian for Many-Body Gravitating Systems in the Post-Minkowskian Approximation,''
  Phys.\ Rev.\ Lett.\  {\bf 100}, 251101 (2008)
[arXiv:0807.0214 [gr-qc]].

\bibitem{Agon:2011mz}
  C.~A.~Agon, J.~F.~Pedraza and J.~Ramos-Caro,
``Kinetic Theory of Collisionless Self-Gravitating Gases: Post-Newtonian Polytropes,''
  Phys.\ Rev.\ D {\bf 83}, 123007 (2011)
[arXiv:1104.5262 [gr-qc]].

\bibitem{RamosCaro:2012rz}
  J.~Ramos-Caro, C.~A.~Agon and J.~F.~Pedraza,
``Kinetic Theory of Collisionless Self-Gravitating Gases: II. Relativistic Corrections in Galactic Dynamics,''
  Phys.\ Rev.\ D {\bf 86}, 043008 (2012)
[arXiv:1206.5804 [gr-qc]].

\bibitem{Bowers}
  R.~L.~Bowers and E.~P.~T.~Liang,
``Anisotropic Spheres in General Relativity,''
  Astrophys.\ J.\  {\bf 188}, 3657 (1974).

\bibitem{Bayin:1982vw}
  S.~S.~Bayin,
``Anisotropic Fluid Spheres In General Relativity,''
  Phys.\ Rev.\ D {\bf 26}, 1262 (1982).

\bibitem{Singh:1992zz}
  T.~Singh, G.~P.~Singh, and R.~S.~Srivastava,
``Static anisotropic fluid spheres in General Relativity,''
  Int.\ J.\ Theor.\ Phys.\  {\bf 31}, 545 (1992).

\bibitem{Coley:1994pk}
  A.~A.~Coley and B.~O.~J.~Tupper,
``Spherically symmetric anisotropic fluid ICKV spacetimes,''
  Class.\ Quant.\ Grav.\  {\bf 11}, 2553 (1994).

\bibitem{Singh:1995zz}
  T.~Singh, G.~P.~Singh, and A.~M.~Helmi,
``New solutions for charged anisotropic fluid spheres in general relativity,''
  Nuovo Cim.\ B {\bf 110}, 387 (1995).

\bibitem{Martinez:1996jq}
  J.~Martinez,
``Transport processes in anisotropic gravitational collapse,''
  Phys.\ Rev.\ D {\bf 53}, 6921 (1996)
[arXiv:astro-ph/9602081].

\bibitem{Das:1997xz}
  A.~Das, N.~Tariq, D.~Aruliah, and T.~Biech,
``Spherically symmetric collapse of an anisotropic fluid body into an exotic black hole,''
  J.\ Math.\ Phys.\  {\bf 38}, 4202 (1997).

\bibitem{Herrera Santos 1997}
  L. Herrera and N. O. Santos,
``Local anisotropy in self-gravitating systems''
  Phys. Rep. {\bf 286}, 53 (1997).

\bibitem{Herrera:1997si}
  L.~Herrera, A.~Di Prisco, J.~L.~Hernandez-Pastora and N.~O.~Santos,
``On the role of density inhomogeneity and local anisotropy in the fate of spherical collapse,''
  Phys.\ Lett.\ A {\bf 237}, 113 (1998)
 [arXiv:gr-qc/9711002].

\bibitem{Corchero:1997tc}
  E.~S.~Corchero,
``Equilibrium of spheres with local anisotropy in postNewtonian gravity: Application to white dwarfs,''
  Class.\ Quant.\ Grav.\  {\bf 15}, 3645 (1998)
[arXiv:gr-qc/9712081].

\bibitem{Das:2000ym}
  A.~Das and S.~Kloster,
``Analytical solutions of a spherically symmetric collapse of an anisotropic fluid body into a regular black hole,''
  Phys.\ Rev.\ D {\bf 62}, 104002 (2000).

\bibitem{Dev:2000gt}
  K.~Dev and M.~Gleiser,
``Anisotropic stars: Exact solutions,''
  Gen.\ Rel.\ Grav.\  {\bf 34}, 1793 (2002)
[arXiv:astro-ph/0012265].

\bibitem{Herrera:2001vg}
  L.~Herrera, A.~Di Prisco, J.~Ospino, and E.~Fuenmayor,
``Conformally flat anisotropic spheres in general relativity,''
  J.\ Math.\ Phys.\  {\bf 42}, 2129 (2001)
 [arXiv:gr-qc/0102058].

\bibitem{Hernandez:2001nr}
  H.~Hernandez and L.~A.~Nunez,
``Nonlocal equation of state in anisotropic static fluid spheres in general relativity,''
  Can.\ J.\ Phys.\  {\bf 82}, 29 (2004)
 [arXiv:gr-qc/0107025].

\bibitem{Mak:2001eb}
  M.~K.~Mak and T.~Harko,
``Anisotropic stars in general relativity,''
  Proc.\ Roy.\ Soc.\ Lond.\ A {\bf 459}, 393 (2003)
 [arXiv:gr-qc/0110103].

\bibitem{Krisch:2001ay}
  J.~P.~Krisch and E.~N.~Glass,
 ``Adding twist to anisotropic fluids,''
  J.\ Math.\ Phys.\  {\bf 43}, 1509 (2002)
 [arXiv:gr-qc/0112019].

\bibitem{Herrera:2002bm}
  L.~Herrera, J.~Martin, and J.~Ospino,
 ``Anisotropic geodesic fluid spheres in general relativity,''
  J.\ Math.\ Phys.\  {\bf 43}, 4889 (2002)
 [arXiv:gr-qc/0207040].

\bibitem{Dev:2003qd}
  K.~Dev and M.~Gleiser,
 ``Anisotropic stars. 2. Stability,''
  Gen.\ Rel.\ Grav.\  {\bf 35}, 1435 (2003)
 [arXiv:gr-qc/0303077].

\bibitem{Herrera:2004xc}
  L.~Herrera, A.~Di Prisco, J.~Martin, J.~Ospino, N.~O.~Santos, and O.~Troconis,
 ``Spherically symmetric dissipative anisotropic fluids: A General study,''
  Phys.\ Rev.\ D {\bf 69}, 084026 (2004)
 [arXiv:gr-qc/0403006].

\bibitem{Chaisi:2005rb}
  M.~Chaisi and S.~D.~Maharaj,
 ``Compact anisotropic spheres with prescribed energy density,''
  Gen.\ Rel.\ Grav.\  {\bf 37}, 1177 (2005)
 [arXiv:gr-qc/0504098].

\bibitem{Naidu:2005pj}
  N.~F.~Naidu, M.~Govender, and K.~S.~Govinder,
 ``Thermal evolution of a radiating anisotropic star with shear,''
  Int.\ J.\ Mod.\ Phys.\ D {\bf 15}, 1053 (2006)
 [arXiv:gr-qc/0509088].

\bibitem{Maharaj:2005vb}
  S.~D.~Maharaj and M.~Chaisi,
 ``New anisotropic models from isotropic solutions,''
  Math.\ Methods Appl.\ Sci.\  {\bf 29}, 67 (2006)
 [arXiv:gr-qc/0510073].

\bibitem{Chaisi:2006sc}
  M.~Chaisi and S.~D.~Maharaj,
 ``A New algorithm for anisotropic solutions,''
  Pramana {\bf 66}, 313 (2006)
 [arXiv:gr-qc/0601030].

\bibitem{Barreto:2006cr}
  W.~Barreto, B.~Rodriguez, L.~Rosales, and O.~Serrano,
 ``Self-similar and charged radiating spheres: An Anisotropic approach,''
  Gen.\ Rel.\ Grav.\  {\bf 39}, 23 (2007); {\bf 39}, 537(E) (2007)
 [arXiv:gr-qc/0611089].

\bibitem{Herrera:2007kz}
  L.~Herrera, J.~Ospino, and A.~Di Prisco,
 ``All static spherically symmetric anisotropic solutions of Einstein's equations,''
  Phys.\ Rev.\ D {\bf 77}, 027502 (2008)
 [arXiv:0712.0713 [gr-qc]]

\bibitem{Herrera:2008bt}
  L.~Herrera, N.~O.~Santos, and A.~Wang,
 ``Shearing Expansion-free Spherical Anisotropic Fluid Evolution,''
  Phys.\ Rev.\ D {\bf 78}, 084026 (2008)
 [arXiv:0810.1083 [gr-qc]]

\bibitem{Lake:2009cd}
  K.~Lake,
 ``Generating static spherically symmetric anisotropic solutions of Einstein's equations from isotropic Newtonian solutions,''
  Phys.\ Rev.\ D {\bf 80}, 064039 (2009)
 [arXiv:0905.3546 [gr-qc]].

\bibitem{Sharma:2012vc}
  R.~Sharma and R.~Tikekar,
 ``Space-time inhomogeneity, anisotropy and gravitational collapse,''
  Gen.\ Rel.\ Grav.\  {\bf 44}, 2503 (2012)
 [arXiv:1206.6011 [gr-qc]].

\bibitem{Mimoso:2013iga}
  J.~\'{e} P.~Mimoso, M.~Le Delliou and F.~C.~Mena,
 ``Local conditions separating expansion from collapse in spherically symmetric models with anisotropic pressures,''
  Phys. Rev. D {\bf 88}, 043501 (2013)
 [arXiv:1302.6186 [gr-qc]].

\bibitem{Herrera:2013dfa}
  L.~Herrera and W.~Barreto,
 ``Newtonian polytropes for anisotropic matter: General framework and applications,''
  Phys.\ Rev.\ D {\bf 87}, 087303 (2013)
 [arXiv:1304.2824 [astro-ph.IM]].

\bibitem{Sharma:2013fsa}
  R.~Sharma and S.~Das,
 ``Collapse of a relativistic self-gravitating star with radial heat flux: Impact of anisotropic stresses,''
  J. Grav. 2013 (2013) 659605
 [arXiv:1304.7765 [gr-qc]].

\bibitem{Sgro:2013tia}
  M.~A.~Sgr\'{o}, D.~J.~Paz and M.~E.~Merch\'{a}n,
 ``Anisotropic Halo Model,''
  arXiv:1305.0563.

\bibitem{WB} S.~Weinberg,
{\em Gravitation and Cosmology: Principles and Applications of
the General Theory of Relativity} (John Wiley \& Sons, New York, 1972).

\bibitem{Ein}
  A.~Einstein,
``The foundation of the general theory of relativity,''
  Annalen der Physik, {\bf 354}, 769 (1916).

\bibitem{Balasin:2006cg}
  H.~Balasin and D.~Grumiller,
``Non-Newtonian behavior in weak field general relativity for extended rotating sources,''
  Int.\ J.\ Mod.\ Phys.\ D {\bf 17}, 475 (2008)
  [astro-ph/0602519].

\bibitem{Rendall:1996gx}
  A.~D.~Rendall,
``An Introduction to the Einstein-Vlasov system,''
  gr-qc/9604001.

\bibitem{Rendall:2002xa}
  A.~D.~Rendall,
``The Einstein-Vlasov system,''
  gr-qc/0208082.

\bibitem{Andreasson:2011ng}
  H.~Andr\'{e}asson,
``The Einstein-Vlasov System/Kinetic Theory,''
  Living Rev.\ Rel.\  {\bf 14}, 4 (2011)
  [arXiv:1106.1367 [gr-qc]].

\bibitem{RamosCaro:2008zz}
  J.~Ramos-Caro and G.~A.~Gonzalez,
``Fokker-Planck-Rosenbluth-type equations for self-gravitating systems in the 1PN approximation,''
  Class.\ Quant.\ Grav.\  {\bf 25}, 045011 (2008)
  [arXiv:0806.4281 [gr-qc]].

\bibitem{jeans} J.~H.~Jeans,
``On the theory of star-streaming and the structure of the universe,''
  Mon. Not. Roy. Astron. Soc. {\bf 76}, 70 (1915).

\bibitem{Schaeffer} J.~Schaeffer,
``A Class of Counterexamples to Jeans' Theorem for the Vlasov-Einstein System,''
  Commun. Math. Phys. {\bf 204}, 313 (1999).

\bibitem{Chandrasekhar} S.~Chandrasekhar, {\em An Introduction to the Theory of Stellar Structure}
(University of Chicago Press, Chicago,1939; Reissued by Dover 1973).

\bibitem{polybook} G.~P.~Horedt,
{\em Polytropes. Applications in Astrophysics and Related Fields} (Kluwer, Dordrecht,2004).

\bibitem{henon} M.~Henon,
``Numerical Experiments on the Stability of Spherical Stellar Systems,''
    Astron.\ Astrophys.\ {\bf 24}, 229 (1973).

\bibitem{Polyanin} A.~D.~Polyanin and V.~F.~Zaitsev,
{\em Handbook of Exact Solutions for Ordinary Differential Equations} (Chapman \& Hall/CRC, Boca Raton, 2003), 2nd ed.

\bibitem{Evans:2005zv}
  N.~W.~Evans and J.~An,
``Hypervirial models of stellar systems,''
  Mon.\ Not.\ Roy.\ Astron.\ Soc.\  {\bf 360}, 492 (2005)
 [astro-ph/0501091].

\bibitem{plummer} H.~C.~Plummer,
``On the problem of distribution in globular star clusters,''
    Mon. Not. Roy. Astron. Soc. \textbf{71}, 460 (1911).

\bibitem{Dejonghe87} H.~Dejonghe,
``A completely analytical family of anisotropic Plummer models,''
    Mon. Not. Roy. Astron. Soc. \textbf{224}, 13 (1987).

\bibitem{BT}  J.~Binney and S.~Tremaine,
{\em Galactic Dynamics} (Princeton University Press, New Jersey, 2008), 2nd ed.

\bibitem{Mahdavi:2002ds}
  A.~Mahdavi,
``A gravitational potential with adjustable slope: A Hydrostatic alternative to cluster cooling flows,''
  arXiv:astro-ph/0212300

\bibitem{Hernquist:1990be}
  L.~Hernquist,
``An Analytical Model For Spherical Galaxies And Bulges,''
  Astrophys.\ J.\  {\bf 356}, 359 (1990).

\bibitem{Baes:2002tw}
  M.~Baes and H.~Dejonghe,
``The Hernquist model revisited: Completely analytical anisotropic dynamical models,''
  Astron.\ Astrophys.\  {\bf 393}, 485 (2002)
  [astro-ph/0207233].

\bibitem{Tolman:1939jz}
  R.~C.~Tolman,
``Static solutions of Einstein's field equations for spheres of fluid,''
  Phys.\ Rev.\  {\bf 55}, 364 (1939).

\bibitem{Oppenheimer:1939ne}
  J.~R.~Oppenheimer and G.~M.~Volkoff,
``On Massive neutron cores,''
  Phys.\ Rev.\  {\bf 55}, 374 (1939).

\bibitem{Lynden Bell Eggleton}
  D. Lynden-Bell and P. P. Eggleton,
``On the consequences of the gravothermal catastrophe,''
  Mon. Not. Roy. Astron. Soc. {\bf 191}, 483 (1980).

\bibitem{Spitzer Thuan}
  L. Spitzer Jr. and T. X. Thuan,
``Random Gravitational Encounters and the Evolution of Spherical Systems. IV Isolated Systems of Identical Stars,''
  Astrophys. J. {\bf 175}, 31 (1972).

\bibitem{Merritt:1995}
  D.~Merritt and T.~Fridman,
  \emph{Fresh Views of Elliptical Galaxies},
  A. S. P. Conf. Ser. Vol. 86, edited by A. Buzzoni, A. Renzini, and A. Serrano (Provo, Astronomical Society of the Pacific, 1995).

\bibitem{Merritt:1995hp}
  D.~Merritt and T.~Fridman,
``Triaxial galaxies with cusps,''
  Astrophys.\ J.\ {\bf 460}, 136 (1996)
  [astro-ph/9511021].

\bibitem{Gebhardt:1996ik}
  K.~Gebhardt, D.~Richstone, E.~A.~Ajhar, T.~R.~Lauer, Y.~-I.~Byun, J.~Kormendy, A.~Dressler and S.~M.~Faber {\it et al.},
``The Centers of early-type galaxies with HST. 3. Non-parametric recovery of stellar luminosity distributions,''
  Astron.\ J.\  {\bf 112}, 105 (1996)
  [astro-ph/9604092].

\bibitem{Navarro:1995iw}
  J.~F.~Navarro, C.~S.~Frenk and S.~D.~M.~White,
``The Structure of cold dark matter halos,''
  Astrophys.\ J.\  {\bf 462}, 563 (1996)
 [astro-ph/9508025].

\bibitem{Navarro:1996gj}
  J.~F.~Navarro, C.~S.~Frenk and S.~D.~M.~White,
``A Universal density profile from hierarchical clustering,''
  Astrophys.\ J.\  {\bf 490}, 493 (1997)
 [astro-ph/9611107].

\bibitem{Vogt:2009gs}
  D.~Vogt and P.~S.~Letelier,
``Newtonian and General Relativistic Models of Spherical Shells,''
  Mon.\ Not.\ Roy.\ Astron.\ Soc.\  {\bf 402}, 1313 (2010)
 [arXiv:0911.4822 [gr-qc]].

\bibitem{Merrifield1990}
  M.~R.~Merrifield and S.~M.~Kent,
``Fourth moments and the dynamics of spherical systems,''
  Astron.\ J.\  {\bf 99}, 1548 (1990).

\bibitem{Gerhard1993}
  O.~E.~Gerhard,
``Line-of-sight velocity profiles in spherical galaxies: breaking the degeneracy between anisotropy and mass,''
  Mon. Not. Roy. Astron. Soc. \textbf{265}, 213 (1993).

\bibitem{Lokas:2003ks}
  E.~L.~Lokas and G.~A.~Mamon,
``Dark matter distribution in the Coma cluster from galaxy kinematics: Breaking the mass - anisotropy degeneracy,''
  Mon.\ Not.\ Roy.\ Astron.\ Soc.\  {\bf 343}, 401 (2003)
 [astro-ph/0302461].

\bibitem{Hunter2001}
  C.~Hunter,
``Series solutions for polytropes and the isothermal sphere,''
  Mon. Not. Roy. Astron. Soc. {\bf 328}, 839 (2001).

\bibitem{BatemanBook}
 H.~Bateman,
{\em Higher Transcendental Functions}
(McGraw-Hill Book Company, Inc., New York, 1955).

\bibitem{Rein}
  H. Andr\'{e}asson and G. Rein,
``On the steady states of the spherically symmetric EinsteinVlasov system''
  Class. Quantum Grav. {\bf 24} 1809 (2007).

\bibitem{Persic:1995ru}
  M.~Persic, P.~Salucci, and F.~Stel,
``The Universal rotation curve of spiral galaxies: 1. The Dark matter connection,''
  Mon.\ Not.\ Roy.\ Astron.\ Soc.\  {\bf 281}, 27 (1996)
 [astro-ph/9506004].

\bibitem{Begeman:1991iy}
  K.~G.~Begeman, A.~H.~Broeils, and R.~H.~Sanders,
``Extended rotation curves of spiral galaxies: Dark haloes and modified dynamics,''
  Mon.\ Not.\ Roy.\ Astron.\ Soc.\  {\bf 249}, 523 (1991).

\bibitem{Milgrom:1983ca}
  M.~Milgrom,
``A Modification of the Newtonian dynamics as a possible alternative to the hidden mass hypothesis,''
  Astrophys.\ J.\  {\bf 270}, 365 (1983).

\bibitem{Bekenstein:1984tv}
  J.~Bekenstein and M.~Milgrom,
``Does the missing mass problem signal the breakdown of Newtonian gravity?,''
  Astrophys.\ J.\  {\bf 286}, 7 (1984).

\bibitem{Mannheim:1988dj}
  P.~D.~Mannheim and D.~Kazanas,
``Exact Vacuum Solution To Conformal Weyl Gravity And Galactic Rotation Curves,''
  Astrophys.\ J.\  {\bf 342}, 635 (1989).

\bibitem{Mannheim:1992vj}
  P.~D.~Mannheim,
``Linear potentials and galactic rotation curves,''
  Astrophys.\ J.\  {\bf 419}, 150 (1993)
 [hep-ph/9212304].

\bibitem{Sofue:2000jx}
  Y.~Sofue and V.~Rubin,
``Rotation curves of spiral galaxies,''
  Ann.\ Rev.\ Astron.\ Astrophys.\  {\bf 39}, 137 (2001)
 [astro-ph/0010594].

\bibitem{Nucamendi:2000jw}
  U.~Nucamendi, M.~Salgado and D.~Sudarsky,
``An Alternative approach to the galactic dark matter problem,''
  Phys.\ Rev.\ D {\bf 63}, 125016 (2001)
 [gr-qc/0011049].

\bibitem{Lake:2003tr}
  K.~Lake,
``Galactic potentials,''
  Phys.\ Rev.\ Lett.\  {\bf 92}, 051101 (2004)
 [gr-qc/0302067].

\bibitem{Mak:2004hv}
  M.~K.~Mak and T.~Harko,
``Can the galactic rotation curves be explained in brane world models?,''
  Phys.\ Rev.\ D {\bf 70}, 024010 (2004)
 [gr-qc/0404104].

\bibitem{Capozziello:2004us}
  S.~Capozziello, V.~F.~Cardone, S.~Carloni, and A.~Troisi,
``Can higher order curvature theories explain rotation curves of galaxies?,''
  Phys.\ Lett.\ A {\bf 326}, 292 (2004)
 [gr-qc/0404114].

\bibitem{Reuter:2004nv}
  M.~Reuter and H.~Weyer,
``Running Newton constant, improved gravitational actions, and galaxy rotation curves,''
  Phys.\ Rev.\ D {\bf 70}, 124028 (2004)
 [hep-th/0410117].

  \bibitem{Moffat:2004bm}
  J.~W.~Moffat,
``Gravitational theory, galaxy rotation curves and cosmology without dark matter,''
  J. Cosmol. Astropart. Phys. {\bf 05} (2005) 003
 [astro-ph/0412195].

\bibitem{Martins:2007uf}
  C.~F.~Martins and P.~Salucci,
``Analysis of Rotation Curves in the framework of R**n gravity,''
  Mon.\ Not.\ Roy.\ Astron.\ Soc.\  {\bf 381}, 1103 (2007)
 [arXiv:astro-ph/0703243].

\bibitem{Chang:2008yv}
  Z.~Chang and X.~Li,
``Modified Newton's gravity in Finsler Space as a possible alternative to dark matter hypothesis,''
  Phys.\ Lett.\ B {\bf 668}, 453 (2008)
 [arXiv:0806.2184 [gr-qc]].

\bibitem{Alvarez:2012gx}
  A.~Herrera-Aguilar, U.~Nucamendi, E.~Santos, O.~Corradini, and C.~Alvarez,
``On the galactic rotation curves problem within an axisymmetric approach,''
  Mon.\ Not.\ Roy.\ Astron.\ Soc.\  {\bf 432}, 301 (2013)
 [arXiv:1206.6788 [astro-ph.GA]].

\bibitem{Lu:2013bt}
  J.~-A.~Lu and C.~-G.~Huang,
``Weak field approximation in a model of de Sitter gravity: Schwarzschild solutions and galactic rotation curves,''
  Gen. Relativ. Gravit. {\bf 45}, 691 (2013).
  [arXiv:1301.5796 [gr-qc]].

\bibitem{Stabile:2013jon}
  A.~Stabile and S.~Capozziello,
``Galaxy rotation curves in $f(R,\phi)$-gravity,''
  Phys.\ Rev.\ D {\bf 87}, 064002 (2013)
 [arXiv:1302.1760 [gr-qc]].

\bibitem{Alexandre:2013ura}
  J.~Alexandre and M.~Kostacinska,
``Galaxy Rotation Curves in Covariant Horava-Lifshitz Gravity,''
  arXiv:1303.1394.

\bibitem{Meierovich:2013jha}
  B.~E.~Meierovich,
``Galaxy rotation curves. The theory,''
  arXiv:1303.7062.

\bibitem{Chang:2013twa}
  Z.~Chang, M.~-H.~Li, X.~Li, H.~-N.~Lin, and S.~Wang,
``Effects of spacetime anisotropy on the galaxy rotation curves,''
  Eur. Phys. J. C {\bf 73}, 2447 (2013)
 [arXiv:1305.2314 [astro-ph.GA]].

\bibitem{Cooperstock-Tieu}
  F.~I.~Cooperstock and S.~Tieu,
``General relativity resolves galactic rotation without exotic dark matter,''
 arXiv:astro-ph/0507619.

\bibitem{Pro-Cooperstock1}
  F.~I.~Cooperstock and S.~Tieu,
``Perspectives on galactic dynamics via general relativity,''
  arXiv:astro-ph/0512048.

\bibitem{Pro-Cooperstock2}
  S.~Capozziello, V.~F.~Cardone, and A.~Troisi,
``Dark energy and dark matter as curvature effects,''
  J. Cosmol. Astropart. Phys. {\bf 08} (2006), 001
  [arXiv:astro-ph/0602349].

\bibitem{Pro-Cooperstock3}
  J.~C.~Jackson and M.~Dodgson,
``Deceleration without dark matter,''
  Mon.\ Not.\ Roy.\ Astron.\ Soc.\  {\bf 285}, 806 (1997)
  [arXiv:astro-ph/0605102].

\bibitem{Pro-Cooperstock4}
  M.~D.~Maia, A.~J.~S.~Capistrano, and D.~Mulller,
``Velocity Curves for Stars in Disk Galaxies: A case for Nearly Newtonian Dynamics,''
  arXiv:astro-ph/0605688.

\bibitem{Anticoperstok1}
  M.~Korzynski,
``Singular disk of matter in the cooperstock and tieu galaxy model,''
  arXiv:astro-ph/0508377.

\bibitem{Anticoperstok2}
  D.~Vogt and P.~S.~Letelier,
``Presence of exotic matter in the cooperstock and tieu galaxy model,''
  arXiv:astro-ph/0510750.

\bibitem{Anticoperstok3}
  D.~Garfinkle,
``The Need for dark matter in galaxies,''
  Class.\ Quant.\ Grav.\  {\bf 23}, 1391 (2006)
 [gr-qc/0511082].

\bibitem{Anticoperstok4}
  D.~Menzies and G.~J.~Mathews,
``General Relativistic Galaxy Rotation Curves: Implications for Dark Matter Distribution,''
  arXiv:gr-qc/0604092.

\bibitem{Anticoperstok5}
  B.~Fuchs and S.~Phleps,
``Comment on `General Relativity Resolves Galactic Rotation Without Exotic Dark Matter' by F.I. Cooperstock and S. Tieu,''
  New Astron.\  {\bf 11}, 608 (2006)
 [arXiv:astro-ph/0604022].

\bibitem{Barnes}
  J.~Barnes, G.~Goodman and P.~Hut,
``Dynamical instabilities in spherical stellar systems,''
  Astrophys.\ J.\ {\bf 300}, 112 (1986).

\bibitem{Guo}
  Y.~Guo,
``On the generalized Antonov's stability criterion,''
  Contemp.\ Math.\ {\bf 150}, 85 (2000).

\bibitem{Kandrup & Sygnet}
  H.~E.~Kandrup and J.~F.~Sygnet,
``A simple proof of dynamical stability for a class of spherical clusters,''
  Astrophys.\ J.\ {\bf 298}, 27 (1985).

\bibitem{Merritt & Aguilar}
  D.~Merritt and L.~Aguilar,
``A numerical study of the stability of spherical galaxies,''
  Mon. Not. Roy. Astron. Soc. \textbf{217}, 787 (1985).

\bibitem{Polyachenko}
  E.~V.~Polyachenko, V.~L.~Polyachenko, and I.~G.~Shukhman,
``Notes on the stability threshold for radially anisotropic polytropes,''
  Mon. Not. Roy. Astron. Soc. \textbf{416}, 1836 (2011)
 arXiv:1104.0741 [astro-ph.SR].

\bibitem{Lord}
  L.~Rayleigh,
``On the dynamics of revolving fluids''
  Proc. R. Soc. Lond. Ser. A \textbf{93}, 148 (1917).

\bibitem{Chandra64}
S.~Chandrasekhar,
``The Dynamical Instability of Gaseous Masses Approaching the Schwarzschild Limit in General Relativity,''
Astrophys.\ J.\ {\bf 140}, 417 (1964).

\bibitem{Le Delliou}
  M. Le Delliou, R.~N. Henriksen, and J.~D. MacMillan,
``Black Holes and Galactic Density Cusps : From Black Hole to Bulge,''
  Astron. Astrophys. {\bf 526}, A13 (2011)
 [arXiv:0911.2238 [astro-ph.GA]].

\bibitem{Bahcall Wolf}
  J. N. Bahcall and R. A. Wolf,
``Star distribution around a massive black hole in a globular cluster,''
  Astrophys. J. {\bf 209}, 214 (1976).


\end{thebibliography}
\end{document}